\documentclass[fleqn,usenatbib]{mnras}
\usepackage[T1]{fontenc}
\usepackage{ae,aecompl}
\usepackage{graphicx}	
\usepackage{amsmath}	
\usepackage{amssymb}
\usepackage{pdflscape}
\usepackage{siunitx}
\usepackage{dcolumn}
\usepackage{multicol} 
\usepackage{hyperref}

\title[IR properties of star-forming galaxies]{{On the Dust properties of the UV galaxies in the redshift range $z \sim 0.6-1.2$}}

\author[M. Sharma et al.]
       {{M. Sharma$^{1,2}$\thanks{E-mail: mnushv@gmail.com (MS)},
           M. J. Page$^{1}$,
           M. Symeonidis$^{1}$},
           I. Ferreras$^{3,4,5}$\\
         $^1$Mullard Space Science Laboratory,
         University College London,
         Holmbury St Mary, Dorking, Surrey, RH5 6NT, UK\\
         $^2$Isaac Newton Group of Telescopes,
         C. Álvarez Abreu, 70, E38700 Santa Cruz de La Palma, 
         La Palma, Spain\\
         $^3$Instituto de Astrofísica de Canarias, 
         Calle Vía Láctea s/n, E38205, 
         La Laguna, Tenerife, Spain\\
         $^4$Departamento de Astrofísica, 
         Universidad de La Laguna, E38206, 
         La Laguna, Tenerife, Spain\\
         $^5$Department of Physics and Astronomy, 
         University College London, 
         Gower Street, London WC1E 6BT, UK}
         
\date{Accepted 2024 January 09; Received 2024 January 08; in original form 2023 November 14}
       
\pubyear{2024}

\begin{document}
\label{firstpage}
\pagerange{\pageref{firstpage}--\pageref{lastpage}}
\maketitle

\begin{abstract}
Far-infrared observations from the \textit{Herschel Space Observatory} are used to estimate the infrared (IR) properties of ultraviolet-selected galaxies. We stack the PACS (100,
160 $\mu \mathrm{m}$) and SPIRE (250, 350 and 500$\mu \mathrm{m}$)
maps of the Chandra deep field south (CDFS) on a source list of galaxies
selected in the rest-frame ultraviolet (UV) in a redshift range of $0.6-1.2$. 
This source list is created using observations from the XMM-OM telescope survey in the CDFS using the UVW1 (2910 {\AA}) filter.
The stacked data are binned according to the UV luminosity function of these sources,
and the average photometry of the UV-selected galaxies is estimated. By fitting modified black bodies and IR model templates to the stacked photometry, average dust temperatures and total IR luminosity are determined. 
The luminosity-weighted average temperatures are consistent with a weak trend of increasing temperature with redshift found by previous studies.
Infrared excess, unobscured, and obscured star formation rate (SFR) values are obtained from the UV and IR luminosities. We see a trend in which dust attenuation increases as UV luminosity decreases. It remains constant as a function of IR luminosities at fixed redshift across the luminosity range of our sources.
In comparison to local luminous infrared galaxies with similar SFRs, the higher redshift star-forming galaxies in the sample show a lesser degree of dust attenuation. Finally, the inferred dust attenuation is used to correct the unobscured SFR density in the redshift range 0.6-1.2. The dust-corrected SFR density is consistent with measurements from IR-selected samples at similar redshifts.

\end{abstract}


\section{Introduction}
\label{sec:1}

Star formation is controlled by various fundamental processes and is one of the key global mechanisms for galaxy evolution. It has shaped the galaxies as we observe them today. 
In other words, it plays a crucial role in the evolutionary history of galaxies, and constraining the star formation rate (SFR) is important for understanding this evolution.

The ultraviolet (UV) continuum in the spectral energy distribution (SED) of star-forming galaxies
is produced by young massive stars, and it is widely used as one of the most important indicators of the SFR \citep[][]{2012ARA&A..50..531K}. It has been employed by studies constraining the luminosity density in the
nearby Universe \citep[e.g.][]{2005ApJ...619L..15W,2005ApJ...619L..31B} as well as
at low \citep[e.g.][]{2000MNRAS.312..442S},
intermediate \citep[e.g.][]{2010ApJ...725L.150O,2021MNRAS.506..473P,2022MNRAS.511.4882S} and high redshifts \citep[][]{2016MNRAS.456.3194P,2015ApJ...803...34B,2023MNRAS.518.6011D}.
The UV continuum can also be produced by AGN, which makes it important to consider their 
identification in a sample of star-forming galaxies under consideration.

Rest frame UV radiation is particularly susceptible to being obscured by dust in star-forming regions of a galaxy. 
The UV flux is scattered and/or absorbed by dust particles, which then reemit this energy as thermal black-body radiation in the far-infrared (FIR) wavelength range.
In the local Universe, for near-UV (NUV) selected sources \citet{2005ApJ...619L..51B} and \citet{2006MNRAS.365..352B} found the typical dust attenuation in far-UV (FUV) to be 1.1 and 1.4 mags, respectively. 
The effect becomes more severe with an increase in
redshift and a larger fraction of UV radiation is absorbed by dust at higher redshifts compared to the local Universe. 
According to a study by \citet{2005A&A...440L..17T}, the portion of the far-UV SFR that is obscured by dust increases from 56 per cent in the nearby Universe to 84 per cent at an average redshift of 1. Therefore, correcting UV luminosities for dust attenuation is essential before it can be used to estimate the SFR.

One of the suggested methods to solve this issue of dust attenuation involves utilizing the empirical correlation between 
UV dust attenuation ($A_{\mathrm{FUV}}$) and the slope ($\beta$) of the UV continuum described by a power law \citep[$f_{\lambda} \propto \lambda^{\beta}$; ][]{1994ApJ...429..582C,2000ApJ...533..682C,1999ApJ...521...64M,Overzier_2010}.
To quantify the UV attenuation, this approach involves the ratio of the FIR to the UV
luminosity, also known as the infrared (IR) excess or $\mathrm{IRX}$.
Many studies that rely solely on UV data have utilised this relationship as a standard practise in order to correct their estimates of SFR or luminosity density (LD) for the effects of dust attenuation \citep[e.g.][]{2005ApJ...619L..47S,2009ApJ...705..936B,2012ApJ...756..164F,2015ApJ...803...34B,2020ApJ...902..112B}.

The central concept underlying this approach is that all galaxies possess the same intrinsic spectral slope ($\beta$), which can only be altered by the presence of dust obscuration. 
While this assumption may seem to hold in the nearby Universe \citep[][]{1999ApJ...521...64M}, 
the spectral slope can also be influenced by a variety of other factors, including the redshift, the initial mass function (IMF), the metallicity, and other quantities, in addition to dust obscuration \citep[][]{2012MNRAS.424.1522W,2012MNRAS.427.1490W,2018MNRAS.475.2363T}.
In addition, the UV continuum slope is generally bluer than the assumed inherent value in the Meurer relation \citep[][]{2013MNRAS.430.2885W}.
As a result of these various factors, this relation, which was initially calibrated for starburst galaxies \citep[][]{1999ApJ...521...64M}, may be subject to modification when applied to normal galaxies, depending on various global factors such as age \citep[e.g.][]{2012ApJ...744..154R,2018MNRAS.474.1718N}, stellar mass \citep{2010ApJ...712.1070R,2018ApJ...853...56R,2020MNRAS.491.4724F}, luminosity, and SFR of the galaxy \citep[e.g.][]{2014ApJ...796...95C}.
Furthermore, this relationship has been found to be influenced by local properties such as the geometry of the dust-emitting region \citep[e.g.][]{2000ApJ...528..799W,2018MNRAS.474.1718N}, and it has been discovered that the specific shape of this relationship can depend on the extinction law \citep[e.g.][]{2018MNRAS.474.1718N} and the source selection criteria \citep[][]{2005ApJ...619L..51B}.

The other more direct and reliable way to measure dust attenuation is through $\mathrm{IRX}$, which is calculated as the ratio of the luminosity in the IR region to the luminosity in the ultraviolet region of the SEDs of the galaxy
\citep[][]{1999ApJ...521...64M}. 
This method works because dust absorbs UV light and re-emits it as thermal radiation in the IR bands. By measuring the IR emission, it is possible to determine
how much UV light has been absorbed by dust. This information can be used to correct UV/optical observations for dust attenuation, and the FIR flux can also be used as a proxy for the amount of obscured star formation in a galaxy. \citep[][]{1992A&A...264..444B,1995A&A...293L..65X,1995AJ....110.2665M,1998ApJ...503..646H,2000ApJ...533..236G}.
Additionally, the use of both UV and IR observations provides a way to trace both attenuated and unattenuated star formation in a galaxy as a composite measure of star formation \citep[][]{2007ApJ...666..870C}.

The $\mathrm{IRX}$ ratio is based on the relative brightness of ultraviolet (UV) radiation that has not been absorbed by dust in a system compared to the FIR radiation that has been absorbed and reemitted by dust. The basic premise behind the $\mathrm{IRX}$ ratio is that there should be a balance between the UV/optical light absorbed by dust and the FIR radiation emitted \citep[][]{1996A&A...306...61B,1999A&A...352..371B}. However, this balance may not always be straightforward in practise and may be influenced by the age of the dust heating system.
Although this method has the advantage of being independent of other IR properties and star-dust geometry \citep[][]{2000ApJ...533..236G,2000ApJ...528..799W,2008MNRAS.386.1157C}, it is not as widely used as the previous method based on the Meurer relation due to the ease of accessing data to calculate the UV spectral slope and the lack of deep FIR data at high redshifts. 
However, if deep FIR data is available, the $\mathrm{IRX}$ method can be a powerful tool for studying dust attenuation in different types of systems \citep[][]{1999A&A...352..371B} and may provide more robust results than the other method that relies on more uncertain assumptions.

The method of using the $\mathrm{IRX}$ ratio to estimate dust attenuation in galaxies has been widely studied in the literature. 
A study using this method, conducted by \citet{2005ApJ...619L..51B}, analysed a sample of galaxies selected by the \textit{Galaxy Evolution Explorer (GALEX)} in the near-ultraviolet (NUV) band and \textit{Infrared Astronomy Satellite (IRAS)} data at 60 $\mu \mathrm{m}$, and found that the mean dust attenuation in the FUV was 1.6 magnitudes in the nearby Universe.
Other studies have extended this to higher redshifts, finding that the $\mathrm{IRX}$ ratio as a function of bolometric luminosity ($L_{\mathrm{IR}}+L_{\mathrm{UV}}$) of the galaxies evolves to redshift 1 for Lyman break galaxies \citep[][]{2007MNRAS.380..986B} and redshift 2 for BM/BX galaxies \citep[][]{2006ApJ...653.1004R}. 
However, \citet{2009A&A...507..693B} did not see a clear evolution in the $\mathrm{IRX}$ ratio at fixed bolometric luminosity up to a redshift of 1 in their homogeneously selected sample of galaxies from \textit{GALEX}, and suggested that it might be more useful to look at the $\mathrm{IRX}$ ratio as a function of UV luminosity rather than bolometric luminosity.

One challenge in studying the dust attenuation in galaxies at high redshifts is the availability of FIR data, as current IR telescopes have limited sensitivity and resolution, making these observations scarce and restricted to the most massive galaxies. Stacking analysis, in which multiple data with a lower signal-to-noise ratio are combined to increase the overall signal-to-noise ratio \citep[][]{2006A&A...451..417D,2009ApJ...707.1729M,2010A&A...516A..43B,2010AJ....139.1592K,2012MNRAS.419.2758R,2013ApJ...779...32V}, has been used in some studies to estimate the UV attenuation due to dust at higher redshifts.

Some examples of studies that have used stacking analysis to estimate UV attenuation due to dust at higher redshifts include those by \citet{2007ApJS..173..432X}, who extended $\mathrm{IRX}$-based dust attenuation estimates to redshift 0.6 using data from the \textit{GALEX} survey and the \textit{Spitzer Space Telescope} at 24 $\mu \mathrm{m}$, and \citet{2013MNRAS.429.1113H}, who used data from the Canada-France-Hawaii Telescope (CFHT) $u^*$-band imaging in the Cosmic Evolution Survey (COSMOS) field to estimate $\mathrm{IRX}$ at redshift $\sim 1.5$. 
They found mean $\mathrm{IRX}$ to be 6.6 and 6.9 respectively.
These measurements were taken to redshifts around 2 by \citet{2012ApJ...744..154R}, who used UV-selected galaxies from the Low-Resolution Imaging Spectrograph (LRIS) on the Keck telescope \citep[][]{Steidel_2004,2006ApJ...653.1004R} and obtained an $\mathrm{IRX}$ value of 7.1, indicating that only 20 per cent of the star formation is not dust obscured. 

Other works have also studied the $\mathrm{IRX}$ ratio at higher redshifts with stacking methods on different data sets.
\citet{2016A&A...587A.122A} calculated the $\mathrm{IRX}$ ratio at redshift $\sim 3$ using $u^*$, $V_{\mathrm{J}}$ and $i^+$ band imaging and \textit{Herschel Space Observatory} (hereafter \textit{Herschel}) maps in the COSMOS field and estimated an average value of 7.9. 
Using Lyman break galaxy candidates at average redshifts 3.8 from the NOAO Deep Wide-Field Survey of the Boötes field, \citet{2012ApJ...758L..31L} found $\mathrm{IRX}$ values of 3 to 4, implying that 30 to 40 per cent of the star formation occurs without any dust attenuation. 
More recently, \citet{2018ApJ...853...56R} used the \textit{Hubble Space Telescope (HST)} data from the 3D-HST survey \citep[][]{2014ApJS..214...24S} and the Hubble Deep UV (HDUV) Legacy Survey \citep[][]{2018ApJS..237...12O}, along with \textit{Spitzer} MIPS 24 $\mu \mathrm{m}$ and \textit{Herschel} PACS 100 and 160 $\mu \mathrm{m}$ data, to calculate an average $\mathrm{IRX}$ value of 2.94 for redshifts between 1.5 and 2.5 in the Great Observatories Origins Deep Survey (GOODS) fields.

For redshifts up to 1, similar studies have been performed using only \textit{GALEX} data for UV 
selection, which can
be subjected to issues related to source confusion due to its poor spatial resolution.
More studies are needed to revisit this redshift range with better data and determine whether the trend observed in previous studies holds at these intermediate redshifts.

In this study, we used the UV-selected sample \citep[][]{2022MNRAS.511.4882S} of galaxies from the Chandra Deep Field South (CDFS) survey of the \textit{XMM-Newton} Optical Monitor \citep[XMM-OM, Optical Monitor;][]{2001A&A...365L..36M} onboard the \textit{XMM-Newton} observatory. 
The UVW1 filter ($\lambda_{\mathrm{eff}}=2910$ {\AA}) of the XMM-OM telescope provides rest-frame 1500 {\AA} imaging in the redshift 
range of our interest (0.6-1.2),
over a field of view of 17 $\times$ 17 sq. arcminutes.
These galaxies are stacked on the FIR maps at 100, 160, 250, 350 and 500 $\mu \mathrm{m}$
from \textit{Herschel} to obtain the average FIR flux of the galaxies in bins of the UV luminosity function created for these galaxies by \citet{2022MNRAS.511.4882S}.
Using the integrated IR luminosity we calculate the $\mathrm{IRX}$ ratio and then use it to calculate the dust attenuation of the FUV radiation, which in turn is used to
correct the SFR density (SFRD) calculated using the UV measurements
of \citet{2022MNRAS.511.4882S}.

This paper is structured as follows. 
The data used in this work are explained in Section \ref{sec:2}. 
We describe our methods in Section \ref{sec:3}. In particular the deblending process for the SPIRE maps in Section \ref{sec:3.1}, stacking the FIR maps and
extraction of the average photometry from the stacks in Section \ref{sec:3.2},
fitting the IR model templates to the average IR flux densities and the estimation of
dust properties like total IR luminosity and dust temperature in Section \ref{sec:3.3}
We discuss the methods to obtain the star formation rates (un-obscured and obscured using the UV and FIR tracers) and the average dust attenuation of the UV light in Sections \ref{sec:3.4}
and \ref{sec:3.5}.
We summarise our results in Section \ref{sec:4} and discuss their implications in Section \ref{sec:5}. 
Finally, we conclude this paper in Section \ref{sec:6}.
Throughout the paper, we adopt a flat cosmology with $\Omega_{\Lambda}=0.7$, $\Omega_{M}=0.3$ and Hubble's constant $H_0=70$\,km\,s$^{-1}$\,Mpc$^{-1}$. The distances (and volumes) are calculated in comoving coordinates in Mpc (and Mpc$^{3}$). We use the AB system of magnitudes \citep{1983ApJ...266..713O}. The FIR luminosities are integrated over $8-1000\mu m$, and the FUV luminosities are defined as $\lambda L_\lambda$, where $\lambda = 1500$ {\AA}.

\section{Data}
\label{sec:2}

\begin{figure}
    \centering
    \includegraphics[width=0.95\columnwidth]{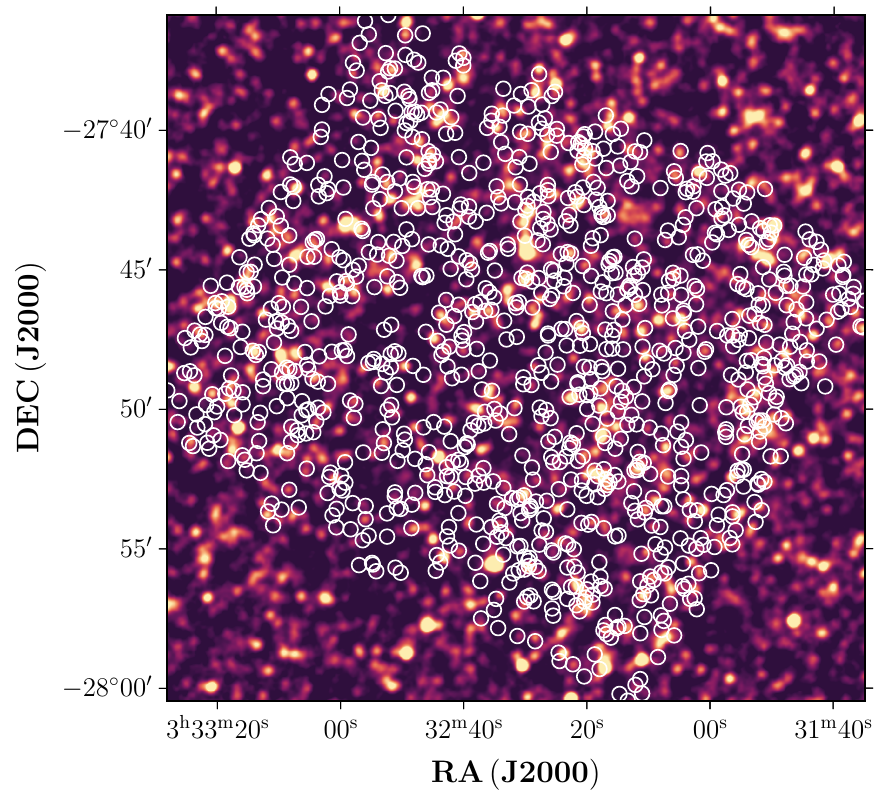}
    \caption{The UV source positions overlaid on the 250 $\mu \mathrm{m}$
    SPIRE map. A total of 1070 sources (ignoring UV bins with < 25 sources) with a S/N above 3, from an area $\sim 400$ sq. arcmins in the CDFS are used in this study.}
    \label{fig:spire250_uvlist}
\end{figure}

\begin{figure}
    \centering
    \hspace*{-0.15cm}\includegraphics[width=\columnwidth]{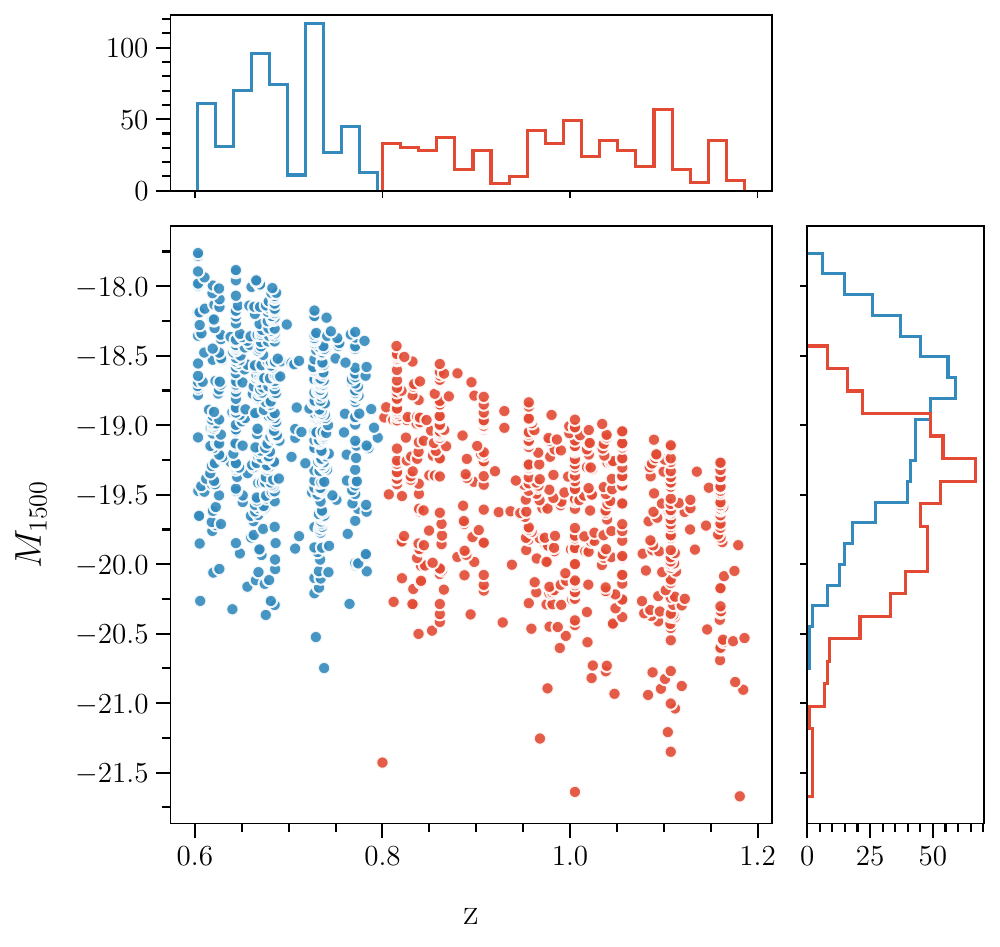}
    \caption{Rest frame UV ($M_{1500}$) absolute magnitudes and redshift distribution of the UVW1 sources stacked in this study. Blue and red colours represent the sources in the redshift bin centered at 0.7 and 1.0 respectively.
    The faint magnitude end of the distribution is truncated because of the apparent magnitude limit of our flux-limited survey in the CDFS. The apparent magnitude limit is UVW1$=24.5$ mag.}
    \label{fig:m1500z}
\end{figure}

\begin{table}
  \setlength{\tabcolsep}{8pt}
  \centering
  \caption{Properties of the \textit{Herschel} SPIRE and PACS data used in this study. The $1\sigma$ noise for the SPIRE maps is the total noise including both the instrumental and confusion
  components.}
  \label{tab:data_prop}
  \begin{tabular}{lclr}
    \hline\hline
    \noalign{\vskip 0.5mm}
    Central $\lambda$ &
    Pixel Scale &
    FWHM &
    $3 \sigma$ depth \\
    ($\mu$m) &
    (arcsec) &
    (arcsec) &
    (mJy) \\
    \hline
    \noalign{\vskip 0.5mm}
    100  & 1.2        & 7.2     & 4.50$^{a}$ \\
    160  & 2.4        & 12      & 8.50$^{a}$ \\
    250  & 6.0        & 18.15   & 6.72$^{b}$ \\
    350  & 8.33       & 25.15   & 5.58$^{b}$ \\
    500  & 12.0       & 36.30   & 8.04$^{b}$ \\
    \hline
    \noalign{\vskip 0.5mm}
  \end{tabular}\\
  \begin{minipage}{7.0cm}
      \raggedright
      \textsuperscript{$a$}{From \citet{2013MNRAS.432...23G}.}\\
      \textsuperscript{$b$}{From \citet{2013ApJ...779...32V}.}
  \end{minipage}
\end{table}

\begin{figure*}
    \centering
    \hspace*{-0.4cm}\includegraphics[width=1.06\textwidth]{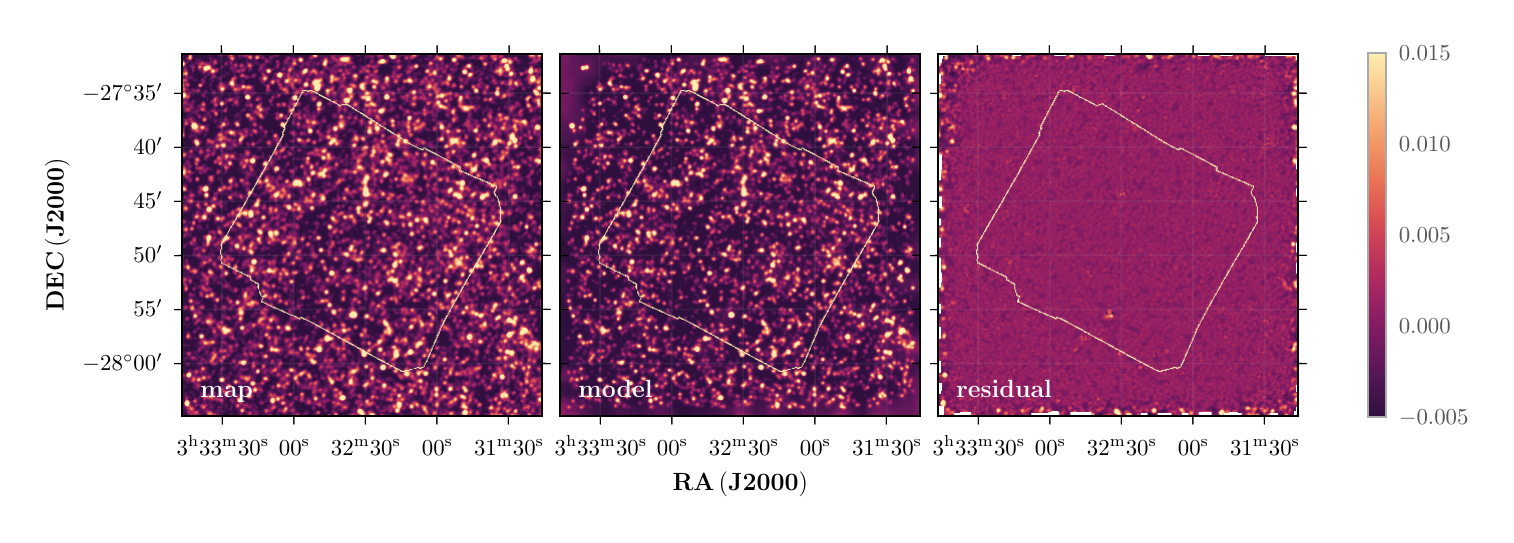}
    \caption{We depict the deblending process here as described in Section \ref{sec:3} with an example of the 250 $\mu \mathrm{m}$ map from \textit{Herschel} SPIRE. The \textit{left} panel shows the original map. In the \textit{middle} panel we show our model for the 250 $\mu \mathrm{m}$ map, and the \textit{ right} hand panel shows the residual map created by subtracting the model from the original map.
    The colour bar represents the pixel values of the flux density calibrated in units of Jy/beam.}
    \label{fig:deb250}
\end{figure*}

For this study, we have employed the data products obtained from the XMM-OM and \textit{Herschel} telescopes, specifically focusing on the Chandra Deep Field South (CDFS). This particular field, which is centred at RA 3h 32m 28.0s DEC $-27^{\circ} 48' 30"$ (J2000.0) \citep{2002ApJ...566..667R} in the southern sky, has been the primary target of observation for the Chandra X-ray Observatory \citep{2008ApJS..179...19L}.
Over the last two decades, this field has been the subject of extensive observation through a variety of multi-wavelength surveys, and as such, a plethora of ancillary information has been accumulated.

\subsection{UV selected galaxies in the CDFS}
\label{sec:2.1}

The UVW1 filter, characterised by an effective wavelength of 2910 {\AA}, of the XMM-OM, can be used to generate samples that comprise star-forming galaxies in the redshift range of 0.6 to 1.2.
This has been demonstrated in studies conducted in the 13 Hr field \citep{2021MNRAS.506..473P}, as well as in the CDFS \citep{2022MNRAS.511.4882S} and COSMOS \citep[][submitted]{sharma2023brightend} fields. 
In the aforementioned study \citet{2022MNRAS.511.4882S}, we used the UV imaging capabilities of the XMM-OM. The CDFS was observed over a decade, from 2001 to 2010 using the UVW1 filter. 
The data from this extensive observing campaign enabled us to create a deep ultraviolet image of the CDFS, which covers an area of approximately 400 sq. arcminutes. 
This image was subsequently used to create a source list of galaxies by extracting photometry in the rest frame 1500 {\AA}, spanning a redshift range of $0.6-1.2$.
The UVW1 filter can also select stars and AGNs due to their UV emissions. Quasars in particular, where the UV radiation from the accretion disc around the supermassive black hole outshines the stars in the host galaxy, could contaminate our samples. Such AGN as well as the stars have been removed using their X-ray emission.
By leveraging supplementary data from other catalogues within the CDFS, a UV catalogue that comprised 1079 galaxies, with a signal-to-noise of $>3$, was compiled. 
The sources of the supplimentary data are mentioned in Section 4.2 and a list of catalogues
used for photometric and spectroscopic redshifts is provided in Table 2 of \citet{2022MNRAS.511.4882S}.
The sample produced through this process is used in this investigation to study the average IR properties of these star-forming galaxies.
Figure \ref{fig:m1500z} shows this sample in the rest-frame magnitude-redshift space.

\subsection{FIR observations of the CDFS}
\label{sec:2.2}

The data used in this analysis are sourced from \textit{Herschel} \citep{2010A&A...518L...1P}, specifically utilising data taken by the Spectral and Photometric Imaging Receiver \citep[SPIRE;][]{2010A&A...518L...3G} and the Photodetector Array Camera and Spectrometer \citep[PACS;][]{2010A&A...518L...2P} instruments.

\subsubsection{\textit{Herschel} SPIRE}

The SPIRE data were obtained at 250 $\mu \mathrm{m}$, 350 $\mu \mathrm{m}$, and 500 $\mu \mathrm{m}$ as part 
of the \textit{Herschel} Multi-tiered Extragalactic Survey \citep[HerMES;][]{2012MNRAS.424.1614O}. 
The maps used in this analysis were taken from HeDam\footnote{\url{https://hedam.lam.fr/HerMES/index/dr4}} and were observed to a $3\sigma$ depth of 6.72, 5.58, and 8.04 mJy, respectively, without taking into account confusion noise \citep[][]{2013ApJ...779...32V}. 
The confusion noise for these SPIRE maps, as calculated by \citet{2010A&A...518L...5N}, is determined to be 5.8, 6.3, and 6.8 mJy at $1\sigma$ level for the 250 $\mu \mathrm{m}$, 350 $\mu \mathrm{m}$, and 500 $\mu \mathrm{m}$ filters, respectively.
Therefore, these maps, due to their larger beam size, are limited by confusion noise. 
In Figure \ref{fig:spire250_uvlist}, we present a plot of our UV source list, which is overlaid on top of the 250 $\mu \mathrm{m}$ map that was obtained from the SPIRE instrument.

\subsubsection{\textit{Herschel} PACS}

The PACS data were obtained as part of the PACS Evolutionary Probe survey \citep[PEP\footnote{\url{https://www.mpe.mpg.de/ir/Research/PEP/DR1}};][]{2011A&A...532A..90L}, at wavelengths of 100 $\mu \mathrm{m}$ and 160 $\mu \mathrm{m}$. 
The particular area of the sky that is the focus of our analysis, characterised by UVW1 sources, is observed as part of the Extended Chandra Deep Field South leg of the PEP survey.  
The overall field has been observed to a $3\sigma$ depths of 4.5 and 8.5 mJy \citep{2013MNRAS.432...23G}. 
It is worth noting that a portion of the field covered by our sources, specifically the GOODS-S region, is observed to deeper fluxes (1.2 and 2.4 mJy at the $3\sigma$ level). However, for the purpose of maintaining uniformity in terms of depth for all sources, these deeper data have not been included in the present analysis.
In contrast to the SPIRE maps, which are limited by confusion noise, the PACS maps, due to their small beam size, are limited by instrumental noise.

\section{Methods}
\label{sec:3}
\subsection{Deblending SPIRE maps}
\label{sec:3.1}

When there are a large number of sources situated in close proximity to each other, it can be challenging to accurately distinguish and identify them as individual entities. 
This situation can arise when the sources are so close to each other that they appear to blend together and appear as a single source.
This can have a significant impact on source identification and compromise the accuracy of the identified source positions, which in turn affects the cross-matching with other catalogues.
When two or more sources are blended together, the measurements of flux density can be overestimated, which can skew the calculations of derived estimates such as dust temperatures and total IR luminosities.
In addition, when sources appear separated but are still close together, the emission from the wings of one source may be incorrectly attributed to another nearby source. 
The SPIRE instrument, in particular, is affected by this blending due to its coarse beam, and, as a result, its FIR maps need to be corrected before they can be used to calculate flux densities.
Furthermore, the clustering of sources in the IR sky can have a significant impact on the stacking measurements performed on such sources, as it has the potential to contribute at the wings of the stacked signals and thus boost the overall flux \citep[][]{2006A&A...451..417D,2010A&A...516A..43B,2010AJ....139.1592K,2012A&A...542A..58B,2013MNRAS.429.1113H,2016A&A...587A.122A}.
In particular, \citet{2012A&A...542A..58B} carried out an estimation of clustering contribution for a sample selected at 24 $\mu \mathrm{m}$, and found that stacked flux measurements are boosted by approximately 8, 10, and 19 per cent at 250, 350 and 500 $\mu \mathrm{m}$, respectively.

To address these issues, we employ a technique known as deblending to correct the SPIRE maps. The basic concept behind this process is relatively straightforward. We model the SPIRE maps using the positions of sources in the 24 $\mu \mathrm{m}$ and radio catalogues, under the assumption that the majority of sources detected in the SPIRE (250, 350 and 500 $\mu \mathrm{m}$) bands should have a corresponding detection in these bands. 
To this end, a comprehensive prior catalogue is created utilising the 24 $\mu \mathrm{m}$ catalogue of the CDFS as part of the Far-Infrared Deep Extra-galactic Legacy (FIDEL) Survey. For the area of the CDFS that overlaps with GOODS-South, we use a more detailed and deep catalogue from \citet{2011A&A...528A..35M} in place of the sources from the CDFS. 
However, it is important to note that some of the SPIRE sources may not have been detected in the 24 $\mu \mathrm{m}$ band, thus, to make the prior catalogue as complete as possible, radio catalogues from \citet{2013ApJS..205...13M} and \citet{2015MNRAS.453.4020F} are also employed.
For the 250 and 350 $\mu \mathrm{m}$ bands, we use sources that are brighter than 30 $\mu$Jy and have a signal-to-noise ratio of at least $3$ and $5$, respectively, from the FIDEL and GOODS-S catalogues. 
To avoid over-deblending of the 500 $\mu \mathrm{m}$ maps, it is necessary to use a source list with a relatively low number density of sources. To accomplish this, we create a separate prior source list specifically for the 500 $\mu \mathrm{m}$  maps by applying more stringent constraints to the 24 $\mu \mathrm{m}$ FIDEL and GOODS-S 24 $\mu$m catalogues.
For the 500 $\mu \mathrm{m}$ prior list, we use sources with fluxes > 40 $\mu$Jy at $5\sigma$
for both the FIDEL and GOODS-S catalogues.

The deblending process is performed in two steps, with the first step using the prior catalogue to produce an initial set of models for the SPIRE maps. 
However, it is likely that some sources may still be missed due to the incompleteness of the prior catalogues. 
So, in order to improve our models, we undertake a second run of the process, this time utilising a modified prior catalogue. Any sources missing from the initial catalogue are identified through source detection in the residual maps produced in the first step. 
This is accomplished using SExtractor \citep[][]{1996A&AS..117..393B} with a detection threshold of 3, 3.5 and $4\sigma$, respectively, for the 250, 350 and 500 $\mu \mathrm{m}$ maps.
These newly detected sources are then added to the original catalogue, and the entire modelling process is re-run. The resulting models, along with the original and residual maps from the second run, are illustrated in Figure \ref{fig:deb250} (for the 250 $\mu \mathrm{m}$ band). 
The equivalent Figures for 350 and 500 $\mu \mathrm{m}$ are shown in Appendix~\ref{sec:deb_maps_stacks}.
Similar methods have been used in several previous studies, such as those conducted by \citet{2014MNRAS.438.1267S}, \citet{2017ApJ...838..119T}, and \citet{2018ApJ...853..172L}.

The sources in the final prior catalogue that are also present in the UV source list are added back to the residual maps to preserve their FIR contribution because, as explained in the next Section, the stacking and average photometry are performed on the residual maps.

\begin{figure*}
    \centering
    \hspace*{-0.25cm}\includegraphics[width=0.5\textwidth]{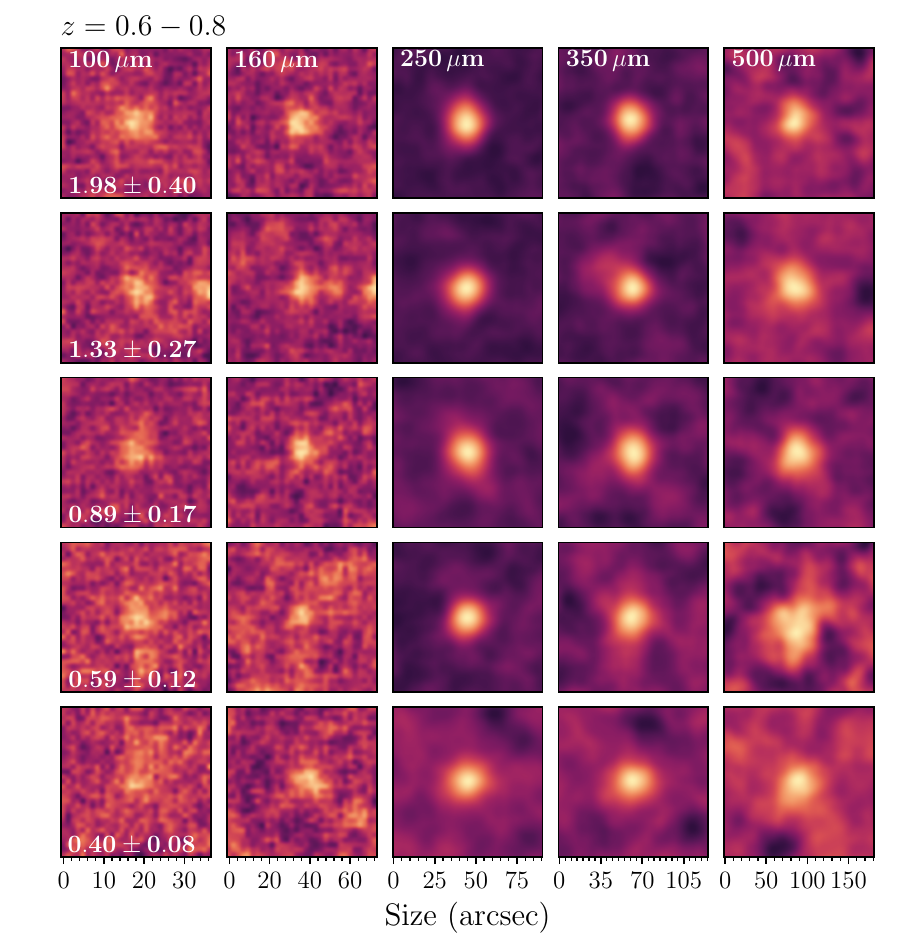}
    \hspace*{0.0cm}\includegraphics[width=0.5\textwidth]{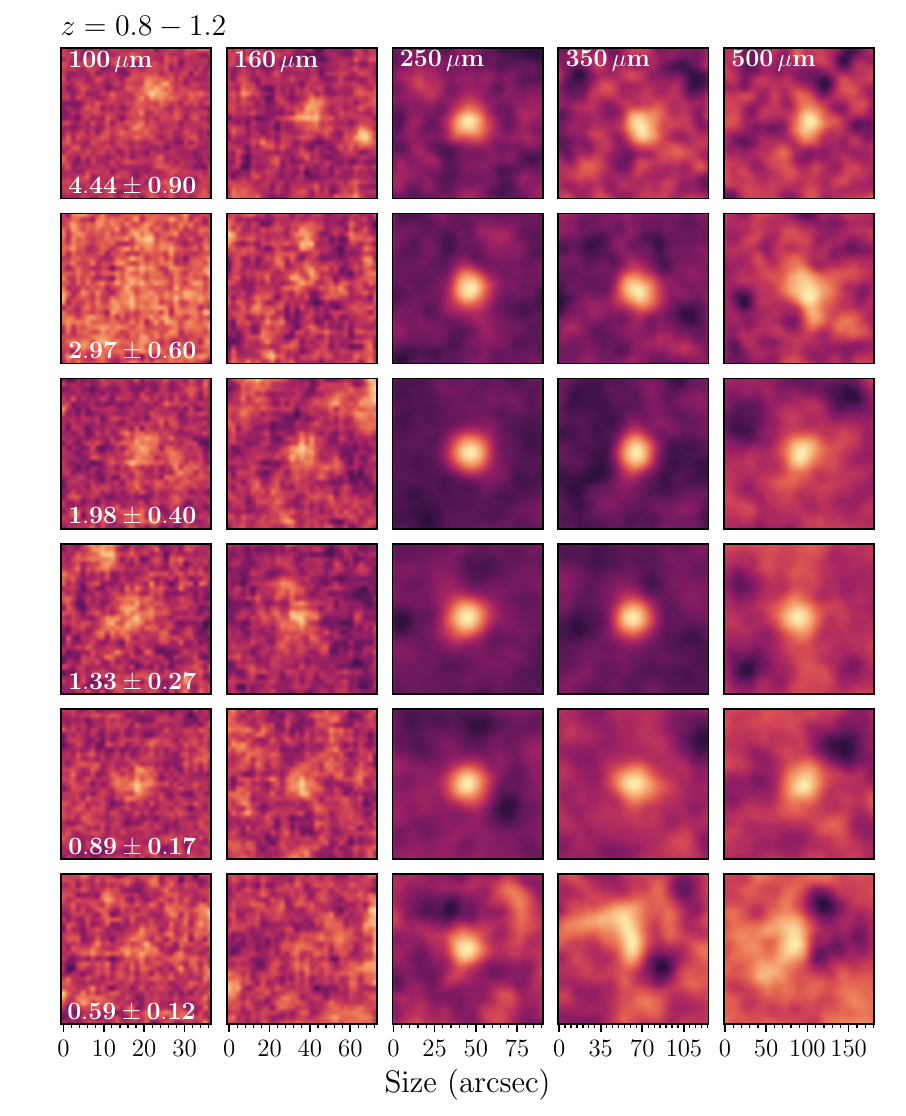}
    \caption{The PACS and SPIRE stacks used in this work for redshift range 0.6-0.8 (on the left) and 0.8-1.2 (on the right). The first two
    columns represent the PACS 100 and 160 $\mu$m band and the last three columns represent
    respectively the SPIRE 250, 350 and 500 $\mu$m wavelength bands. The top stack of each column is labelled with the \textit{Herschel} waveband (from 100 to 500 $\mu$m) they represent. 
    Each row represents a UV luminosity bin in descending order of UV luminosity from top to bottom.
    These UV luminosities of the UV LF bins are labelled in the first (leftmost) stack of each row in units of $\times 10^{10}\, \mathrm{L_\odot}$.
    Table \ref{tab:res} lists the UV luminosity, and IR luminosity of each UV luminosity bin along with the number of sources stacked to produce these stacks.
    The size of the stacks in arcseconds is labelled on the bottom edge of each stack in the bottom row. Every stack has a total width roughly 5 times the FWHM. Due to the 
    different pixel scales of the maps each map has a different size 
    in arcseconds.}
    \label{fig:stacks}
\end{figure*}

\subsection{Stacking FIR maps}
\label{sec:3.2}

Astronomical imaging at long wavelengths, such as FIR and sub-millimeter, is often hindered by high levels of noise. 
Additionally, the point spread function (PSF) is large at these wavelengths, which means that individual sources appear to spread out and blend together. This makes it difficult to resolve individual sources and determine the fluxes of each source. 
One way to overcome this is to use stacking analysis, a technique that utilises the improved positional accuracy of short-wavelength catalogues from the same region of the sky as the long-wavelength imaging. 
These catalogues can be used to identify the positions of sources that are likely to be present in the long-wavelength images. Using this positional information, one can extract fluxes at the prior source positions from the long-wavelength maps.
By averaging the fluxes of many sources together, the stacking process results in an improved signal-to-noise ratio (S/N) by a factor of $1/\sqrt{N}$, where $N$ is the number of sources averaged and assuming that the individual sources have the same S/N. This improvement in the S/N ratio makes it possible to extract information about faint sources that would otherwise be hidden in the noise, and impossible to detect.

\begin{figure*}
    \centering
    \hspace*{-0.1cm}\includegraphics[width=0.65\textwidth]{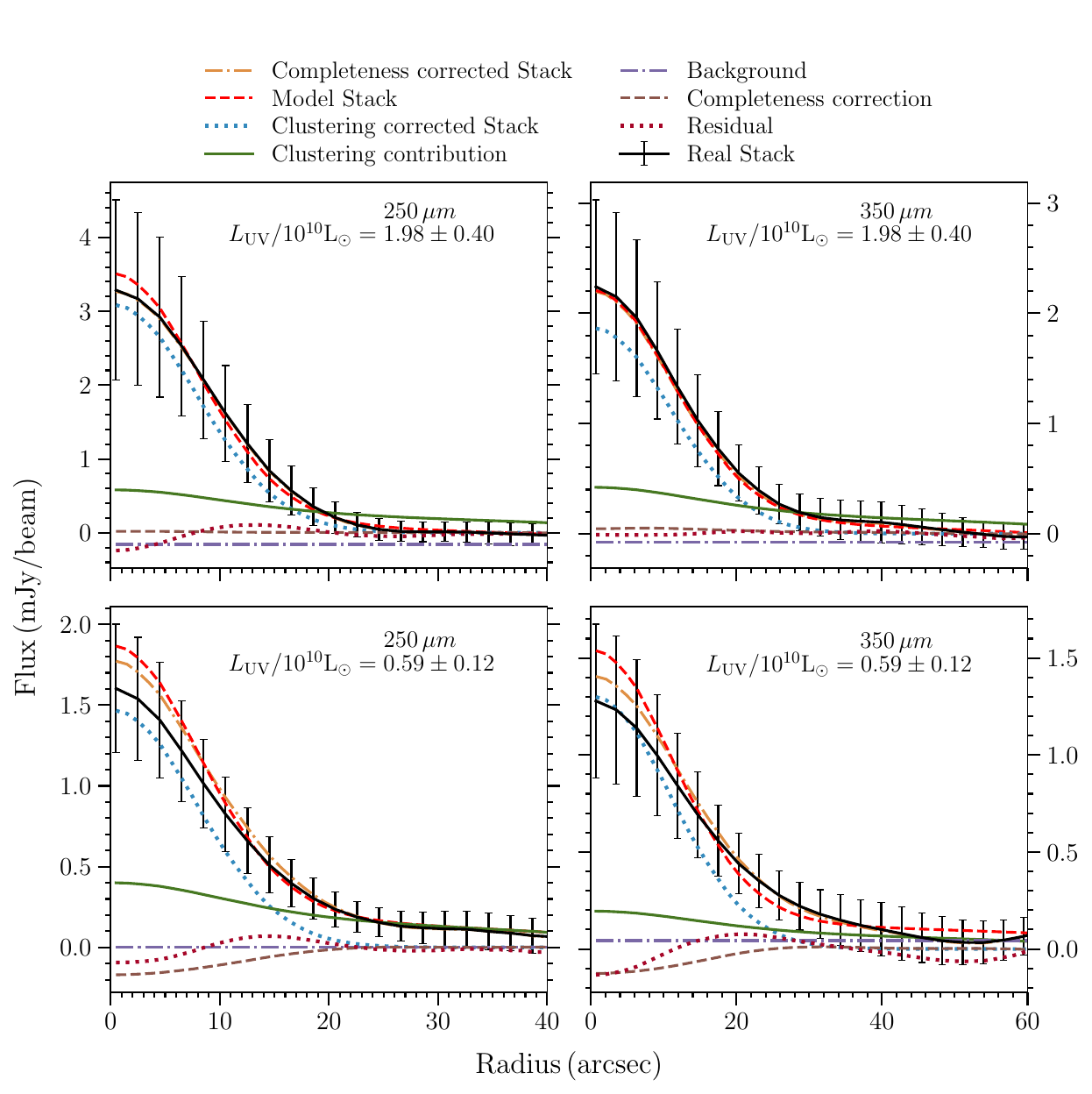}
    \caption{Here we show the radial profiles of various components in the \textit{Herschel} SPIRE photometry described by eq. \ref{eqn:clus_corr},
    by using examples of the 250 and 350 $\mu \mathrm{m}$ stacks at redshift $0.6-0.8$ in the first and fourth luminosity bins (rows) and the third and fourth columns of the left panel of Figure \ref{fig:stacks}).
    The IR band and UV luminosity are labelled on the top right of each panel. 
    The original stack is shown as black solid line with error bars. The stack is corrected for stacking bias and this correction is represented by brown dashed line and the yellow dot-dashed line shows the stacking bias corrected stack.
    The radial profile of the model fitted to the bias-corrected stack is shown by the dashed red line. 
    We show the components of this model: the extracted flux, clustering contribution, and constant background by yellow, solid green, and dot-dashed purple, respectively.
    The dotted brown line represents the residual between the bias-corrected stack and its model.}
    \label{fig:clu_corr}
\end{figure*}

In our case, we used the prior positions of the UV-selected galaxies from the UVW1 source list from \citet{2022MNRAS.511.4882S}. We stack our UVW1 sources on
the (stacking) bias-corrected PACS maps. However, for SPIRE maps, we use the residual maps coming out of the deblending process, which are corrected for issues related to confusion noise and clustering. 
The stacking of residual maps should give results consistent with the stacking of the actual maps, as demonstrated by \citet{2012ApJ...744..154R},
with the advantage of a lower noise level or uncertainty than the stacking of the actual maps.

\subsubsection{Stacking Process}

We take small square Sections (``stamps") from the SPIRE and PACS maps at the predefined location of each UV-selected source. 
The dimensions of these square stamps, which are centred on the prior position of each UV-selected source, are $S$ $\times$ $S$, where $S$ is approximately 5 times the full width at half-maximum (FWHM) of the corresponding \textit{Herschel} map.
Once the stamps have been extracted, we then proceed to sort them into UV luminosity bins based on the UV luminosity function from \citet{2022MNRAS.511.4882S}. 
To ensure that our statistics are robust and reliable, we remove bins that contain less than 25 sources. 
As a result of this process, we are left with a collection of data cubes, each cube corresponding to a specific bin in the UV LF.
Subsequently, we collapse these data cubes by averaging the pixel values of all the stacked stamps contained within each bin. This process yields a stacked average image for every bin. 

During this stacking procedure, a rotation of $\pi/2$ clockwise with respect to the preceding stamp is applied to each stamp, to cancel out any potential wing-like structures of bright sources located in proximity to the stacked signal. 
We repeat this process for the SPIRE (250, 350 and 500 $\mu \mathrm{m}$) and for PACS (100 and 160 $\mu \mathrm{m}$) maps, resulting in stacked images for each FIR waveband in each UV luminosity bin. 
Figure \ref{fig:stacks} shows these stacked images for each FIR waveband in the redshift ranges $0.6-0.8$
and $0.8-1.2$, sorted according to their UV luminosities. 
A clear signal can be seen at the centres of most stacks.

\begin{table*}
\setlength{\tabcolsep}{9pt}
\centering
\caption{The flux densities we obtain for the \textit{Herschel} PACS and SPIRE bands after making the clustering correction as explained in Section \ref{sec:3.2}. The top half of the Table corresponds to the redshift bin 0.6-0.8, and the bottom half represents the bin 0.8-1.2.}
\label{tab:stacked_fluxes}
  \begin{tabular}{lcccccr}
    \hline\hline
    \noalign{\vskip 0.75mm}
    $^{a}$\# &
    $^{b}L_{\mathrm{FUV}}$ &
    $S_{100 \,\mu \mathrm{m}}$ &
    $S_{160 \,\mu \mathrm{m}}$ &
    $S_{250 \,\mu \mathrm{m}}$ &
    $S_{350 \,\mu \mathrm{m}}$ &
    $S_{500 \,\mu \mathrm{m}}$ \\
    \noalign{\vskip 0.25mm}
    sources &
    ($\times 10^{10}\, \mathrm{L_\odot}$) &
    (mJy) &
    (mJy) &
    (mJy) &
    (mJy) &
    (mJy) \\
    \noalign{\vskip 0.25mm}
    \hline
    \noalign{\vskip 0.5mm}
    \multicolumn{1}{@{}l}{z = 0.6-0.8}\\
    \noalign{\vskip 0.5mm}
    53  &  $1.98\pm0.40$ & $1.96\pm0.55$ & $4.22\pm0.98$   & $3.08\pm0.58$    & $1.86\pm0.49$ & $1.05\pm0.31$ \\
    111 &  $1.33\pm0.27$ & $1.44\pm0.32$ & $2.19\pm0.62$   & $1.86\pm0.33$    & $1.27\pm0.26$ & $1.04\pm0.28$ \\
    135 &  $0.89\pm0.17$ & $0.94\pm0.18$ & $0.95\pm0.57$   & $1.41\pm0.33$    & $0.84\pm0.27$ & $0.34\pm0.12$ \\
    159 &  $0.59\pm0.12$ & $0.72\pm0.17$ & $1.80\pm0.44$   & $1.46\pm0.23$    & $1.30\pm0.22$ & $0.70\pm0.17$ \\
    66  &  $0.40\pm0.08$ & $0.61\pm0.20$ & $0.73\pm0.67$   & $1.23\pm0.47$    & $1.15\pm0.39$ & $0.61\pm0.19$ \\
  
    \hline
    \noalign{\vskip 0.5mm}
    \multicolumn{1}{@{}l}{z = 0.8-1.2}\\
    \noalign{\vskip 0.5mm}
    25  & $4.44\pm0.90$ & $0.75\pm0.33$ & $<2.16$ & $1.94\pm0.57$    & $1.57\pm0.43$ & $0.41\pm0.36$ \\
    85 & $2.97\pm0.60$ & $1.04\pm0.24$ & $1.81\pm0.63$ & $1.96\pm0.36$    & $1.61\pm0.28$ & $0.64\pm0.23$ \\
    115 & $1.98\pm0.40$ & $0.96\pm0.25$ & $2.54\pm0.68$ & $1.57\pm0.29$    & $1.78\pm0.29$ & $0.89\pm0.23$ \\
    165 & $1.33\pm0.27$ & $0.64\pm0.18$ & $0.98\pm0.42$ & $1.49\pm0.22$    & $0.94\pm0.21$ & $0.73\pm0.17$ \\
    111 & $0.89\pm0.17$ & $0.33\pm0.19$ & $<0.86$ & $0.93\pm0.22$    & $0.51\pm0.27$ & $0.32\pm0.23$ \\
    25  & $0.59\pm0.12$ &  $0.44\pm0.33$ & $1.23\pm0.81$ & $0.96\pm0.33$    & $0.37\pm0.24$ & ... \\ 
    \hline
    \noalign{\vskip 0.5mm}
  \end{tabular}
  \begin{minipage}{0.80\textwidth}
      \textsuperscript{$a$}{We ignore UV luminosity bins with $<$ 25 sources.}\\
      \textsuperscript{$b$}{The bin centers of the UV LF of \citet{2022MNRAS.511.4882S}.}
  \end{minipage}
\end{table*}

\subsubsection{Stacking bias}

The stacking procedure, by its very nature, is prone to a certain degree of bias, particularly toward sources that are relatively brighter and located in regions of the sky that are less densely populated with other objects.
This bias is closely related to the catalogue incompleteness \citep[as described in Section 3 of][]{2022MNRAS.511.4882S}. Galaxies that are either too faint to be detected or are situated in close proximity to a particularly bright source may be missed during the detection process. This results in their exclusion from the final catalogue, making it incomplete.
If we stack this incomplete catalogue on the FIR maps, the contribution of undetected sources to the local background of the stacks is not included.

To address this issue and recover the accurate local background of the stacks, we can use the completeness simulations from \citet{2022MNRAS.511.4882S}. These simulations involve the introduction of synthetic sources into the UVW1 image, followed by an attempt to recover them using the same detection method applied to our actual source list. The stacking of the recovered sources in the completeness simulations on the FIR maps generates correction maps to mitigate this bias.
In Figure \ref{fig:clu_corr}, we illustrate the stacking process for 250 and 350 $\mu \mathrm{m}$ maps in the first and fourth luminosity bins within the redshift range of $0.6-0.8$. The corresponding radial profiles of these correction maps are depicted as brown dashed lines in Figure \ref{fig:clu_corr}.
The top panels, which represent bright UV luminosity bins at 250 and 350 $\mu \mathrm{m}$, reveal there is not much impact of stacking bias on these bright UV bins. On the contrary, in the two bottom panels, representing faint bins, the profiles turn negative as the distance from the centre of the stack decreases, indicating a substantial bias.
To correct for stacking bias in the local background, these correction maps are subtracted from the stacks.

\subsubsection{Stacked Photometry}

The process of creating the maps using the \textit{Herschel}-PACS and \textit{Herschel}-SPIRE instrumentation is carried out by using various pipelines and techniques, which are independently developed and implemented by separate teams. 
As a result of these distinct methods, the maps produced by these instruments exhibit variations in their units and calibrations. 
So, in order to extract accurate flux densities from the PACS and SPIRE stacks, different approaches are employed.

The aperture photometry technique is particularly well-suited for the \textit{Herschel}-PACS maps, as they are provided in Jy/pixel units. This technique involves measuring fluxes by using a circular aperture of a certain size to enclose the source of interest and integrate its pixel values. For the 100 $\mu \mathrm{m}$ maps, we use an aperture radius of 7.2 arcseconds, while for the 160 $\mu \mathrm{m}$ maps we use a radius of 12 arcseconds.
This happens because the size of the PSF varies between the different bands and the aperture size needs to be adjusted accordingly.
However, it is important to note that these extracted fluxes are not necessarily the true fluxes of the sources. 
To correct this, we apply corrections for the fraction of the PSF that falls outside the aperture and for any losses resulting from high-pass filtering of the data. 
These corrections are determined using empirical results from the PEP Data Release 1 (DR1) notes\footnote{\url{https://www.mpe.mpg.de/resources/PEP/DR1_tarballs/readme_PEP_global.pdf}}.

\begin{figure*}
    \centering
    \includegraphics[width=0.70\textwidth]{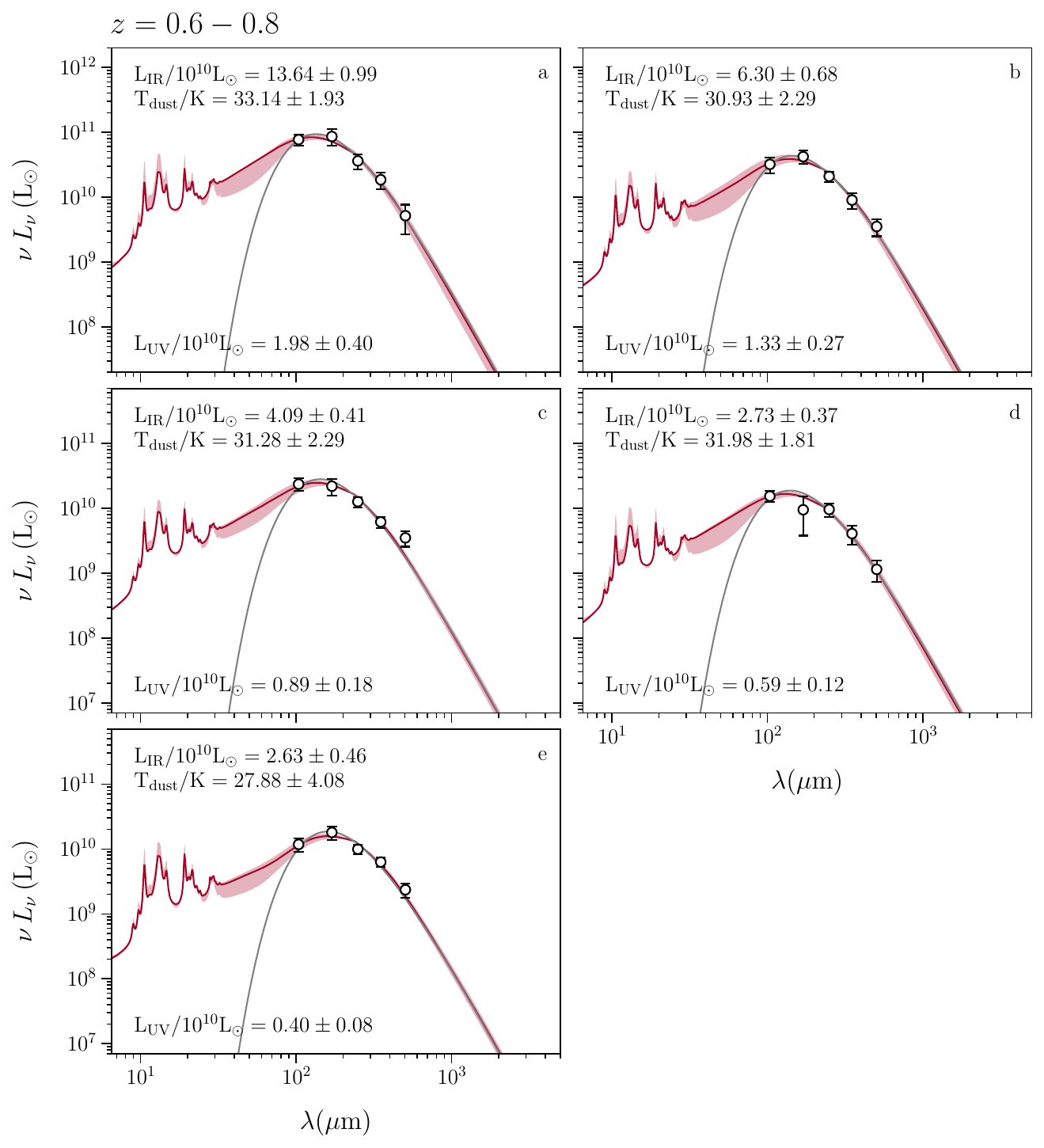}
    \caption{IR SED fits for the redshift bins $0.6-0.8$. Each panel represents the UV luminosity bins from brightest to faintest (from a to e). The UV luminosity of each UV LF bin is labelled at the bottom left of each panel. The black hollow circles show the 
    FIR flux densities from \textit{Herschel} PACS at 100 and 160 $\mu \mathrm{m}$ and \textit{Herschel} SPIRE at 250, 350 and 500 $\mu \mathrm{m}$.
    The red curve represents the best fit \citet{2021A&A...653A.149B} template to the FIR data.
    The shaded region represents the range of templates within $|\chi^2 - \chi^2_\mathrm{min}| \leq 1$.
    The grey curve is the best-fit modified black body curve.
    The estimates for the dust temperature and integrated IR luminosity are labelled at the top-left of each panel. Note that the IR luminosity on the x-axis is in $\nu L\nu$.}
    \label{fig:sed68}
\end{figure*}

\begin{figure*}
    \centering
    \includegraphics[width=0.70\textwidth]{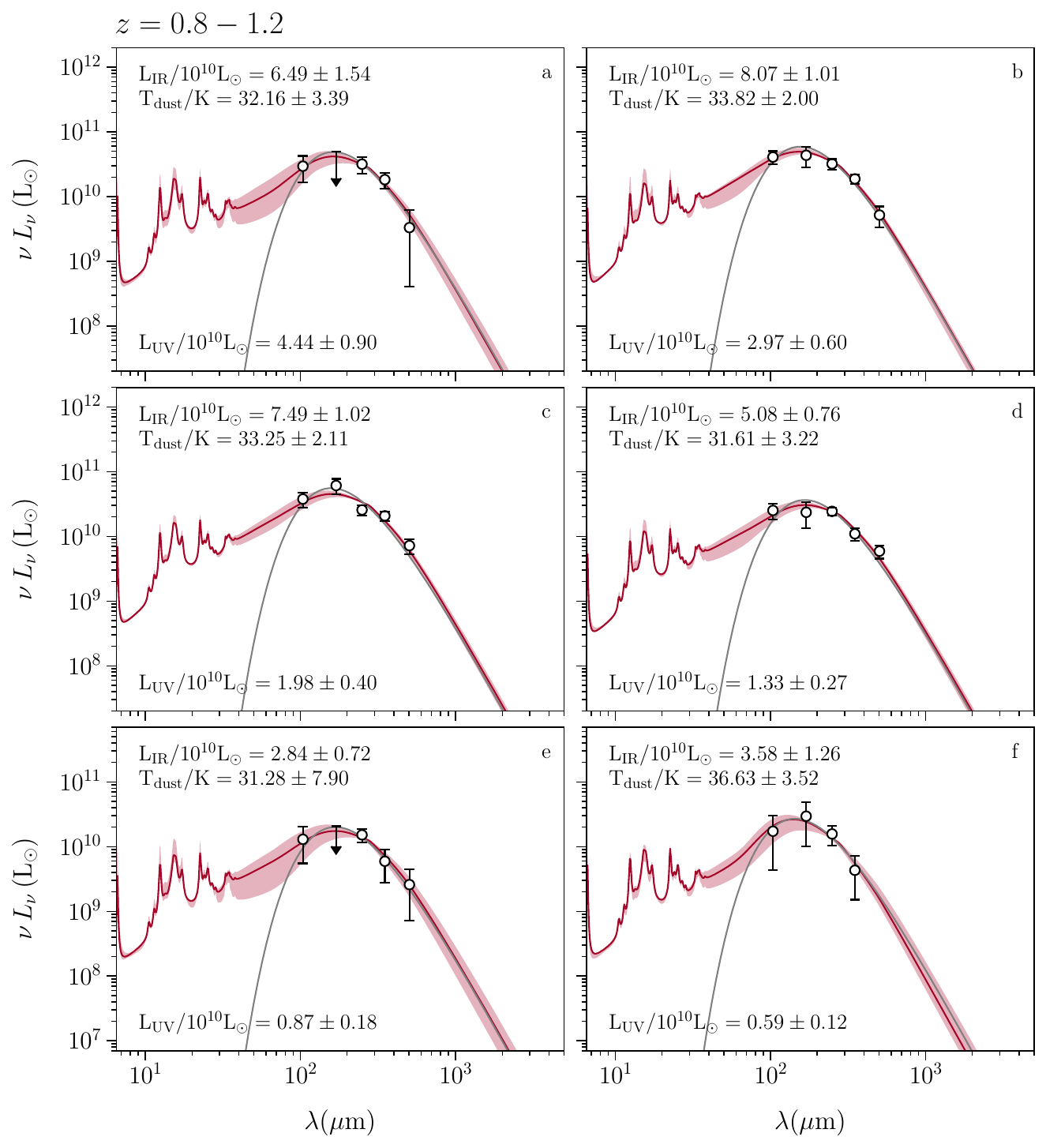}
    \caption{IR SED fits for the redshift bins $0.8-1.2$. Each panel represents a 
    luminosity bin, going from the brightest to the faintest UV luminosity (a-f for $z=0.8-1.2$), also 
    labelled at the bottom-left of each panel. 
    The colour coding is the same as in Figure \ref{fig:sed68}.
    The dust temperature and the integrated IR luminosity are labelled in the upper left of each panel.}
    \label{fig:sed812}
\end{figure*}

The SPIRE maps are expressed in units of Jy/beam, making PSF fitting an effective method for determining the photometry of these stacks. 
This involves determining the flux densities of the SPIRE stacks which are equal to the peak of the PSF models fitted to the central pixels of the stacks. 
However, it is important to note that the SPIRE stacks can be susceptible to clustering effects, which can result in confusion and overestimation of the stacked photometry.
To mitigate this, we use a deblending approach (in Section \ref{sec:3}) that enables us to overcome the confusion limit and minimize the clustering contribution from sources present in the prior catalogue. 
However, while stacking on deblended residual images reduces the flux contribution of bright off-centre sources in the prior catalogue, this method does not take into account the clustering of objects that are not part of the prior catalogue 
or are too faint to be detected in our residual maps. Such sources might be clustered in the IR imaging along with our UV-selected galaxies. This inherent clustering of sources can still result in an overestimation of the stacked photometry.
To solve this problem, a method prescribed by \citet{2010A&A...516A..43B} is used. The method involves fitting the final stack as a linear sum of the PSF and its convolution with the angular correlation function. This is expressed mathematically as follows:
\begin{equation}
  \mathcal{S}(\theta, \phi) + \mathcal{B} \, = \,
  \alpha \times \mathcal{P(\theta, \phi)} \,
  + \, \beta \times \
  \left[\mathcal{P}(\theta, \phi)\, \ast\, \mathit{w}(\theta, \phi)\right].
  \label{eqn:clus_corr}
\end{equation}
Here, $\mathcal{S}(\theta, \phi)$ represents the stacked stamp, $\mathcal{P(\theta, \phi)}$ represents the PSF, and $\mathit{w}(\theta, \phi)$ represents the angular correlation function of the galaxies under consideration. The fixed background level is represented by $\mathcal{B}$.
The best-fit values for $\alpha$, $\beta$ and $\mathcal{B}$ are found for each FIR band and UV luminosity bin, and the value of $\alpha$ is considered to be the final flux value for each SPIRE stack. This method provides a more comprehensive approach to account for the inherent clustering of sources and ensures more accurate photometry results.

Figure \ref{fig:clu_corr} shows an example of different components of this process.
For this particular example, we find that clustering contributions are 18.8 and 27.3 
per cent of the actual flux in the first and fourth luminosity bins for the 250 $\mu \mathrm{m}$
map. Corresponding values for the 350 map are 22.6 and 15.0 per cent. 
The average value of this fraction, for the 250, 350 and 500 $\mu \mathrm{m}$ maps were found to be 16, 18 and 22 per cent respectively in the redshift bin $0.6-0.8$ and 14, 6 and 41 
per cent respectively in the redshift bin $0.8-1.2$.

\subsubsection{Errors}
We use standard bootstrap to calculate the statistical errors on stacked flux densities in
each bin. In each bin, N stamps are selected at random with replacement and 
stacked. The flux densities are calculated from these error stacks in the same
fashion as the original stacks (i.e. aperture photometry for PACS and PSF photometry for SPIRE). Using 1000 bootstraps, we calculate the 68 per cent 
confidence intervals around the measured values. In Table \ref{tab:stacked_fluxes} we show the resulting average fluxes extracted from
the stacks at all FIR bands considered in this study.

\subsection{IR SED fits}
\label{sec:3.3}

Now that we have obtained the stacked photometry for our galaxy sample, the next task is to extract the average IR properties from the stacked flux densities. We fit two different types of model to the SEDs, one for determining the IR luminosity and the other for the dust temperatures.

To estimate the total IR luminosity, we fit FIR model templates to our dataset. Specifically, we utilised the two-parameter dust templates from \citet{2021A&A...653A.149B}. These templates are parameterised in the total IR luminosity and specific star-formation rate, and they are built upon physically motivated dust models by \citet{2007ApJ...657..810D}.
These templates are well suited to star-forming galaxies in the luminosity range of our sample.
The resulting fits are shown in Figures \ref{fig:sed68} and \ref{fig:sed812}.
We measure the integrated IR luminosity by integrating the rest-frame flux density from 8 to 1000 $\mu$m range, determining the fluxes, and subsequently employing the luminosity distance to calculate the IR luminosity
\begin{equation}
  L_{\mathrm{8-1000\mu \mathrm{m}}} =\,
  4 \,\pi \,d^{2}_{\mathrm{L}} \left(z\right)\,
  \int_{\nu_{1}}^{\nu_{2}} \, S_{\nu} \,\mathrm{d}\nu
  \label{eqn:irlum}
\end{equation}
where $\nu_{1}$ and $\nu_{2}$ are the rest-frame frequencies corresponding 
to $8-1000 \,\mu$m limit and $d_{\mathrm{L}}$ is the luminosity distance.

We determine the dust temperatures by fitting the isothermal grey bodies.
Most of the FIR originates from large grains that radiate as isothermal grey bodies at temperatures $10 - 50$ K and are in equilibrium with the ambient interstellar radiation field.
An isothermal black-body model can be adjusted to account for variable source emissivities and opacities, resulting in a grey-body or modified black-body model.
It takes the form,
\begin{equation}
  S(\nu) \propto \,
  \nu^{\alpha}\, B_{\nu}(T_{\mathrm{d}}) = \,
  \frac{\nu^{\alpha + 3}}{\mathrm{exp}\displaystyle\left(\frac{h\nu}{k_{B}T_{\mathrm{d}}}\right)-1}
  \label{eqn:mbb}
\end{equation}
for the approximation of optically thin media. Here $\alpha$ is the source emissivity and $T_{\mathrm{d}}$ is the characteristic dust temperature.
The typical values of $\alpha$ fall within the range of 1.5 to 2 \citep[][]{2003MNRAS.338..733B,2011MNRAS.411..505C,2011MNRAS.415.2723C,2012MNRAS.421.2161V} and for this work we use $\alpha = 1.5$ following \citet{2003MNRAS.338..733B,2011MNRAS.415.2723C}.
We fit these grey bodies to our stacked SEDs, with the amplitude and the dust temperature
as the free parameters.

\begin{table*}
\setlength{\tabcolsep}{8pt}
\centering
\caption{The main results of this chapter are summarised in this Table. The top and bottom halves of the Table correspond to the redshift bins 0.6-0.8 and 0.8-1.2 respectively.}
\label{tab:res}
  \begin{tabular}{lccccccr}
    \hline\hline
    \noalign{\vskip 0.5mm}
    $^{a}$\# &
    $^{b}L_{\mathrm{FUV}}$ &
    $^{c}L_{\mathrm{FUV}}$ &
    $^{d}L_{\mathrm{IR}}$ &
    $^{e}T_{\mathrm{dust}}$ &
    $^{f}A_{\mathrm{FUV}}$ &
    $^{g}\mathrm{SFR}_{\mathrm{FIR}}$ &
    $^{h}\mathrm{SFR}_{\mathrm{FIR+FUV}}$ \\
    \noalign{\vskip 0.5mm}
    \noalign{\vskip 0.5mm}
    sources &
    ($\times 10^{10}\, \mathrm{L_\odot}$) &
    ($\times 10^{10}\, \mathrm{L_\odot}$) &
    ($\times 10^{10}\, \mathrm{L_\odot}$) &
    (K) &
    (mag) &
    $(\mathrm{M_\odot}\, \mathrm{yr}^{-1})$ &
    $(\mathrm{M_\odot}\, \mathrm{yr}^{-1})$ \\
    \hline
    \noalign{\vskip 0.5mm}
    \multicolumn{1}{@{}l}{z = 0.6-0.8}\\
    \noalign{\vskip 0.5mm}    
    53  &  $1.98\pm0.40$ & $1.95\pm0.24$ &   $6.30\pm0.67$    & $30.93\pm2.70$ & $1.15\pm0.39$& $9.42\pm1.00$& $15.18\pm2.33$ \\
    111 &  $1.33\pm0.27$ & $1.31\pm0.15$ &   $4.09\pm0.42$     & $31.28\pm2.20$ & $1.13\pm0.38$& $6.11\pm0.64$& $9.97\pm1.54$ \\
    135 &  $0.89\pm0.17$ & $0.89\pm0.10$ &   $2.73\pm0.35$     & $31.98\pm1.79$ & $1.13\pm0.39$& $4.08\pm0.52$& $6.66\pm1.07$ \\
    159 &  $0.59\pm0.12$ & $0.60\pm0.07$ &   $2.63\pm0.44$     & $27.88\pm3.68$ & $1.39\pm0.57$& $3.87\pm0.66$& $5.59\pm0.91$ \\
    66  &  $0.40\pm0.08$ & $0.41\pm0.04$ &   $2.14\pm0.65$     & $27.04\pm4.13$ & $1.53\pm0.84$& $3.08\pm0.97$& $4.23\pm1.06$ \\
  
    \hline
    \noalign{\vskip 0.5mm}
    \multicolumn{1}{@{}l}{z = 0.8-1.2}\\
    \noalign{\vskip 0.5mm}  
    25  & $4.44\pm0.90$ & $4.43\pm0.58$ & $6.49\pm1.54$ &  $32.16\pm3.39$ & $0.68\pm0.26$ & $9.71\pm2.30$& $22.60\pm5.23$ \\
    85 & $2.97\pm0.60$ & $2.92\pm0.31$ & $8.07\pm1.01$ &  $33.82\pm2.00$ & $1.04\pm0.35$ & $12.07\pm1.50$& $20.69\pm3.49$ \\
    115 & $1.98\pm0.40$ & $2.02\pm0.21$ & $7.49\pm1.02$ &  $33.25\pm2.11$ & $1.28\pm0.47$ & $11.20\pm1.52$& $16.96\pm2.60$ \\
    165 & $1.33\pm0.27$ & $1.33\pm0.16$ & $5.08\pm0.76$ &  $31.61\pm3.22$ & $1.29\pm0.49$ & $7.60\pm1.13$& $11.45\pm1.81$ \\
    111 & $0.89\pm0.17$ & $0.92\pm0.10$ & $2.84\pm0.72$ &  $31.28\pm7.09$ & $1.16\pm0.52$ & $4.25\pm1.07$& $6.82\pm1.42$ \\
    25 & $0.59\pm0.12$ & $0.64\pm0.05$ & $3.58\pm1.26$ &  $36.63\pm3.52$ & $1.66\pm0.98$ & 
    $5.36\pm1.88$ & $7.07\pm1.98$ \\    
    \hline
    \noalign{\vskip 0.5mm}
  \end{tabular}
  \begin{minipage}{0.94\textwidth}
      \textsuperscript{$a$}{This column shows the number of sources in each UV luminosity bin. UV LF bins with < 25 sources have been ignored.} \\
      \textsuperscript{$b$}{The bin centers of the UV LF in \citet{2022MNRAS.511.4882S}. These values are used as labels in this work.} \\
      \textsuperscript{$c$}{Mean of the UV luminosity of the sources inside the UV LF
      bins. We use these values for all the calculations in this paper.} \\
      \textsuperscript{$d$}{Average integrated IR luminosity obtained from the stacked flux densities.} \\
      \textsuperscript{$e$}{Average dust temperature of galaxies in each UV LF bin, obtained from the MBB fits.} \\
      \textsuperscript{$f$}{Average dust attenuation from eq. \ref{eqn:IRX}.}\\
      \textsuperscript{$g$}{The SFR estimated from the IR luminosity.}\\
      \textsuperscript{$h$}{The total SFR, calculated as the sum of the IR and UV components.} \\
  \end{minipage}
\end{table*}

\subsection{Star Formation Rate}
\label{sec:3.4}

The integrated IR luminosity calculated from the IR templates and the UV
luminosity can be used to calculate obscured and un-obscured SFR
respectively by using the following scaling relation \citep{1998ARA&A..36..189K,2011ApJ...737...67M,2012ARA&A..50..531K},
\begin{equation}
  \log \left(\frac{\mathrm{SFR}_\mathrm{FUV}}{M_{\odot}\, \mathrm{yr}^{-1}}\right) = \,
  \log \left(\frac{L_{\mathrm{FUV}}/\nu_{\mathrm{FUV}}}{\mathrm{ergs}\, \mathrm{s}^{-1}\, \mathrm{Hz}^{-1}}\right)
  - 27.85
  \label{eqn:sfruv}
\end{equation}
assuming a continuous and constant SFR with the \citet{1955ApJ...121..161S} initial mass function (IMF) over timescales longer than $10^8$ yr and a mass range from 0.1 to 100 $M_\odot$ and
\begin{equation}
  \log \left(\frac{\mathrm{SFR}_\mathrm{FIR}}{M_{\odot}\, \mathrm{yr}^{-1}}\right) = \,
  \log \left(\frac{L_{\mathrm{FIR}}}{\mathrm{ergs}\, \mathrm{s}^{-1}}\right)
  - 43.41
  \label{eqn:sfrir}
\end{equation}
assuming continuous bursts from 10 to 100 Myr, adopting the \citet{2001MNRAS.322..231K} IMF. 
Total SFR ($\mathrm{SFR}_\mathrm{Tot}$) is obtained as the sum of the contributions from the luminosities of
FIR ($\mathrm{SFR}_\mathrm{FIR}$) and FUV ($\mathrm{SFR}_\mathrm{FUV}$).
The $\mathrm{SFR}_\mathrm{FUV}$ value based on the \citet{1955ApJ...121..161S} IMF is transformed into the equivalent value corresponding to the \citet{2001MNRAS.322..231K} IMF through a multiplication by a factor of 1.8.

\subsection{Dust Attenuation}
\label{sec:3.5}

Using the IR and UV luminosities, we calculate the IRX \citep{1999ApJ...521...64M}, such that $\mathrm{\mathrm{IRX}} = L_{\mathrm{IR}} / L_{\mathrm{UV}}$.
Different relations are used in the literature \citep{1999ApJ...521...64M,2005ApJ...619L..55S,2011ApJ...741..124H,2013ApJ...762..125N} to convert the $\mathrm{IRX}$ ratio into the dust attenuation in the UV luminosity. For the sake of comparison, we apply the relation commonly used in previous studies \citep[e.g.][]{2013ApJ...762..125N,2013MNRAS.429.1113H,2016A&A...587A.122A},
\begin{equation}
  A_{\mathrm{FUV}} = \,
  2.5 \log \ \left(1 + 0.59 \times \mathrm{IRX}\right).
  \label{eqn:IRX}
\end{equation}
This can be used to correct for the UV light absorbed by the dust and calculate the total SFR in the next Section. In order to make this correction to the unobscured SFR
we use the relation from \citet{2013ApJ...762..125N} given by
\begin{equation}
	\log \left(\mathrm{SFR}_\mathrm{Tot}\right) = \,
	\log \left({\mathrm{SFR}_\mathrm{FUV}}\right) + 0.4 \times A_{\mathrm{FUV}}
	\label{eqn:dust_corr_sfr}
\end{equation}

\section{Results}
\label{sec:4}

\begin{figure*}
    \centering
    \hspace*{-0.8cm}\includegraphics[width=1.1\textwidth]{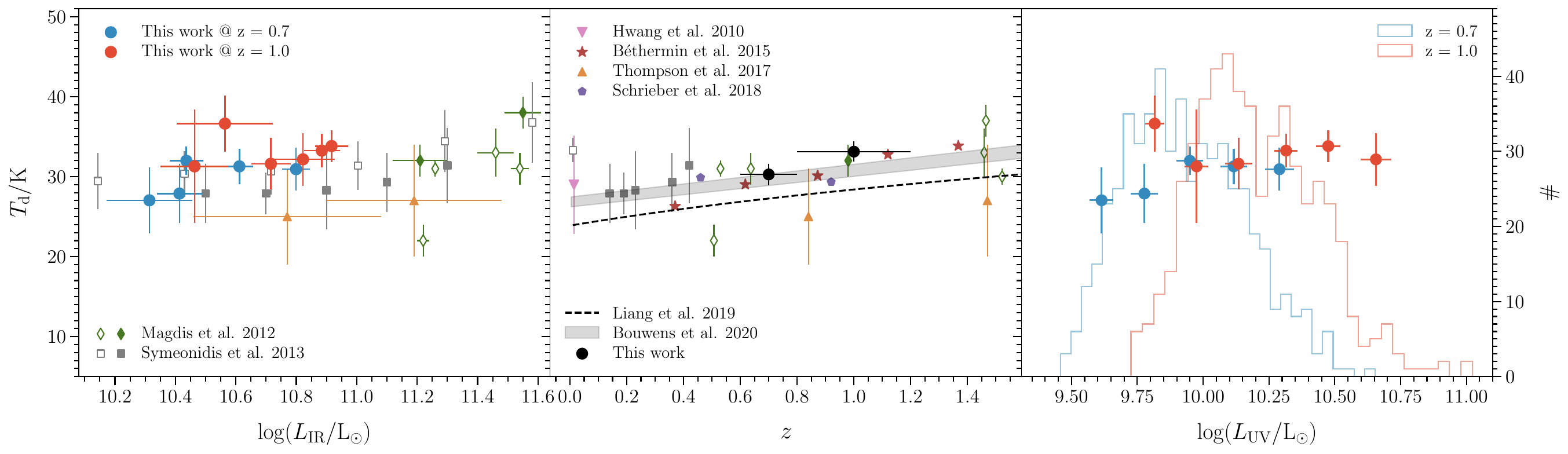}
    \caption{Dust temperature as a function of total IR luminosity, redshift
    and UV luminosity.  
    \textit{Left panel:} dust temperature in logarithmic IR luminosity bins.
    The results from this work at redshifts bins centred at 0.7 and 1.0 are shown in blue and red circles, respectively. For comparison, we also plot the literature results from \citet{2012ApJ...760....6M} as green diamond symbols. The solid and hollow symbols represent the values from stacking analysis and individual star-forming galaxies, respectively. 
    The grey squares represent the median temperatures of the IR-selected galaxies
    from \citet{2013MNRAS.431.2317S}. The hollow squares represent the values for the galaxies in their local sample ($z < 0.1$).
    The yellow triangles represent the results for the $H_{\alpha}$ selected samples from \citet{2017ApJ...838..119T}.
    \textit{Middle panel:} average temperature as a function of redshift.
    The solid circles in black show the values obtained in this work by luminosity weighing the temperatures in redshift bins centred at 0.7 and 1.0.
    We show the redshift weighted mean dust temperature of the galaxies in the
    local Universe from \citet{2010MNRAS.409...75H} as a pink downward triangle.
    The brown stars and purple pentagons represent the estimates of \citet{2015A&A...573A.113B} and \citet{2018A&A...609A..30S}. These values are taken from the compilation of \citet{2020ApJ...902..112B}.
    The grey shaded area shows the fit from \citet{2020ApJ...902..112B}.
    The dashed black line is the predicted trend for the $T_{\mathrm{d}}-z$ relation from \citet{2019MNRAS.489.1397L}.
    \textit{Right panel:} dust temperature of the grey body fits as a function of UV luminosity. 
    The blue and red histograms represent the distributions of the star-forming galaxies in the UV luminosity space for redshift bins centred at 0.7 and 1.0 respectively. 
    The y-axis on the right represents the number of sources.
    The data points share the redshift colour scheme with the histograms.}
    \label{fig:T_LIR_z_LUV}
\end{figure*}

We stack maps of FIR emission obtained from \textit{Herschel} on ultraviolet (UV) selected sources that lie in the redshift range of 0.6 to 1.2. 
To determine the average stacked photometry in the FIR for these sources, we fit the IR model templates from \citet{2021A&A...653A.149B} and integrating the results over the wavelength range of 8 to 1000 $\mu \mathrm{m}$.
The results indicate that the typical IR luminosities of the stacked galaxies fall within the range of \num{2.15e10} $\mathrm{L_\odot}$ to \num{6.30e10} $\mathrm{L_\odot}$ at redshift 0.7 and \num{3.58e10} $\mathrm{L_\odot}$ to \num{6.49e10} $\mathrm{L_\odot}$ at redshift 1.0. On average, our sample is composed of galaxies belonging to the normal (sub-luminous; $L_\mathrm{IR} < 10^{11}{\mathrm{L}_\odot}$) infrared galaxies.

In order to obtain the average dust temperatures, we fit isothermal grey bodies to the average FIR photometry in each UV luminosity bin. 
These temperatures are presented in the left panel of Figure \ref{fig:T_LIR_z_LUV} as functions of IR luminosity.
Additionally, we have plotted the luminosity-weighted dust temperatures as a function of redshift in the middle panel of Figure \ref{fig:T_LIR_z_LUV}, along with literature values for comparison.
Individual temperatures, calculated in the UV Luminosity Function (UV LF) bins, are shown as a function of UV luminosity in the right panel of Figure \ref{fig:T_LIR_z_LUV}.

Using the estimated FIR luminosities, we calculated the IRX in each UV luminosity bin. In Figure \ref{fig:AFUV_Z}, we have plotted the average value of $\mathrm{IRX}$ as a function of the redshift. $\mathrm{IRX}$ as a function of the FIR and FUV luminosities are plotted in Figures \ref{fig:AFUV_LIR} and \ref{fig:AFUV_UVLF}, respectively.
The top panel of Figure \ref{fig:AFUV_UVLF} also shows the UV LF from \citet{2022MNRAS.511.4882S}.
This dust attenuation ($A_\mathrm{FUV}$) in the UV radiation is parametrised in terms of the $\mathrm{IRX}$ ratio, as described by eq. \ref{eqn:IRX}, and labelled as the secondary y-axis on the right-hand side of the panels in Figures \ref{fig:AFUV_Z} and \ref{fig:AFUV_LIR}. 
We estimate the SFR of our galaxies in Section \ref{sec:3.4}, and present the results
in Figure \ref{fig:SFR_LUV} comparing the estimates from the UV and IR luminosities
as well as the total SFR.
Then in Figure \ref{fig:irx_sfr}, we show $\mathrm{IRX}$ as a function of the total SFR.

Finally, we use these values of $A_\mathrm{FUV}$ to correct the SFR density (SFRD), which is calculated from dust-attenuated UV radiation.
The SFRD is estimated from the luminosity density estimates, as explained in Section \ref{sec:3.4}, where the estimates of luminosity density calculated using rest-frame UV radiation come from \citet{2022MNRAS.511.4882S}. 
In Figure \ref{fig:sfrd}, we present the estimates for the total SFRD after it has been corrected for dust attenuation, providing a more accurate picture of the star formation activity in these galaxies.

\section{Discussion}
\label{sec:5}

The aim of this paper is to study the dust properties of UV-selected galaxies in
the redshift range of 0.6-1.2.
The galaxies selected through the UVW1 filter on XMM-OM are stacked on the
FIR imaging from \textit{Herschel} PACS and SPIRE instruments, and the dust properties
are constrained in UV luminosity bins of the UV LF in the same redshift range.
We considered only luminosity bins with at least 25 sources in each redshift bin (at 0.7 and 1.0), to obtain robust statistics.

\subsection{Dust Temperature and Infrared Luminosities}
We start with the dust temperature and total IR luminosities of galaxies and explore
their correlation if any.
The relationship between dust temperature and IR luminosity is related to the physical conditions within star-forming regions, from which the IR emission originates. In this context, higher temperatures indicate either more compact or more luminous star-forming regions. The equilibrium temperature essentially depends on the UV flux that impinges on the dust grains.
A correlation between dust temperature and IR luminosity has been observed in some previous studies \citep[e.g.][]{1987ARA&A..25..187S,2000MNRAS.315..115D,2001ApJ...549..215D,2003ApJ...588..186C,2009MNRAS.397.1728S,2014A&A...561A..86M} and is suggested to be likely the result of the transition of
galaxies into starburst phase \citep[][]{2014A&A...561A..86M}.

In our study, we use the isothermal graybody and IR model templates to fit the stacked \textit{Herschel} PACS and SPIRE photometry, allowing us to calculate the dust temperature and IR luminosities. Our results are directly comparable to other studies in the literature that have adopted similar definitions of dust temperature and used MBBs for calculations. However, if a different definition or technique is used, it will be specifically noted.
Our findings do not suggest any significant trend in dust temperature with IR luminosity for a fixed redshift (as seen in the left panel of Figure \ref{fig:T_LIR_z_LUV}).

If averaged over the UV LF bins, the average temperature increases very slightly with redshift within the range explored in this study. However, the difference is not very significant and the values are within the 2$\sigma$ distance. 
The range of redshifts explored in this work is not wide enough to make any conclusive remarks, so we include measurements at other redshifts from previous studies \citep[from redshift 0.15-1.5;][]{2012ApJ...760....6M,2013MNRAS.431.2317S,2015A&A...573A.113B,2017ApJ...838..119T,2018A&A...609A..30S}. In this case, now we observe a weak trend, which is also confirmed by the fit from \citet{2020ApJ...902..112B}.
The fit is mainly driven by values from \citet{2015A&A...573A.113B} and \citet{2018A&A...609A..30S}, but as we can see (middle panel of Figure \ref{fig:T_LIR_z_LUV}) it is also somewhat consistent with other works considered in our study.
Our values seem to be in agreement with these previous studies and the fit from \citet{2020ApJ...902..112B}. 
In the same plot, we show the trend found by \citet{2019MNRAS.489.1397L} for galaxies at redshifts 2 and higher, extrapolated to redshift 0.01. This trend is offset towards lower temperatures from the values obtained in this and previous works.
We compare our results with the values of dust temperature calculated by \citet{2010MNRAS.409...75H} 
and \citet{2013MNRAS.431.2317S} for their samples of IR galaxies.
About 90 per cent of \citet{2010MNRAS.409...75H} galaxies have $\mathrm{log}(L_{\mathrm{IR}}/\mathrm{L_\odot}) \leq 11.2$, which is roughly the upper limit of the highest IR luminosity bin.
They estimated the median dust temperature to be 28.98 K, with a 16-84th percentile range of 24.78 K to 37.13 K. These values as a function of the median redshift of the galaxies are plotted in the middle panel of Figure \ref{fig:T_LIR_z_LUV} along with the mean dust temperature for the \citet{2013MNRAS.431.2317S} local sample.
On comparison, it can be seen that the average dust temperature does not evolve significantly from redshift 1 to the present time.   

\begin{figure}
    \centering
    \includegraphics[width=\columnwidth]{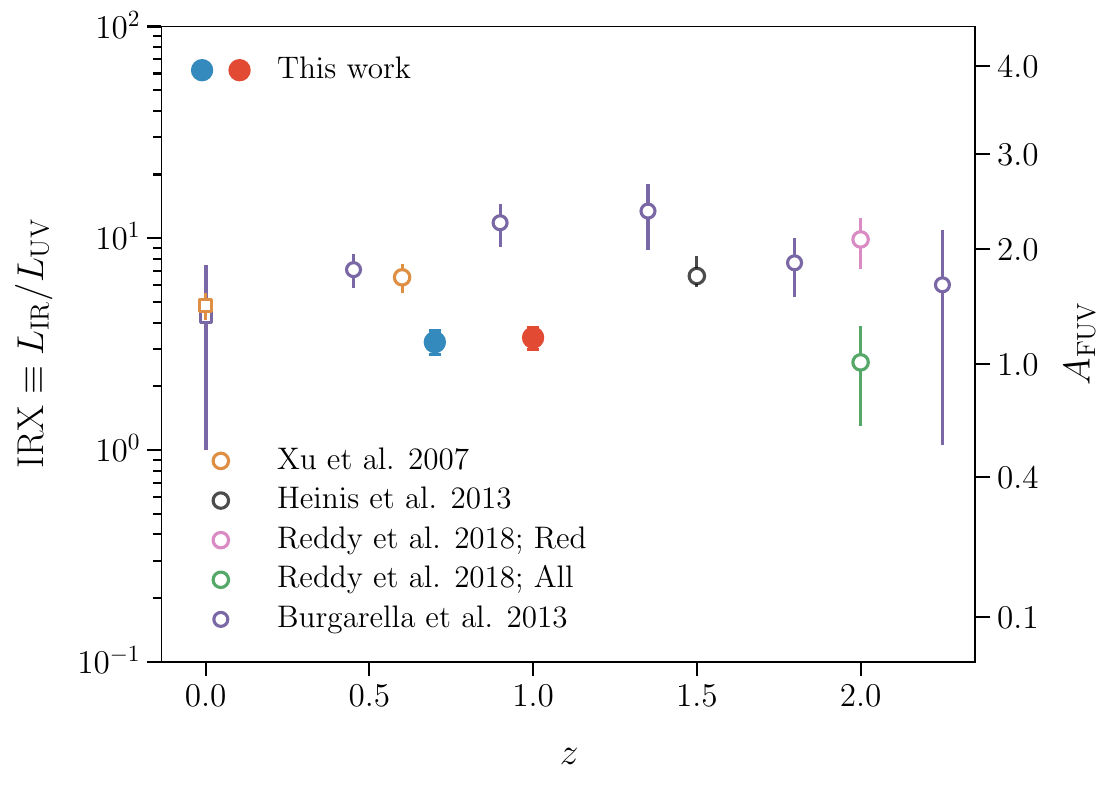}
    \caption{Here we show the $\mathrm{IRX}$ ratio as a function of redshift.
    On the right side of the panel, we plot the attenuation $A_\mathrm{FUV}$ in the UV radiation due to dust, which is parameterised as a function of the $\mathrm{IRX}$.
    The blue and red solid circles show the values obtained in this work. The hollow circular symbols represent the literature estimates from \citet{2007ApJS..173..432X}, \citet{2013MNRAS.429.1113H}, 
    \citet{2018ApJ...853...56R} and \citet{2013A&A...554A..70B} in yellow, black, green and purple colours respectively.
    The pink hollow circles represent the subset of \citet{2018ApJ...853...56R} 
    sample, referred to as ``Red" galaxies.
    The redshift 0 data from \citet{2007ApJS..173..432X} and \citet{2013A&A...554A..70B}.}
    \label{fig:AFUV_Z}
\end{figure}

Despite much effort, determining the precise relationship between dust temperature and redshift is a difficult task, especially in light of recent results, which often appear contradictory.
Some studies observe that the dust temperature increases from the local Universe to high redshifts \citep[e.g.][]{2012ApJ...760....6M,2014A&A...561A..86M,2015A&A...573A.113B,2018A&A...609A..30S}.
Conversely, others argue for a colder dust temperature at higher redshifts \citep[e.g.][]{2002ApJ...573...66C,2010MNRAS.409...75H,2009MNRAS.397.1728S,2013MNRAS.431.2317S,2012ApJ...759..139K,2017ApJ...843...71K}. 
There is a third group of studies that find no compelling evidence for a redshift-dependent evolution of dust temperature \citep[e.g.][]{2018ApJ...862...77C,2022ApJ...930..142D}.
These conflicting outcomes can be partly attributed to selection bias in flux-limited samples \citep[][]{2019MNRAS.489.1397L},
as well as to the influence of various factors capable of significantly affecting dust temperature.
These factors include, but are not limited to, the specific SFR \citep[][]{2014A&A...561A..86M}, the amount and opacity of dust, the gas metallicities etc. \citep[][]{2019MNRAS.489.1397L}.
In fact, the correlation between dust temperature and specific SFR is suggested to be more robust and statistically significant than that with redshift \citep[][]{2014A&A...561A..86M,2018A&A...609A..30S}.
Our results for the UV-selected galaxies agree with the findings of the third group of studies mentioned above. However, we remark here that all of these studies are conducted on IR-selected galaxies samples.

The dust temperature and UV luminosity, are plotted in the right panel of Figure \ref{fig:T_LIR_z_LUV}. From the plot we observe a modest correlation between these two parameters only for redshift 0.7.
The Spearman correlation coefficients for these variables are 0.60 and -0.03, at redshifts of 0.7 and 1.0; however, the significance for any correlations is very low.

\subsection{Dust Attenuation}

The $\mathrm{IRX}$, which is a ratio of total IR to UV luminosity, is commonly used to estimate the amount of dust attenuation of UV light. We calculated the median dust attenuation values to be 1.15 and 1.22 magnitudes (equivalent to IRX of 3.17 and 3.49) in the redshifts bins centered at 0.7 and 1.0 respectively. These values suggest that the dust content of our galaxies does not change significantly over this redshift range.

\begin{figure}
    \centering
    \includegraphics[width=\columnwidth]{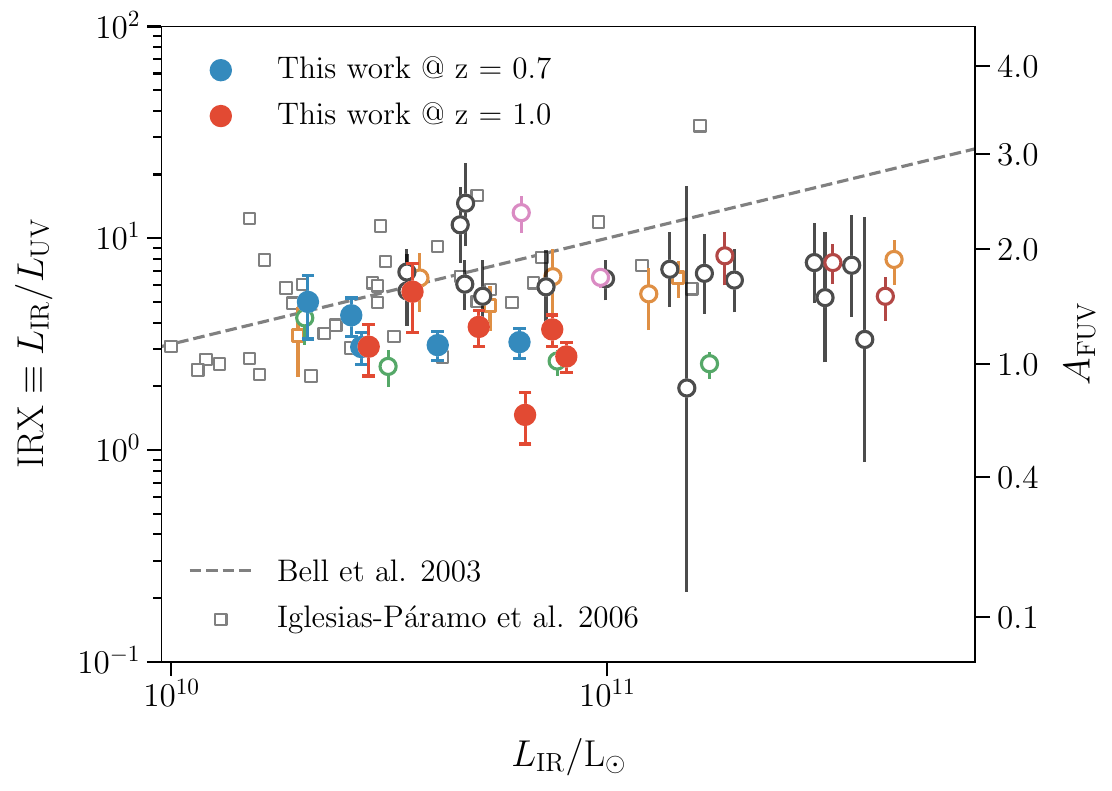}
    \caption{We plot the $\mathrm{IRX}$ ratio as a function of FIR luminosity in this Figure. 
    The symbols and colour coding is the same as in Figure \ref{fig:AFUV_Z}.
    We plot the individual galaxies in the nearby Universe, from the NUV selected sample (with a median redshift of 0.013) from \citet{2006ApJS..164...38I} as grey hollow squares, in addition to other literature values mentioned in Figure \ref{fig:AFUV_Z}.}
    \label{fig:AFUV_LIR}
\end{figure}

As depicted in Figure \ref{fig:AFUV_Z}, the average dust attenuation appears to remain constant from the local Universe up to a redshift of 2.5 \citep{2007ApJS..173..432X,2013MNRAS.429.1113H,2013A&A...554A..70B}.
We note here that \citet{2007ApJS..173..432X} used 24 $\mu$m data to estimate their average dust attenuation, while \citet{2013A&A...554A..70B} used the 60 $\mu$ m LF of \citet{2005A&A...440L..17T} for their $z=0$ estimates.
Our results at redshifts of 0.7 and 1.0 are offset below other work at similar redshifts \citep{2007ApJS..173..432X,2013A&A...554A..70B}, but are in good agreement with studies in the local Universe \citep{2013A&A...554A..70B} or redshifts higher than those explored in our study \citep[][]{2013A&A...554A..70B,2018ApJ...853...56R}.

We do observe a weak trend between the dust attenuation and the total IR luminosity for our UV-selected galaxies in the redshift range $0.6-1.2$ (Figure \ref{fig:AFUV_LIR}), wherein the dust attenuation decreases with IR luminosity at redshifts 0.7 and 1.0.
The correlation coefficients are -0.61 and -0.43. 
However, these correlations have low significance (p values of 0.28 and 0.36 at z = 0.7 and 1.0), and the range of IR luminosities of our galaxies is not wide enough to draw any definitive
conclusions.

Previous studies at redshifts ranging from 0.6 to 3 do not report any trends with IR luminosity \citep{2007ApJS..173..432X,2013MNRAS.429.1113H,2016A&A...587A.122A,2018ApJ...853...56R}. 
In Figure \ref{fig:AFUV_LIR}, we plot the NUV-selected galaxies from \textit{GALEX}
surveys in the local Universe of \citet{2006ApJS..164...38I}, who used the \textit{IRAS} 60 $\mu$m data to estimate $L_\mathrm{IR}$. 
Their galaxies follow a relation (grey dashed line in Figure \ref{fig:AFUV_LIR}), obtained by \citet{2003ApJ...586..794B} for a local compilation which contains sources from the literature with FUV, optical, IR (60 and 100 $\mu$m) and radio wavelengths.
The local galaxies thus show a correlation wherein galaxies
with high IR luminosity are more dust attenuated, which is expected as the increased dust attenuation results in a larger fraction of UV radiation being absorbed by dust, consequently leading to a higher IR luminosity.
It is interesting why this behaviour stops as we go past redshift 0.6.
The majority of galaxies in the local datasets described in \citet{2006ApJS..164...38I} and \citet{2003ApJ...586..794B} do not appear to exceed a luminosity of $2 \times 10^{11} \mathrm{L_\odot}$.
Below this IR luminosity, the results from the local samples of \citet{2003ApJ...586..794B}, \citet{2006ApJS..164...38I}, and the high redshift values from the literature are consistent. 
However, above this limit, there seems to be a noticeable discrepancy. The discrepancy may imply that the \citet{2003ApJ...586..794B} relation does not apply beyond the luminosity range of local galaxies from which it was derived, although it is also possible that the relationship between IRX and IR luminosity may change with redshift.
The range of IR luminosities in our sample closely resembles that of the local datasets. However, our findings align with local results only at IR luminosities below $5 \times 10^{10} \mathrm{L_\odot}$. As luminosity increases, our results begin to deviate from the \citet{2003ApJ...586..794B} curve and other high redshift measurements. 

\begin{figure}
    \centering
    \includegraphics[width=\columnwidth]{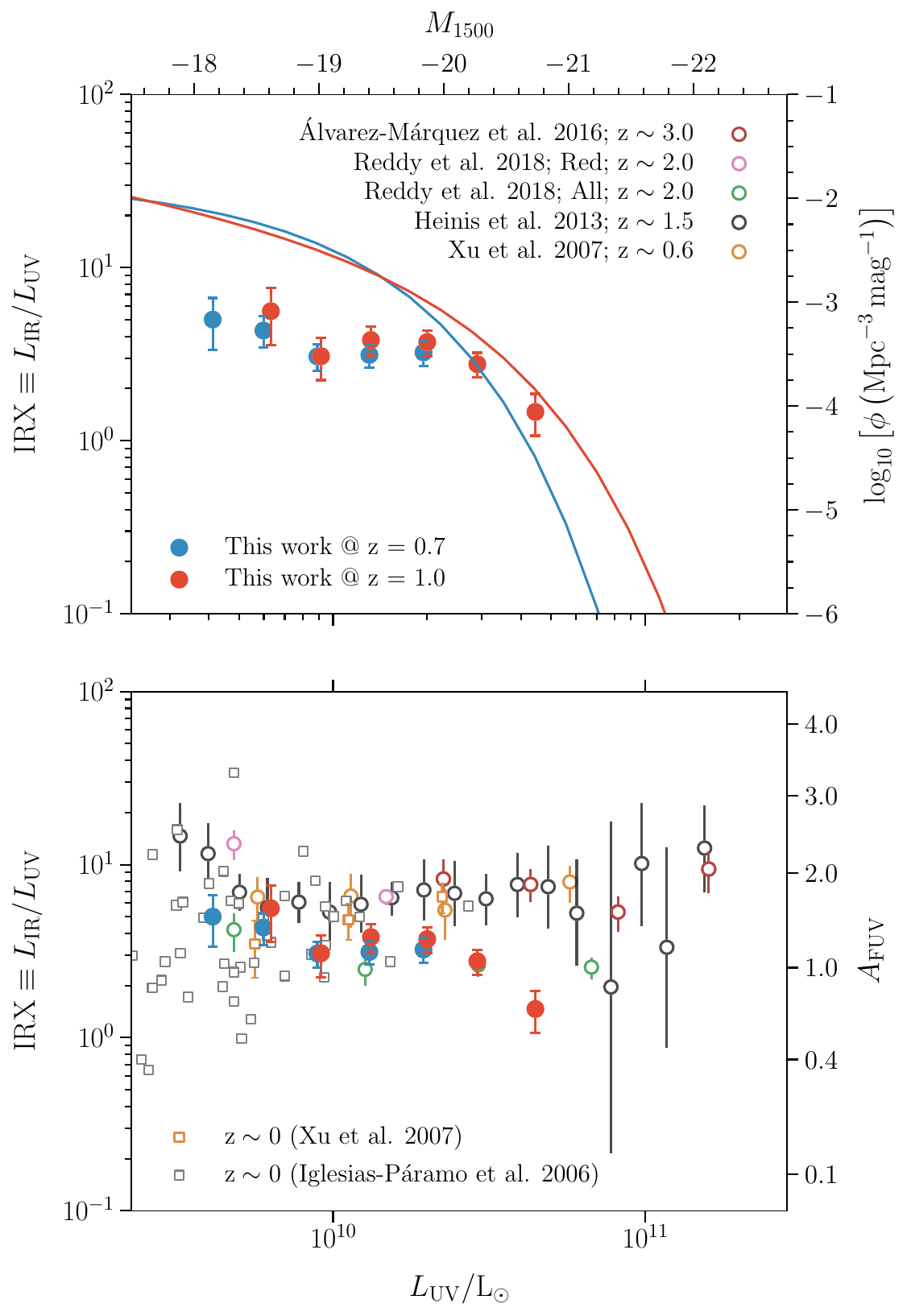}
    \caption{The $\mathrm{IRX}$ ratio as a function of UV luminosity. In the \textit{top panel} we show the values calculated in our study as blue and red solid circles at redshifts 0.7 and 1.0.
    We also show the UV luminosity functions of galaxies at redshifts of 0.7 (blue) and 1.0 (red) from \citet{2022MNRAS.511.4882S} with the corresponding y-axis label on the right hand side.
    Note that the y-axes of the IRX values (datapoints) and the luminosity functions (curves) are independent of each other.
    The \textit{bottom panel} shows our IRX values along with the results from 
    the literature. The symbols and colour coding is the same as in Figures \ref{fig:AFUV_Z} and \ref{fig:AFUV_LIR}.}
    \label{fig:AFUV_UVLF}
\end{figure}

In Figure \ref{fig:AFUV_UVLF}, we present the relationship between $\mathrm{IRX}$ and UV luminosity. 
We observe a trend in both redshift ranges for IRX to be smaller at higher luminosities.
The Spearman correlation coefficients are -0.6 and -0.8 (with p-values of 0.28 and 0.04) at redshifts of 0.7 and 1.0.

In the existing literature, various studies have proposed different behaviours.
There are studies using UV selection from the local Universe up to redshift 8, which report a decreasing trend of $\mathrm{IRX}$ with increasing $L_\mathrm{UV}$ \citep[e.g.][]{2009A&A...507..693B,2009ApJ...705..936B,2014ApJ...793L...5K}. 
Some others report a flat $\mathrm{IRX}-L_\mathrm{UV}$ relationship for UV-selected galaxies with average redshifts in the range $0.6 - 2$ \citep[e.g. see][]{2007ApJS..173..432X,2013MNRAS.429.1113H} 
and for Lyman-break galaxies with average redshifts from 2 to 8 \citep[see][]{2011MNRAS.417..717W,2012ApJ...754...83B,2016A&A...587A.122A,2018ApJ...853...56R}.
We remark here that \citet{2009ApJ...705..936B,2012ApJ...754...83B}, \citet{2011MNRAS.417..717W},
and \citet{2014ApJ...793L...5K}, used the UV spectral slope to estimate dust attenuation, and the results of \citet{2009A&A...507..693B} were based on rather uncertain mid-IR to total-IR calibrations.

Our average IRX values within the two redshift bins ($0.6-0.8$ and $0.8-1.2$) are lower than those reported in previous stacking studies conducted at different redshifts, specifically $z=0.6$ \citep[][]{2007ApJS..173..432X}, $z=1.5$ \citep[][]{2013MNRAS.429.1113H} and $z=3.0$ \citep{2016A&A...587A.122A}.
Furthermore, comparing to the results of \citep{2007ApJS..173..432X}, who stacked the local ($z=0$) UV-selected sample of \citet{2006ApJS..164...38I}, we observe that their average IRX value is also higher than what we find in both our redshift bins (Figures \ref{fig:AFUV_LIR} and \ref{fig:AFUV_UVLF}).
A similar observation was made by \citet{2018ApJ...853...56R}, using a sample that is dominated 
by blue ($\beta \leq -1.4$) star-forming galaxies at redshift $\sim 2$.
Our average results at redshifts 0.7 and 1.0 tend to align more closely with the values reported by \citet{2018ApJ...853...56R} than the other studies.

The discrepancies in the behaviour of the $\mathrm{IRX}-L_\mathrm{UV}$ relation are often attributed to the way samples are selected \citep[][]{2007ApJS..173..404B}.
UV-selected samples, in particular, tend to favour galaxies with lower dust content, resulting in most bright UV galaxies having low IR luminosities. Consequently, $\mathrm{IRX}$ is expected to exhibit a negative correlation with $L_\mathrm{UV}$ for a UV-selected sample.
For our case, the downward trend can be explained if we assume a population of star-forming galaxies and a distribution of extinction in
those galaxies. We would expect the lowest extinction galaxies to have the largest 
contribution in the brightest absolute magnitude bins, so that 
the balance of absorbed sources, or the typical degree of extinction, will change as we move from the bright end to the faint end of the LF.

\begin{figure}
    \centering
    \includegraphics[width=\columnwidth]{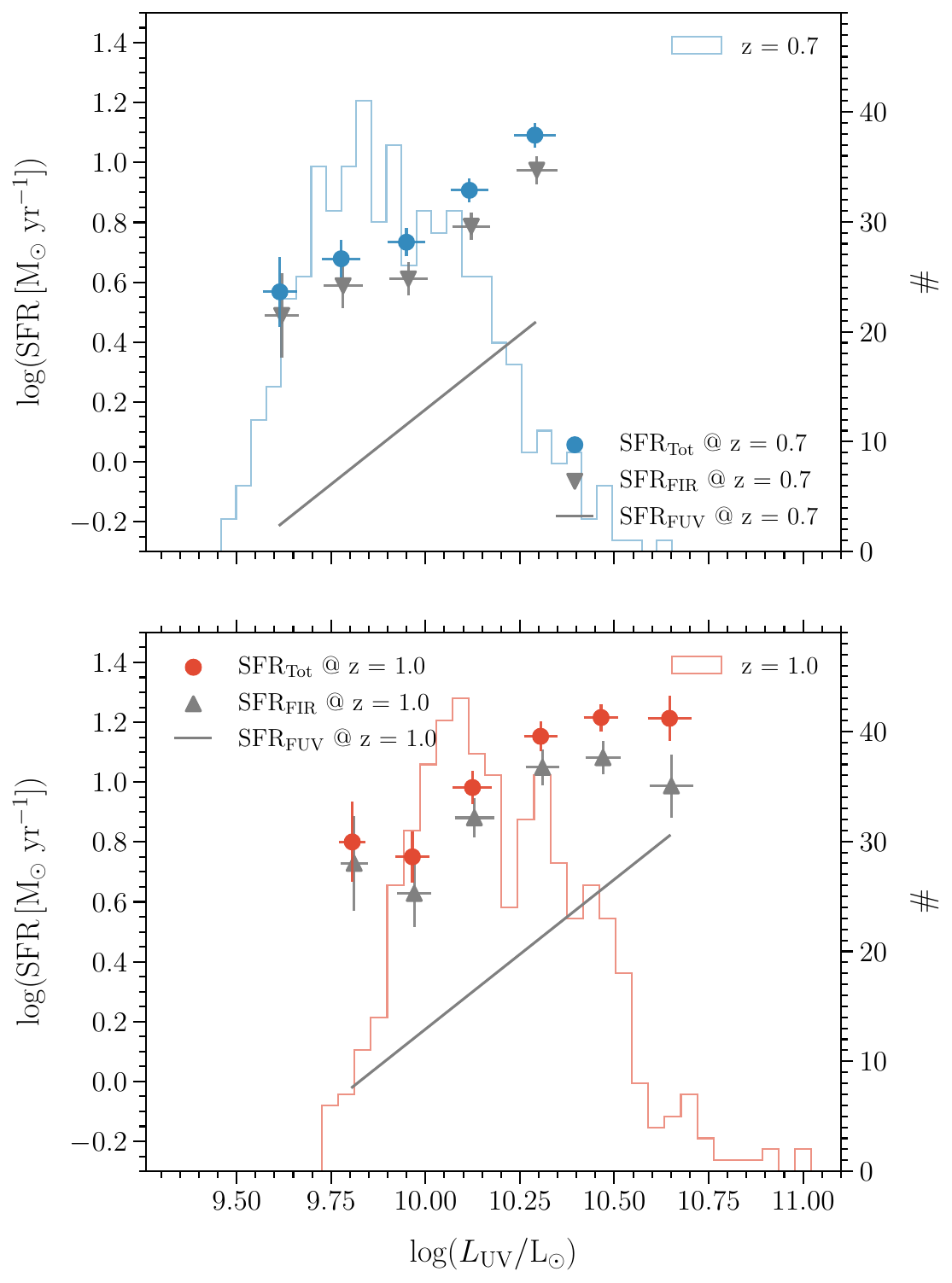}
    \caption{We plot the SFR as a function of the UV luminosity at 
    redshifts 0.7 and 1.0 respectively in the \textit{top} and \textit{bottom} panels of this Figure. 
    The histograms represent the distributions of the star-forming galaxies in the UV luminosity space with the number of galaxies marked on the y-axis on the right axis similar to the right panel of Figure \ref{fig:T_LIR_z_LUV}. 
    The blue and red-filled circles show the total SFR at redshifts 0.7 and 1.0.
    The gray-filled triangles show the SFR from IR luminosity and the lines show the SFR calculated using the UV luminosity.}
    \label{fig:SFR_LUV}
\end{figure}

\begin{figure}
    \centering
    \includegraphics[width=\columnwidth]{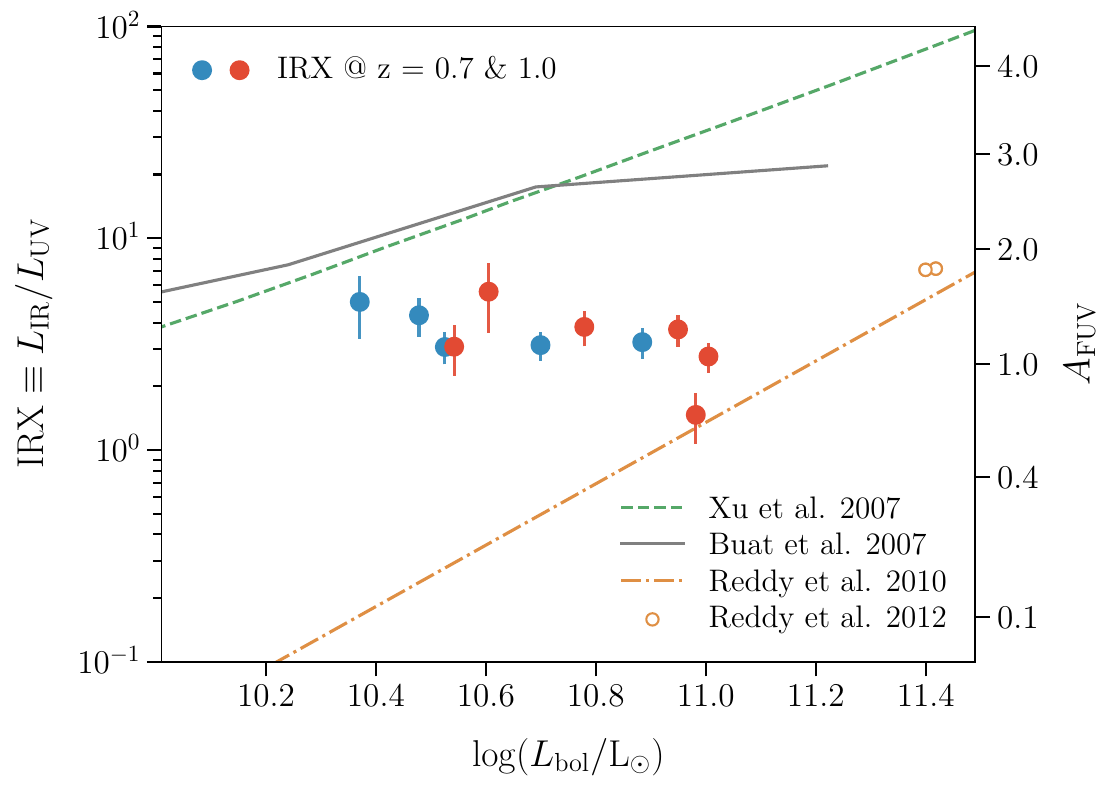}
    \caption{We plot here the $\mathrm{IRX}$ ratio and the dust attenuation 
    ($A_\mathrm{FUV}$) as a function of the bolometric luminosity (which is used as a proxy for the total SFR by the studies, we compare our results to). 
    The red and blue solid circles represent our values at redshift $0.6-0.8$ and $0.8-1.2$. 
    The green dashed line, grey solid line and yellow dot-dashed line show the results from 
    \citet{2007ApJS..173..432X,2007A&A...469...19B} and \citet{2010ApJ...712.1070R}.
    The yellow hollow circles represent the values of average dust attenuation from \citet{2012ApJ...744..154R}.}
    \label{fig:irx_sfr}
\end{figure}

\begin{figure*}
    \centering
    \hspace*{-1.1cm}\includegraphics[width=0.50\textwidth]{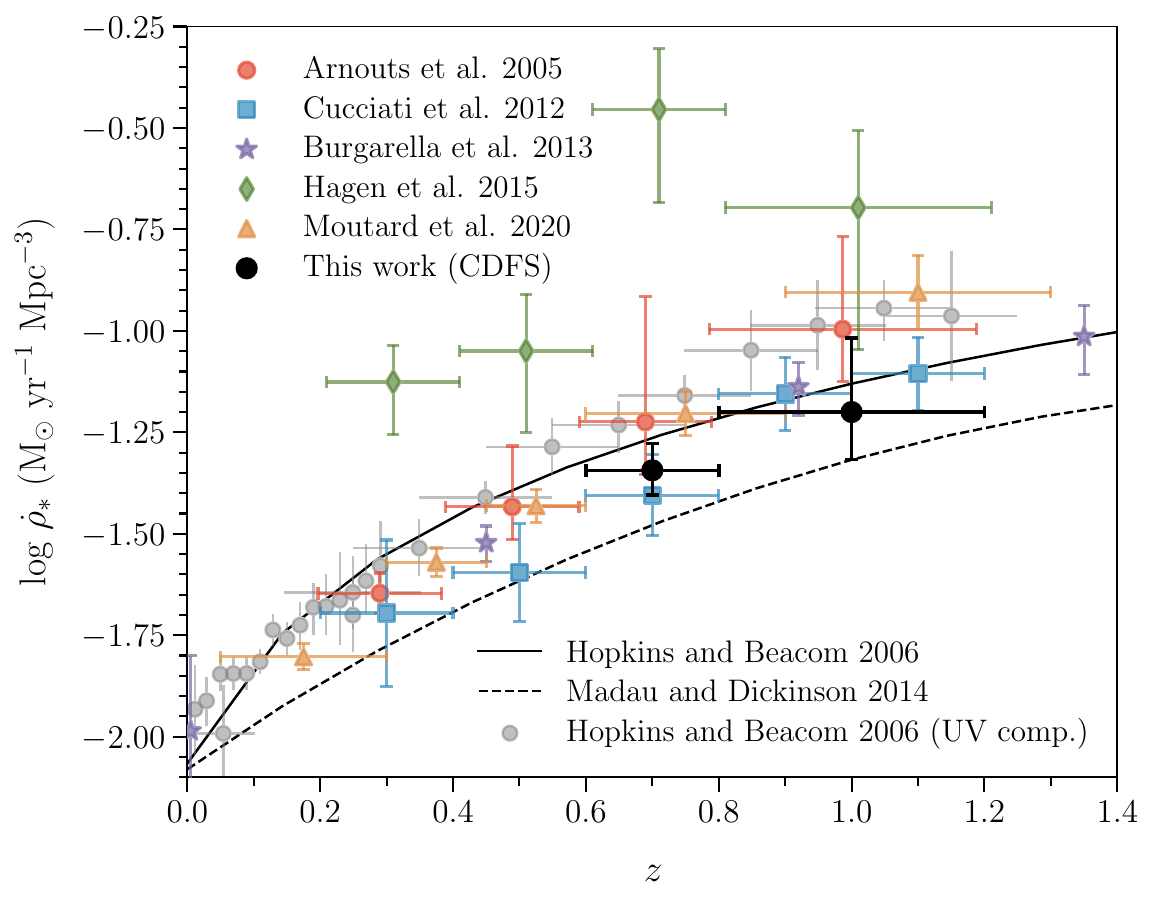}
    \caption{The SFRD of the UV-selected star-forming galaxy sample. Our results are presented as black solid circles. We plot the literature results as blue squares \citep[][]{refId0}, purple stars \citep[][]{2013A&A...554A..70B}, green diamonds \citep[][]{2015ApJ...808..178H} and yellow triangles showing \citet{2020MNRAS.494.1894M} values corrected for dust attenuation using \citet{2005ApJ...632..169L}. The grey circles represent dust-corrected
    UV-based SFRD from the \citet{2006ApJ...651..142H} compilation. From this compilation, we plot the SFRD values calculated using the \citet{2005ApJ...619L..43A} data in red circles.
    Trends obtained from \citet{2006ApJ...651..142H} and \citet{2014ARA&A..52..415M}
    are shown as solid black and dashed lines, respectively.}
    \label{fig:sfrd}
\end{figure*}

\subsection{Total star formation rate}

We estimate the SFR for our UV-selected sample using equations in Section \ref{sec:3.4}. The resulting values of the total SFR, which is the sum of the UV and infrared components of the SFR ($\mathrm{SFR}_\mathrm{Tot} = \mathrm{SFR}_\mathrm{UV} + \mathrm{SFR}_\mathrm{IR}$), are shown in Figure \ref{fig:SFR_LUV} assuming the \citet{2001MNRAS.322..231K} IMF. For comparison, we have also plotted the SFR values that were calculated using the UV luminosity ($\mathrm{SFR}_\mathrm{UV}$) and the IR luminosity ($\mathrm{SFR}_\mathrm{IR}$) separately.

It is evident from Figure \ref{fig:SFR_LUV} that if we relied solely on ultraviolet (UV) indicators, we would be underestimating the mean SFR by approximately a factor of 3 in redshift bins centred at 0.7 and 1.0, respectively.
At redshift 0.7, the underestimate decreases as the UV luminosity increases.
This is expected behaviour, as we observe the same in the IRX vs ${L_\mathrm{UV}}$ plot (Figure \ref{fig:AFUV_UVLF}).

The difference between the UV-derived and total SFR is indicative of the substantial amount of dust present, which attenuates the UV light, thereby obscuring the star formation. 
The relationship between this attenuation and the SFR should be worth investigating. Some previous studies using samples selected using the UV \citep[][]{2007A&A...469...19B,2010ApJ...712.1070R,2012ApJ...744..154R}, the Lyman-break  \citep{2006ApJ...653.1004R} and 24 $\mu$m observation \citep{2007ApJ...670..301Z} have demonstrated a positive correlation between these two quantities i.e. higher dust attenuation for higher SFR (or bolometric luminosity). 
However, we did not find a significant correlation between these quantities in either redshift bin in this study: 
in Figure \ref{fig:irx_sfr}, apart from a single low datapoint, IRX appears roughly constant with bolometric luminosity. We also show results from UV-selected galaxies in the 
local Universe \citep{2007A&A...469...19B}, at redshift of 0.6 \citep{2007ApJS..173..432X} and at redshift of 2 \citep{2012ApJ...744..154R}. For comparison,
the values for Lyman-$\alpha$ galaxies at redshift 2 \citep{2010ApJ...712.1070R} are also plotted. We can see that the \citet{2007A&A...469...19B} 
findings also have a flat trend in the luminosity range of our sources, although their values have higher normalisation. 
Among the redshift bins of our study ($0.6-0.8$ and $0.8-1.2$) we did not observe any significant change in dust attenuation given the bolometric luminosity.

\subsection{The star formation rate density}

We calculate the contribution to the SFRD at redshift 0.7 and 1.0 from the UV sources using our UV-selected galaxy sample. 
To estimate the SFRD, we use the UV luminosity density provided by \citet{2022MNRAS.511.4882S} at these redshifts, which is then converted into the SFRD using eq. \ref{eqn:sfruv}, assuming the \citet{2001MNRAS.322..231K} IMF. 
We take into account the impact of dust on the UV estimates and correct them accordingly by using the dust attenuation inferred from the IRX ratio (see Eq. \ref{eqn:IRX} in Section \ref{sec:3.5}).

Our results are illustrated in Figure \ref{fig:sfrd} along with previous estimates based on UV luminosity.
We do not find any significant evolution of the SFRD from redshift 1.0 to 0.7. 
Compared to previous works such as \citet{2005ApJ...619L..43A} and other UV luminosity-based estimates compiled in \citet{2006ApJ...651..142H}, we observe a good level of agreement at redshifts 0.7 and 1.0. It is worth noting that this fit is primarily driven by a large number of data points at redshifts smaller than 0.5, covers a wider range of redshifts, and takes into consideration the SFRD measured from tracers other than UV.
At redshift 0.7, we notice a deviation of more than $1 \sigma$ from the UV compilation of \citet{2006ApJ...651..142H} and \citet{2020MNRAS.494.1894M}. However, our results are within the error bars of the SFRD calculated using the \citet{2005ApJ...619L..43A} results. 

We calculate the fraction of obscured star formation, using the obscured to total star formation ratio, assuming the energy balance argument. These are estimated to be 65 and 68 per cent at redshifts 0.7 and 1.0, suggesting the dominance of dust-obscured components in the overall SFRD in these redshift bins. These figures also imply that the dust content of our galaxies does not change significantly over this redshift range.
Contrasting these results with those from the local Universe - roughly 75 per cent dust obscured star formation obtained by \citet[][]{2013A&A...553A.132M}, it is evident that there is no significant evolution in the fraction of dust-obscured star formation rate density from redshift 0 to 0.7, and even further to redshift 1 \citep{2005ApJ...632..169L}.

\begin{table}
\setlength{\tabcolsep}{3.5pt}
\centering
\caption{The luminosity and SFR density of the UV-selected galaxies from CDFS at redshifts 0.7 and 1.0.}
\label{tab:sfrd2}
  \begin{tabular}{lcccr}
    \hline\hline
    \noalign{\vskip 1.0mm}
    $^{a}\langle z \rangle$ &
    $^{b}\mathrm{log} \ \rho$ &
    $^{c}A_{\mathrm{FUV}}$ &
    $^{d}\mathrm{log} \ \dot \rho_{*}^\mathrm{UV}$ &
    $^{e}\mathrm{log} \ \dot \rho_{*}^\mathrm{corr}$ \\    
    \noalign{\vskip 1.0mm}
    &
    $(\mathrm{erg} \ \mathrm{s}^{-1} \mathrm{Hz}^{-1} \mathrm{Mpc^{-3}})$ &
    (mag) &
    \multicolumn{2}{c}{$(\mathrm{M_\odot} \ \mathrm{yr}^{-1} \mathrm{Mpc^{-3}})$} \\
    \noalign{\vskip 1.0mm}
    \hline
    \noalign{\vskip 0.5mm}
    0.7 &  $26.31_{-0.06}^{+0.07}$ & $1.15$ & $-1.80_{-0.06}^{+0.07}$ & $-1.34_{-0.06}^{+0.07}$
    \\
    \noalign{\vskip 0.75mm}
    1.0 &  $26.43_{-0.12}^{+0.18}$ & $1.22$ & $-1.68_{-0.12}^{+0.18}$ & $-1.19_{-0.12}^{+0.18}$ \\
    \noalign{\vskip 0.5mm}
    \hline
  \end{tabular}
  \begin{minipage}{0.47\textwidth}
      \textsuperscript{$a$}{The center of the redshift bin.} \\
      \textsuperscript{$b$}{From \citet{2022MNRAS.511.4882S}.} \\
      \textsuperscript{$c$}{The median value for all UV luminosity bins.} \\
      \textsuperscript{$d$}{Un-obscured SFRD Calculated from the UV luminosity density using the
      \citet{1998ARA&A..36..189K} calibration assuming the \citet{2001MNRAS.322..231K} IMF.} \\
      \textsuperscript{$e$}{The un-obscured SFRD corrected for the dust attenuation using 
      $A_\mathrm{FUV}$.} \\
  \end{minipage}
\end{table}

\begin{figure*}
    \centering
    \hspace*{-1.1cm}\includegraphics[width=0.50\textwidth]{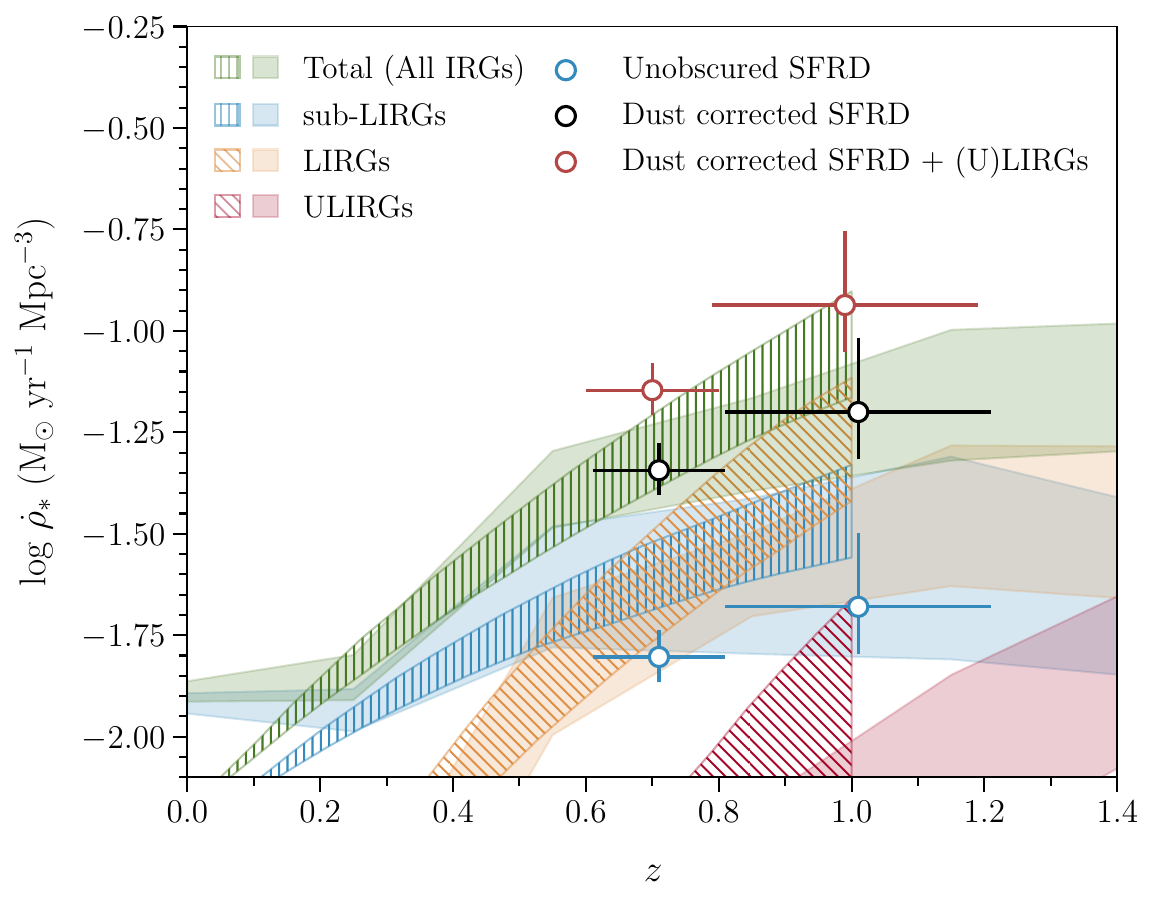}
    \caption{We compare our results for the SFRD with and without dust correction. For reference,
    we also plot the total SFRD calculated from the IR galaxies from the studies by \citet{2005ApJ...632..169L} and \citet{2013A&A...553A.132M} as hatched and shaded regions, respectively.
    These works also considered the contributions of normal IR galaxies (sub-LIRGs), LIRGs, and ULIRGs to the total IR luminosity density (and SFRD). Here, we show the total SFRD from the IR galaxies as green-coloured regions. Contributions from sub-LIRGs, LIRGs, and ULIRGs are shown as blue, yellow, and red regions.
    The data points show the values obtained in this work. The blue hollow circles represent the unobscured SFRD obtained from the UV luminosity density of \citet{2022MNRAS.511.4882S} without correcting for dust attenuation.
    The black hollow circles are the dust-corrected values using the attenuation factor ($A_\mathrm{FUV}$) calculated as shown in Section \ref{sec:3.5}.
    The red hollow circles represent the SFRD obtained if we add the contribution of the ULIRGs from the studies mentioned above to the dust-corrected UV SFRD (black circles).}
    \label{fig:sfrd2}
\end{figure*}

The results presented above may not accurately reflect the total star formation activity in the studied redshift range because of the possibility of missing heavily obscured systems in our UV selection. 
It has previously been shown that these particular galaxies, (U)LIRGs, dominate star formation activity in the redshift range explored in our study \citep[][]{2005ApJ...632..169L}.
So, we test whether or not the contribution from bright IR galaxies (which might not have been detected in our UV-selected catalogue) makes a significant difference to the SRFD estimated in this study. 
To estimate the SFRD of these bright IR galaxies, we used the results from \citet{2005ApJ...632..169L} and \citet{2013A&A...553A.132M}. 
The study conducted by \citet{2005ApJ...632..169L} utilised a sample of 24 $\mu \mathrm{m}$ sources from \textit{Spitzer} MIPS in CDFS to determine the IR luminosity function and the total IR luminosity density at redshift $0 \leq z \leq 1$. On the other hand, \citet{2013A&A...553A.132M} used observations of GOODS fields from \textit{Herschel} PACS to obtain the IR LF and luminosity density. 
Both of these studies also provide insight into the relative contribution of Luminous Infrared Galaxies (LIRGs; $10^{11}{\mathrm{L}\odot} < L_\mathrm{IR} < 10^{12}{\mathrm{L}\odot}$; see \citet[]{1996ARA&A..34..749S}, Ultra-Luminous Infrared Galaxies (ULIRGs; $10^{12}{\mathrm{L}\odot} < L_\mathrm{IR} < 10^{13}{\mathrm{L}\odot}$; \citet[]{1996ARA&A..34..749S,1998ApJ...498..579G}), and the normal (sub-luminous; $L_\mathrm{IR} < 10^{11}{\mathrm{L}_\odot}$) infrared galaxies. 
A summary of their findings is depicted in Figure \ref{fig:sfrd2}, which shows reasonable consistency up to redshift 0.6. However, as we move towards redshift 1, the values obtained by \citet{2005ApJ...632..169L} start to deviate slightly above the ones from \citet{2013A&A...553A.132M}. 
It is important to note here that neither of these studies are corrected for the contribution of the
AGN to the IR galaxies, and AGN are known to have a significant effect on the IR luminosity of galaxies from local Universe up to a redshift 2.5 \citep{2019MNRAS.485L..11S,2021MNRAS.503.3992S}.

We compare the values estimated in our study to the average of these two works in Figure \ref{fig:sfrd2}.
The blue circles show our unobscured SFRD calculated from direct UV observations. 
It is noteworthy that these values are observed to fall on the blue-shaded regions, which represent the contribution of sub-LIRGs to the overall IR contribution. This indicates that our sample may account, to some extent, for the contribution from these sub-LIRGs observed in the 24 $\mu \mathrm{m}$ and FIR samples of \citet{2005ApJ...632..169L} and \citet{2013A&A...553A.132M}, respectively.
This behaviour is somewhat expected, as the IR luminosities of the majority of our UV-selected galaxies fall within the range, which is typically associated with the sub-LIRG population.
Thus, the sub-LIRG populations might receive some contribution to the SFRD from the UV-bright sources.
The black hollow circles in Figure \ref{fig:sfrd2} show the dust-corrected SFRD values. 
These values are generally consistent with the total IR SFRD estimates of \citet{2005ApJ...632..169L} and \citet{2013A&A...553A.132M}.
We added the average of the (U)LIRGs SFRD from these studies to our dust-corrected UV SFRD measurements. 
This yields values (red circles in Figure \ref{fig:sfrd2}) that greatly exceed the total IR luminosity density (and the dust-corrected UV SFRD estimates).
This is an interesting outcome, suggesting that if we consider (U)LIRGs as a distinct population from our UV galaxies and sum their SFRD values to our dust-corrected UV SFRD, we arrive at a value much higher than if we were to correct the UV SFRD for dust extinction using a mean attenuation correction. 
This implies that although these populations (ULIRGs and LIRGs) might not be represented in our sample, we have not missed much extinction from our UV selection, provided that we account for dust extinction. This holds true despite the inherent uncertainties associated with attenuating UV galaxies.

\section{Conclusion}
\label{sec:6}

In this work, we investigate the dust properties of a sample of UV-selected galaxies from the Chandra deep field south (CDFS) in the redshift range $0.6-1.2$. 
The sample under consideration comprises 1070 galaxies, with a magnitude range of $M_\mathrm{UV} = [-21.15, -19.0]$, after removing the UV LF bins containing < 25 sources.
This conservative cut makes sure we have robust enough statistics for our calculations.
To assess the average FIR properties of this UV-selected sample, we make use of the FIR maps of the CDFS, generated by the \textit{Herschel} Multi-tiered Extragalactic Survey (HerMES). The FIR maps are created on the basis of observations from the PACS and SPIRE instruments onboard \textit{Herschel}.

We stack the UV sources from the CDFS dataset onto the FIR maps obtained from \textit{Herschel}-PACS at 100 and 160 $\mu \mathrm{m}$, as well as \textit{Herschel}-SPIRE at 250, 350 and 500 $\mu \mathrm{m}$, in order to determine the average flux densities as a function of the redshift and UV luminosity binned according to the UV LF of these galaxies.
Prior to stacking, we deblend the FIR maps to mitigate the effects of blending and confusion of sources in the \textit{Herschel} IR maps.
Using the stacked fluxes, we determined the average dust temperature and total FIR luminosities (from 8-1000 $\mu \mathrm{m}$) for the galaxies in each bin of the UV LF. These FIR luminosities, along with the UV luminosities, are then employed to estimate the dust attenuation of the galaxies and to characterise the evolution of the comoving SFR density between redshifts $0.6-1.2$. 
The primary conclusions derived from our study can be summarised as follows:

1. The IR luminosities of our UV-selected sources are on average in the range \num{2.15e10} to \num{6.5e10} $\mathrm{L_\odot}$, placing them in the sub-LIRG category. We find that the typical luminosity-weighted dust temperatures at redshifts 0.7 and 1.0 are $30.28 \pm 1.33$ K and $33.12 \pm 1.26$ K, respectively.
We have not observed any significant trends between the average dust temperatures and integrated IR luminosities of these galaxies at a fixed redshift.
Furthermore, our analysis of the temperature (averaged over the UV LF bins) within the redshift range explored in our study has revealed no significant variation in the redshift range of our study. However, it is important to remark here that this study explores a rather limited range in redshift space. When we add data from other studies conducted in the redshift range of 0.1 to 1.5, our values agree with a weak trend between these parameters observed in previous studies.

2. Our UV-selected galaxies have median dust obscuration levels of $\mathrm{IRX} = 3.17 \pm 0.52$ and $3.49 \pm 0.51$, which correspond to dust attenuation of $1.15 \pm 0.24$ and $1.22 \pm 0.23$ magnitudes, at redshifts 0.7 and 1.0, respectively. 
We did not find any changes in the dust attenuation within the redshift range covered by our study, which suggests that the dust content in UV-selected star-forming galaxies does not evolve very much between redshifts of 1.0 to 0.7. 
We do observe a pattern in the values of IRX with IR luminosity, wherein the IRX decreases as the IR luminosity increases for a constant redshift. However, the trend is very weak and cannot be substantiated due to the small range of IR luminosities covered in this study.
No significant trends are detected between IRX and redshift at a constant IR luminosity. However, in the case of local galaxies, there is a positive correlation between IRX and IR luminosity. We speculate that this difference may be due to selection bias.
We see an increase in IRX with decreasing UV luminosity.

3. It is observed that the SFR calculated using UV indicators is underestimated by a factor of 3 at redshifts of 0.7 and 1.0 compared to the total SFR. 
This offset decreases as the UV luminosity increases for both redshift bins.
This indicates that the dust obscuration decreases as the UV luminosity of the galaxies increases, within the range analysed in this study.
It has also been found that the relationship between IRX and bolometric luminosity remains unchanged from redshift 1 to 0.7. The IRX exhibits a roughly constant trend with increasing bolometric luminosity, which is in agreement with the local relations for UV-selected galaxies, where these quantities show a weak correlation within the luminosity range of our sources. 
Overall, the results are consistent with the proposed picture from previous studies that the UV-selected galaxies at higher redshifts exhibit a lesser degree of dust attenuation at a fixed bolometric luminosity compared to those in the local Universe.
However, we did not observe any evolution of the dust attenuation at a given bolometric luminosity from redshift 0.7 to 1 in our sample.

4. We did not find any significant change in the SFR density with the redshift changing from 1.0 to 0.7. The values at both our redshifts agree reasonably well with previous investigations. The ratio of the obscured to the total star formation is in the 65-70 per cent range.

\section{Acknowledgements}

This research makes use of observations taken with the \text{Herschel} observatory.
\textit{Herschel} is an ESA space observatory with science instruments provided by European-led Principal Investigator consortia and with important participation from NASA.
This research has used data from the HerMES project (\url{http://hermes.sussex.ac.uk}). HerMES is a \textit{Herschel} Key Programme utilising Guaranteed Time from the SPIRE instrument team, ESAC scientists, and a mission scientist.
The HerMES data were accessed through the \textit{Herschel} Database in Marseille (HeDaM - \url{http://hedam.lam.fr}) operated by CeSAM and hosted by the Laboratoire d'Astrophysique de Marseille.
MS thanks Unnikrishnan Sureshkumar for the help and discussions
regarding the calculation of the angular correlation functions used in this
work.
MS also extends their gratitude to Benjamin Magnelli for providing the MIPS-24$\mu$m/70$\mu$m ECDFS FIDEL data.
We thank the referee Tom Bakx for their constructive report which further improved this manuscript.

\section{Data Availability}
\label{sec:9}
The \textit{Herschel} maps used in this paper can be obtained from the HeDam database available at 
\url{https://hedam.lam.fr/HerMES/index/dr4} and the PEP pages at \url{https://www.mpe.mpg.de/ir/Research/PEP/DR1}. 
The UVW1 source list is provided as the supplementary Table with the online version of our previous study \citet{2022MNRAS.511.4882S}.
Other supporting material related to this article is available on a 
reasonable request to the corresponding author.

\bibliographystyle{mnras}
\bibliography{References}

\begin{thebibliography}{}
\makeatletter
\relax
\def\mn@urlcharsother{\let\do\@makeother \do\$\do\&\do\#\do\^\do\_\do\%\do\~}
\def\mn@doi{\begingroup\mn@urlcharsother \@ifnextchar [ {\mn@doi@} {\mn@doi@[]}}
\def\mn@doi@[#1]#2{\def\@tempa{#1}\ifx\@tempa\@empty \href {http://dx.doi.org/#2} {doi:#2}\else \href {http://dx.doi.org/#2} {#1}\fi \endgroup}
\def\mn@eprint#1#2{\mn@eprint@#1:#2::\@nil}
\def\mn@eprint@arXiv#1{\href {http://arxiv.org/abs/#1} {{\tt arXiv:#1}}}
\def\mn@eprint@dblp#1{\href {http://dblp.uni-trier.de/rec/bibtex/#1.xml} {dblp:#1}}
\def\mn@eprint@#1:#2:#3:#4\@nil{\def\@tempa {#1}\def\@tempb {#2}\def\@tempc {#3}\ifx \@tempc \@empty \let \@tempc \@tempb \let \@tempb \@tempa \fi \ifx \@tempb \@empty \def\@tempb {arXiv}\fi \@ifundefined {mn@eprint@\@tempb}{\@tempb:\@tempc}{\expandafter \expandafter \csname mn@eprint@\@tempb\endcsname \expandafter{\@tempc}}}

\bibitem[\protect\citeauthoryear{{{\'A}lvarez-M{\'a}rquez} et~al.,}{{{\'A}lvarez-M{\'a}rquez} et~al.}{2016}]{2016A&A...587A.122A}
{{\'A}lvarez-M{\'a}rquez} J.,  et~al., 2016, \mn@doi [\aap] {10.1051/0004-6361/201527190}, \href {https://ui.adsabs.harvard.edu/abs/2016A&A...587A.122A} {587, A122}

\bibitem[\protect\citeauthoryear{{Arnouts} et~al.,}{{Arnouts} et~al.}{2005}]{2005ApJ...619L..43A}
{Arnouts} S.,  et~al., 2005, \mn@doi [\apjl] {10.1086/426733}, \href {https://ui.adsabs.harvard.edu/abs/2005ApJ...619L..43A} {619, L43}

\bibitem[\protect\citeauthoryear{{Bell}}{{Bell}}{2003}]{2003ApJ...586..794B}
{Bell} E.~F.,  2003, \mn@doi [\apj] {10.1086/367829}, \href {https://ui.adsabs.harvard.edu/abs/2003ApJ...586..794B} {586, 794}

\bibitem[\protect\citeauthoryear{{Bertin} \& {Arnouts}}{{Bertin} \& {Arnouts}}{1996}]{1996A&AS..117..393B}
{Bertin} E.,  {Arnouts} S.,  1996, \mn@doi [\aaps] {10.1051/aas:1996164}, \href {https://ui.adsabs.harvard.edu/abs/1996A&AS..117..393B} {117, 393}

\bibitem[\protect\citeauthoryear{{B{\'e}thermin}, {Dole}, {Cousin}  \& {Bavouzet}}{{B{\'e}thermin} et~al.}{2010}]{2010A&A...516A..43B}
{B{\'e}thermin} M.,  {Dole} H.,  {Cousin} M.,   {Bavouzet} N.,  2010, \mn@doi [\aap] {10.1051/0004-6361/200913910}, \href {https://ui.adsabs.harvard.edu/abs/2010A&A...516A..43B} {516, A43}

\bibitem[\protect\citeauthoryear{{B{\'e}thermin} et~al.,}{{B{\'e}thermin} et~al.}{2012}]{2012A&A...542A..58B}
{B{\'e}thermin} M.,  et~al., 2012, \mn@doi [\aap] {10.1051/0004-6361/201118698}, \href {https://ui.adsabs.harvard.edu/abs/2012A&A...542A..58B} {542, A58}

\bibitem[\protect\citeauthoryear{{B{\'e}thermin} et~al.,}{{B{\'e}thermin} et~al.}{2015}]{2015A&A...573A.113B}
{B{\'e}thermin} M.,  et~al., 2015, \mn@doi [\aap] {10.1051/0004-6361/201425031}, \href {https://ui.adsabs.harvard.edu/abs/2015A&A...573A.113B} {573, A113}

\bibitem[\protect\citeauthoryear{{Blain}, {Barnard}  \& {Chapman}}{{Blain} et~al.}{2003}]{2003MNRAS.338..733B}
{Blain} A.~W.,  {Barnard} V.~E.,   {Chapman} S.~C.,  2003, \mn@doi [\mnras] {10.1046/j.1365-8711.2003.06086.x}, \href {https://ui.adsabs.harvard.edu/abs/2003MNRAS.338..733B} {338, 733}

\bibitem[\protect\citeauthoryear{{Boquien} \& {Salim}}{{Boquien} \& {Salim}}{2021}]{2021A&A...653A.149B}
{Boquien} M.,  {Salim} S.,  2021, \mn@doi [\aap] {10.1051/0004-6361/202140992}, \href {https://ui.adsabs.harvard.edu/abs/2021A&A...653A.149B} {653, A149}

\bibitem[\protect\citeauthoryear{{Bouwens} et~al.,}{{Bouwens} et~al.}{2009}]{2009ApJ...705..936B}
{Bouwens} R.~J.,  et~al., 2009, \mn@doi [\apj] {10.1088/0004-637X/705/1/936}, \href {https://ui.adsabs.harvard.edu/abs/2009ApJ...705..936B} {705, 936}

\bibitem[\protect\citeauthoryear{{Bouwens} et~al.,}{{Bouwens} et~al.}{2012}]{2012ApJ...754...83B}
{Bouwens} R.~J.,  et~al., 2012, \mn@doi [\apj] {10.1088/0004-637X/754/2/83}, \href {https://ui.adsabs.harvard.edu/abs/2012ApJ...754...83B} {754, 83}

\bibitem[\protect\citeauthoryear{{Bouwens} et~al.,}{{Bouwens} et~al.}{2015}]{2015ApJ...803...34B}
{Bouwens} R.~J.,  et~al., 2015, \mn@doi [\apj] {10.1088/0004-637X/803/1/34}, \href {https://ui.adsabs.harvard.edu/abs/2015ApJ...803...34B} {803, 34}

\bibitem[\protect\citeauthoryear{{Bouwens} et~al.,}{{Bouwens} et~al.}{2020}]{2020ApJ...902..112B}
{Bouwens} R.,  et~al., 2020, \mn@doi [\apj] {10.3847/1538-4357/abb830}, \href {https://ui.adsabs.harvard.edu/abs/2020ApJ...902..112B} {902, 112}

\bibitem[\protect\citeauthoryear{{Buat}}{{Buat}}{1992}]{1992A&A...264..444B}
{Buat} V.,  1992, \aap, \href {https://ui.adsabs.harvard.edu/abs/1992A&A...264..444B} {264, 444}

\bibitem[\protect\citeauthoryear{{Buat} \& {Xu}}{{Buat} \& {Xu}}{1996}]{1996A&A...306...61B}
{Buat} V.,  {Xu} C.,  1996, \aap, \href {https://ui.adsabs.harvard.edu/abs/1996A&A...306...61B} {306, 61}

\bibitem[\protect\citeauthoryear{{Buat}, {Donas}, {Milliard}  \& {Xu}}{{Buat} et~al.}{1999}]{1999A&A...352..371B}
{Buat} V.,  {Donas} J.,  {Milliard} B.,   {Xu} C.,  1999, \aap, \href {https://ui.adsabs.harvard.edu/abs/1999A&A...352..371B} {352, 371}

\bibitem[\protect\citeauthoryear{{Buat} et~al.,}{{Buat} et~al.}{2005}]{2005ApJ...619L..51B}
{Buat} V.,  et~al., 2005, \mn@doi [\apjl] {10.1086/423241}, \href {https://ui.adsabs.harvard.edu/abs/2005ApJ...619L..51B} {619, L51}

\bibitem[\protect\citeauthoryear{{Buat} et~al.,}{{Buat} et~al.}{2007a}]{2007ApJS..173..404B}
{Buat} V.,  et~al., 2007a, \mn@doi [\apjs] {10.1086/516645}, \href {https://ui.adsabs.harvard.edu/abs/2007ApJS..173..404B} {173, 404}

\bibitem[\protect\citeauthoryear{{Buat}, {Marcillac}, {Burgarella}, {Le Floc'h}, {Takeuchi}, {Iglesias-Par{\`a}mo}  \& {Xu}}{{Buat} et~al.}{2007b}]{2007A&A...469...19B}
{Buat} V.,  {Marcillac} D.,  {Burgarella} D.,  {Le Floc'h} E.,  {Takeuchi} T.~T.,  {Iglesias-Par{\`a}mo} J.,   {Xu} C.~K.,  2007b, \mn@doi [\aap] {10.1051/0004-6361:20066685}, \href {https://ui.adsabs.harvard.edu/abs/2007A&A...469...19B} {469, 19}

\bibitem[\protect\citeauthoryear{{Buat}, {Takeuchi}, {Burgarella}, {Giovannoli}  \& {Murata}}{{Buat} et~al.}{2009}]{2009A&A...507..693B}
{Buat} V.,  {Takeuchi} T.~T.,  {Burgarella} D.,  {Giovannoli} E.,   {Murata} K.~L.,  2009, \mn@doi [\aap] {10.1051/0004-6361/200912024}, \href {https://ui.adsabs.harvard.edu/abs/2009A&A...507..693B} {507, 693}

\bibitem[\protect\citeauthoryear{{Budav{\'a}ri} et~al.,}{{Budav{\'a}ri} et~al.}{2005}]{2005ApJ...619L..31B}
{Budav{\'a}ri} T.,  et~al., 2005, \mn@doi [\apjl] {10.1086/423319}, \href {https://ui.adsabs.harvard.edu/abs/2005ApJ...619L..31B} {619, L31}

\bibitem[\protect\citeauthoryear{{Burgarella}, {Buat}  \& {Iglesias-P{\'a}ramo}}{{Burgarella} et~al.}{2006}]{2006MNRAS.365..352B}
{Burgarella} D.,  {Buat} V.,   {Iglesias-P{\'a}ramo} J.,  2006, \mn@doi [\mnras] {10.1111/j.1365-2966.2005.09830.x}, \href {https://ui.adsabs.harvard.edu/abs/2006MNRAS.365..352B} {365, 352}

\bibitem[\protect\citeauthoryear{{Burgarella}, {Le Floc'h}, {Takeuchi}, {Huang}, {Buat}, {Rieke}  \& {Tyler}}{{Burgarella} et~al.}{2007}]{2007MNRAS.380..986B}
{Burgarella} D.,  {Le Floc'h} E.,  {Takeuchi} T.~T.,  {Huang} J.~S.,  {Buat} V.,  {Rieke} G.~H.,   {Tyler} K.~D.,  2007, \mn@doi [\mnras] {10.1111/j.1365-2966.2007.12063.x}, \href {https://ui.adsabs.harvard.edu/abs/2007MNRAS.380..986B} {380, 986}

\bibitem[\protect\citeauthoryear{{Burgarella} et~al.,}{{Burgarella} et~al.}{2013}]{2013A&A...554A..70B}
{Burgarella} D.,  et~al., 2013, \mn@doi [\aap] {10.1051/0004-6361/201321651}, \href {https://ui.adsabs.harvard.edu/abs/2013A&A...554A..70B} {554, A70}

\bibitem[\protect\citeauthoryear{{Calzetti}, {Kinney}  \& {Storchi-Bergmann}}{{Calzetti} et~al.}{1994}]{1994ApJ...429..582C}
{Calzetti} D.,  {Kinney} A.~L.,   {Storchi-Bergmann} T.,  1994, \mn@doi [\apj] {10.1086/174346}, \href {https://ui.adsabs.harvard.edu/abs/1994ApJ...429..582C} {429, 582}

\bibitem[\protect\citeauthoryear{{Calzetti}, {Armus}, {Bohlin}, {Kinney}, {Koornneef}  \& {Storchi-Bergmann}}{{Calzetti} et~al.}{2000}]{2000ApJ...533..682C}
{Calzetti} D.,  {Armus} L.,  {Bohlin} R.~C.,  {Kinney} A.~L.,  {Koornneef} J.,   {Storchi-Bergmann} T.,  2000, \mn@doi [\apj] {10.1086/308692}, \href {https://ui.adsabs.harvard.edu/abs/2000ApJ...533..682C} {533, 682}

\bibitem[\protect\citeauthoryear{{Calzetti} et~al.,}{{Calzetti} et~al.}{2007}]{2007ApJ...666..870C}
{Calzetti} D.,  et~al., 2007, \mn@doi [\apj] {10.1086/520082}, \href {https://ui.adsabs.harvard.edu/abs/2007ApJ...666..870C} {666, 870}

\bibitem[\protect\citeauthoryear{{Casey} et~al.,}{{Casey} et~al.}{2011}]{2011MNRAS.415.2723C}
{Casey} C.~M.,  et~al., 2011, \mn@doi [\mnras] {10.1111/j.1365-2966.2011.18885.x}, \href {https://ui.adsabs.harvard.edu/abs/2011MNRAS.415.2723C} {415, 2723}

\bibitem[\protect\citeauthoryear{{Casey} et~al.,}{{Casey} et~al.}{2014}]{2014ApJ...796...95C}
{Casey} C.~M.,  et~al., 2014, \mn@doi [\apj] {10.1088/0004-637X/796/2/95}, \href {https://ui.adsabs.harvard.edu/abs/2014ApJ...796...95C} {796, 95}

\bibitem[\protect\citeauthoryear{{Casey} et~al.,}{{Casey} et~al.}{2018}]{2018ApJ...862...77C}
{Casey} C.~M.,  et~al., 2018, \mn@doi [\apj] {10.3847/1538-4357/aac82d}, \href {https://ui.adsabs.harvard.edu/abs/2018ApJ...862...77C} {862, 77}

\bibitem[\protect\citeauthoryear{{Chapin} et~al.,}{{Chapin} et~al.}{2011}]{2011MNRAS.411..505C}
{Chapin} E.~L.,  et~al., 2011, \mn@doi [\mnras] {10.1111/j.1365-2966.2010.17697.x}, \href {https://ui.adsabs.harvard.edu/abs/2011MNRAS.411..505C} {411, 505}

\bibitem[\protect\citeauthoryear{{Chapman}, {Smail}, {Ivison}, {Helou}, {Dale}  \& {Lagache}}{{Chapman} et~al.}{2002}]{2002ApJ...573...66C}
{Chapman} S.~C.,  {Smail} I.,  {Ivison} R.~J.,  {Helou} G.,  {Dale} D.~A.,   {Lagache} G.,  2002, \mn@doi [\apj] {10.1086/340552}, \href {https://ui.adsabs.harvard.edu/abs/2002ApJ...573...66C} {573, 66}

\bibitem[\protect\citeauthoryear{{Chapman}, {Helou}, {Lewis}  \& {Dale}}{{Chapman} et~al.}{2003}]{2003ApJ...588..186C}
{Chapman} S.~C.,  {Helou} G.,  {Lewis} G.~F.,   {Dale} D.~A.,  2003, \mn@doi [\apj] {10.1086/37403810.48550/arXiv.astro-ph/0301233}, \href {https://ui.adsabs.harvard.edu/abs/2003ApJ...588..186C} {588, 186}

\bibitem[\protect\citeauthoryear{{Cortese}, {Boselli}, {Franzetti}, {Decarli}, {Gavazzi}, {Boissier}  \& {Buat}}{{Cortese} et~al.}{2008}]{2008MNRAS.386.1157C}
{Cortese} L.,  {Boselli} A.,  {Franzetti} P.,  {Decarli} R.,  {Gavazzi} G.,  {Boissier} S.,   {Buat} V.,  2008, \mn@doi [\mnras] {10.1111/j.1365-2966.2008.13118.x}, \href {https://ui.adsabs.harvard.edu/abs/2008MNRAS.386.1157C} {386, 1157}

\bibitem[\protect\citeauthoryear{{Cucciati} et~al.,}{{Cucciati} et~al.}{2012}]{refId0}
{Cucciati} et~al., 2012, \mn@doi [A\&A] {10.1051/0004-6361/201118010}, 539, A31

\bibitem[\protect\citeauthoryear{{Dale}, {Helou}, {Contursi}, {Silbermann}  \& {Kolhatkar}}{{Dale} et~al.}{2001}]{2001ApJ...549..215D}
{Dale} D.~A.,  {Helou} G.,  {Contursi} A.,  {Silbermann} N.~A.,   {Kolhatkar} S.,  2001, \mn@doi [\apj] {10.1086/319077}, \href {https://ui.adsabs.harvard.edu/abs/2001ApJ...549..215D} {549, 215}

\bibitem[\protect\citeauthoryear{{Dole} et~al.,}{{Dole} et~al.}{2006}]{2006A&A...451..417D}
{Dole} H.,  et~al., 2006, \mn@doi [\aap] {10.1051/0004-6361:20054446}, \href {https://ui.adsabs.harvard.edu/abs/2006A&A...451..417D} {451, 417}

\bibitem[\protect\citeauthoryear{{Donnan} et~al.,}{{Donnan} et~al.}{2023}]{2023MNRAS.518.6011D}
{Donnan} C.~T.,  et~al., 2023, \mn@doi [\mnras] {10.1093/mnras/stac3472}, \href {https://ui.adsabs.harvard.edu/abs/2023MNRAS.518.6011D} {518, 6011}

\bibitem[\protect\citeauthoryear{{Draine} \& {Li}}{{Draine} \& {Li}}{2007}]{2007ApJ...657..810D}
{Draine} B.~T.,  {Li} A.,  2007, \mn@doi [\apj] {10.1086/511055}, \href {https://ui.adsabs.harvard.edu/abs/2007ApJ...657..810D} {657, 810}

\bibitem[\protect\citeauthoryear{{Drew} \& {Casey}}{{Drew} \& {Casey}}{2022}]{2022ApJ...930..142D}
{Drew} P.~M.,  {Casey} C.~M.,  2022, \mn@doi [\apj] {10.3847/1538-4357/ac6270}, \href {https://ui.adsabs.harvard.edu/abs/2022ApJ...930..142D} {930, 142}

\bibitem[\protect\citeauthoryear{{Dunne}, {Eales}, {Edmunds}, {Ivison}, {Alexander}  \& {Clements}}{{Dunne} et~al.}{2000}]{2000MNRAS.315..115D}
{Dunne} L.,  {Eales} S.,  {Edmunds} M.,  {Ivison} R.,  {Alexander} P.,   {Clements} D.~L.,  2000, \mn@doi [\mnras] {10.1046/j.1365-8711.2000.03386.x10.48550/arXiv.astro-ph/0002234}, \href {https://ui.adsabs.harvard.edu/abs/2000MNRAS.315..115D} {315, 115}

\bibitem[\protect\citeauthoryear{{Finkelstein} et~al.,}{{Finkelstein} et~al.}{2012}]{2012ApJ...756..164F}
{Finkelstein} S.~L.,  et~al., 2012, \mn@doi [\apj] {10.1088/0004-637X/756/2/164}, \href {https://ui.adsabs.harvard.edu/abs/2012ApJ...756..164F} {756, 164}

\bibitem[\protect\citeauthoryear{{Franzen} et~al.,}{{Franzen} et~al.}{2015}]{2015MNRAS.453.4020F}
{Franzen} T.~M.~O.,  et~al., 2015, \mn@doi [\mnras] {10.1093/mnras/stv1866}, \href {https://ui.adsabs.harvard.edu/abs/2015MNRAS.453.4020F} {453, 4020}

\bibitem[\protect\citeauthoryear{{Fudamoto} et~al.,}{{Fudamoto} et~al.}{2020}]{2020MNRAS.491.4724F}
{Fudamoto} Y.,  et~al., 2020, \mn@doi [\mnras] {10.1093/mnras/stz3248}, \href {https://ui.adsabs.harvard.edu/abs/2020MNRAS.491.4724F} {491, 4724}

\bibitem[\protect\citeauthoryear{{Genzel} et~al.,}{{Genzel} et~al.}{1998}]{1998ApJ...498..579G}
{Genzel} R.,  et~al., 1998, \mn@doi [\apj] {10.1086/305576}, \href {https://ui.adsabs.harvard.edu/abs/1998ApJ...498..579G} {498, 579}

\bibitem[\protect\citeauthoryear{{Gordon}, {Clayton}, {Witt}  \& {Misselt}}{{Gordon} et~al.}{2000}]{2000ApJ...533..236G}
{Gordon} K.~D.,  {Clayton} G.~C.,  {Witt} A.~N.,   {Misselt} K.~A.,  2000, \mn@doi [\apj] {10.1086/308668}, \href {https://ui.adsabs.harvard.edu/abs/2000ApJ...533..236G} {533, 236}

\bibitem[\protect\citeauthoryear{{Griffin} et~al.,}{{Griffin} et~al.}{2010}]{2010A&A...518L...3G}
{Griffin} M.~J.,  et~al., 2010, \mn@doi [\aap] {10.1051/0004-6361/201014519}, \href {https://ui.adsabs.harvard.edu/abs/2010A&A...518L...3G} {518, L3}

\bibitem[\protect\citeauthoryear{{Gruppioni} et~al.,}{{Gruppioni} et~al.}{2013}]{2013MNRAS.432...23G}
{Gruppioni} C.,  et~al., 2013, \mn@doi [\mnras] {10.1093/mnras/stt308}, \href {https://ui.adsabs.harvard.edu/abs/2013MNRAS.432...23G} {432, 23}

\bibitem[\protect\citeauthoryear{{Hagen}, {Hoversten}, {Gronwall}, {Wolf}, {Siegel}, {Page}  \& {Hagen}}{{Hagen} et~al.}{2015}]{2015ApJ...808..178H}
{Hagen} L. M.~Z.,  {Hoversten} E.~A.,  {Gronwall} C.,  {Wolf} C.,  {Siegel} M.~H.,  {Page} M.,   {Hagen} A.,  2015, \mn@doi [\apj] {10.1088/0004-637X/808/2/178}, \href {https://ui.adsabs.harvard.edu/abs/2015ApJ...808..178H} {808, 178}

\bibitem[\protect\citeauthoryear{{Hao}, {Kennicutt}, {Johnson}, {Calzetti}, {Dale}  \& {Moustakas}}{{Hao} et~al.}{2011}]{2011ApJ...741..124H}
{Hao} C.-N.,  {Kennicutt} R.~C.,  {Johnson} B.~D.,  {Calzetti} D.,  {Dale} D.~A.,   {Moustakas} J.,  2011, \mn@doi [\apj] {10.1088/0004-637X/741/2/124}, \href {https://ui.adsabs.harvard.edu/abs/2011ApJ...741..124H} {741, 124}

\bibitem[\protect\citeauthoryear{{Heckman}, {Robert}, {Leitherer}, {Garnett}  \& {van der Rydt}}{{Heckman} et~al.}{1998}]{1998ApJ...503..646H}
{Heckman} T.~M.,  {Robert} C.,  {Leitherer} C.,  {Garnett} D.~R.,   {van der Rydt} F.,  1998, \mn@doi [\apj] {10.1086/306035}, \href {https://ui.adsabs.harvard.edu/abs/1998ApJ...503..646H} {503, 646}

\bibitem[\protect\citeauthoryear{{Heinis} et~al.,}{{Heinis} et~al.}{2013}]{2013MNRAS.429.1113H}
{Heinis} S.,  et~al., 2013, \mn@doi [\mnras] {10.1093/mnras/sts397}, \href {https://ui.adsabs.harvard.edu/abs/2013MNRAS.429.1113H} {429, 1113}

\bibitem[\protect\citeauthoryear{{Hopkins} \& {Beacom}}{{Hopkins} \& {Beacom}}{2006}]{2006ApJ...651..142H}
{Hopkins} A.~M.,  {Beacom} J.~F.,  2006, \mn@doi [\apj] {10.1086/506610}, \href {https://ui.adsabs.harvard.edu/abs/2006ApJ...651..142H} {651, 142}

\bibitem[\protect\citeauthoryear{{Hwang} et~al.,}{{Hwang} et~al.}{2010}]{2010MNRAS.409...75H}
{Hwang} H.~S.,  et~al., 2010, \mn@doi [\mnras] {10.1111/j.1365-2966.2010.17645.x}, \href {https://ui.adsabs.harvard.edu/abs/2010MNRAS.409...75H} {409, 75}

\bibitem[\protect\citeauthoryear{{Iglesias-P{\'a}ramo} et~al.,}{{Iglesias-P{\'a}ramo} et~al.}{2006}]{2006ApJS..164...38I}
{Iglesias-P{\'a}ramo} J.,  et~al., 2006, \mn@doi [\apjs] {10.1086/502628}, \href {https://ui.adsabs.harvard.edu/abs/2006ApJS..164...38I} {164, 38}

\bibitem[\protect\citeauthoryear{{Kennicutt}}{{Kennicutt}}{1998}]{1998ARA&A..36..189K}
{Kennicutt} Robert~C. J.,  1998, \mn@doi [\araa] {10.1146/annurev.astro.36.1.189}, \href {https://ui.adsabs.harvard.edu/abs/1998ARA&A..36..189K} {36, 189}

\bibitem[\protect\citeauthoryear{{Kennicutt} \& {Evans}}{{Kennicutt} \& {Evans}}{2012}]{2012ARA&A..50..531K}
{Kennicutt} R.~C.,  {Evans} N.~J.,  2012, \mn@doi [\araa] {10.1146/annurev-astro-081811-125610}, \href {https://ui.adsabs.harvard.edu/abs/2012ARA&A..50..531K} {50, 531}

\bibitem[\protect\citeauthoryear{{Kirkpatrick} et~al.,}{{Kirkpatrick} et~al.}{2012}]{2012ApJ...759..139K}
{Kirkpatrick} A.,  et~al., 2012, \mn@doi [\apj] {10.1088/0004-637X/759/2/139}, \href {https://ui.adsabs.harvard.edu/abs/2012ApJ...759..139K} {759, 139}

\bibitem[\protect\citeauthoryear{{Kirkpatrick} et~al.,}{{Kirkpatrick} et~al.}{2017}]{2017ApJ...843...71K}
{Kirkpatrick} A.,  et~al., 2017, \mn@doi [\apj] {10.3847/1538-4357/aa76dc}, \href {https://ui.adsabs.harvard.edu/abs/2017ApJ...843...71K} {843, 71}

\bibitem[\protect\citeauthoryear{{Kroupa}}{{Kroupa}}{2001}]{2001MNRAS.322..231K}
{Kroupa} P.,  2001, \mn@doi [\mnras] {10.1046/j.1365-8711.2001.04022.x}, \href {https://ui.adsabs.harvard.edu/abs/2001MNRAS.322..231K} {322, 231}

\bibitem[\protect\citeauthoryear{{Kurczynski} \& {Gawiser}}{{Kurczynski} \& {Gawiser}}{2010}]{2010AJ....139.1592K}
{Kurczynski} P.,  {Gawiser} E.,  2010, \mn@doi [\aj] {10.1088/0004-6256/139/4/1592}, \href {https://ui.adsabs.harvard.edu/abs/2010AJ....139.1592K} {139, 1592}

\bibitem[\protect\citeauthoryear{{Kurczynski} et~al.,}{{Kurczynski} et~al.}{2014}]{2014ApJ...793L...5K}
{Kurczynski} P.,  et~al., 2014, \mn@doi [\apjl] {10.1088/2041-8205/793/1/L5}, \href {https://ui.adsabs.harvard.edu/abs/2014ApJ...793L...5K} {793, L5}

\bibitem[\protect\citeauthoryear{{Le Floc'h} et~al.,}{{Le Floc'h} et~al.}{2005}]{2005ApJ...632..169L}
{Le Floc'h} E.,  et~al., 2005, \mn@doi [\apj] {10.1086/432789}, \href {https://ui.adsabs.harvard.edu/abs/2005ApJ...632..169L} {632, 169}

\bibitem[\protect\citeauthoryear{{Lee}, {Alberts}, {Atlee}, {Dey}, {Pope}, {Jannuzi}, {Reddy}  \& {Brown}}{{Lee} et~al.}{2012}]{2012ApJ...758L..31L}
{Lee} K.-S.,  {Alberts} S.,  {Atlee} D.,  {Dey} A.,  {Pope} A.,  {Jannuzi} B.~T.,  {Reddy} N.,   {Brown} M. J.~I.,  2012, \mn@doi [\apjl] {10.1088/2041-8205/758/2/L31}, \href {https://ui.adsabs.harvard.edu/abs/2012ApJ...758L..31L} {758, L31}

\bibitem[\protect\citeauthoryear{{Liang} et~al.,}{{Liang} et~al.}{2019}]{2019MNRAS.489.1397L}
{Liang} L.,  et~al., 2019, \mn@doi [\mnras] {10.1093/mnras/stz213410.48550/arXiv.1902.10727}, \href {https://ui.adsabs.harvard.edu/abs/2019MNRAS.489.1397L} {489, 1397}

\bibitem[\protect\citeauthoryear{{Liu} et~al.,}{{Liu} et~al.}{2018}]{2018ApJ...853..172L}
{Liu} D.,  et~al., 2018, \mn@doi [\apj] {10.3847/1538-4357/aaa600}, \href {https://ui.adsabs.harvard.edu/abs/2018ApJ...853..172L} {853, 172}

\bibitem[\protect\citeauthoryear{{Luo} et~al.,}{{Luo} et~al.}{2008}]{2008ApJS..179...19L}
{Luo} B.,  et~al., 2008, \mn@doi [\apjs] {10.1086/591248}, \href {https://ui.adsabs.harvard.edu/abs/2008ApJS..179...19L} {179, 19}

\bibitem[\protect\citeauthoryear{{Lutz} et~al.,}{{Lutz} et~al.}{2011}]{2011A&A...532A..90L}
{Lutz} D.,  et~al., 2011, \mn@doi [\aap] {10.1051/0004-6361/201117107}, \href {https://ui.adsabs.harvard.edu/abs/2011A&A...532A..90L} {532, A90}

\bibitem[\protect\citeauthoryear{{Madau} \& {Dickinson}}{{Madau} \& {Dickinson}}{2014}]{2014ARA&A..52..415M}
{Madau} P.,  {Dickinson} M.,  2014, \mn@doi [\araa] {10.1146/annurev-astro-081811-125615}, \href {https://ui.adsabs.harvard.edu/abs/2014ARA&A..52..415M} {52, 415}

\bibitem[\protect\citeauthoryear{{Magdis} et~al.,}{{Magdis} et~al.}{2012}]{2012ApJ...760....6M}
{Magdis} G.~E.,  et~al., 2012, \mn@doi [\apj] {10.1088/0004-637X/760/1/610.48550/arXiv.1210.1035}, \href {https://ui.adsabs.harvard.edu/abs/2012ApJ...760....6M} {760, 6}

\bibitem[\protect\citeauthoryear{{Magnelli}, {Elbaz}, {Chary}, {Dickinson}, {Le Borgne}, {Frayer}  \& {Willmer}}{{Magnelli} et~al.}{2011}]{2011A&A...528A..35M}
{Magnelli} B.,  {Elbaz} D.,  {Chary} R.~R.,  {Dickinson} M.,  {Le Borgne} D.,  {Frayer} D.~T.,   {Willmer} C.~N.~A.,  2011, \mn@doi [\aap] {10.1051/0004-6361/200913941}, \href {https://ui.adsabs.harvard.edu/abs/2011A&A...528A..35M} {528, A35}

\bibitem[\protect\citeauthoryear{{Magnelli} et~al.,}{{Magnelli} et~al.}{2013}]{2013A&A...553A.132M}
{Magnelli} B.,  et~al., 2013, \mn@doi [\aap] {10.1051/0004-6361/201321371}, \href {https://ui.adsabs.harvard.edu/abs/2013A&A...553A.132M} {553, A132}

\bibitem[\protect\citeauthoryear{{Magnelli} et~al.,}{{Magnelli} et~al.}{2014}]{2014A&A...561A..86M}
{Magnelli} B.,  et~al., 2014, \mn@doi [\aap] {10.1051/0004-6361/20132221710.48550/arXiv.1311.2956}, \href {https://ui.adsabs.harvard.edu/abs/2014A&A...561A..86M} {561, A86}

\bibitem[\protect\citeauthoryear{{Marsden} et~al.,}{{Marsden} et~al.}{2009}]{2009ApJ...707.1729M}
{Marsden} G.,  et~al., 2009, \mn@doi [\apj] {10.1088/0004-637X/707/2/1729}, \href {https://ui.adsabs.harvard.edu/abs/2009ApJ...707.1729M} {707, 1729}

\bibitem[\protect\citeauthoryear{{Mason} et~al.,}{{Mason} et~al.}{2001}]{2001A&A...365L..36M}
{Mason} K.~O.,  et~al., 2001, \mn@doi [\aap] {10.1051/0004-6361:20000044}, \href {https://ui.adsabs.harvard.edu/abs/2001A&A...365L..36M} {365, L36}

\bibitem[\protect\citeauthoryear{{Meurer}, {Heckman}, {Leitherer}, {Kinney}, {Robert}  \& {Garnett}}{{Meurer} et~al.}{1995}]{1995AJ....110.2665M}
{Meurer} G.~R.,  {Heckman} T.~M.,  {Leitherer} C.,  {Kinney} A.,  {Robert} C.,   {Garnett} D.~R.,  1995, \mn@doi [\aj] {10.1086/117721}, \href {https://ui.adsabs.harvard.edu/abs/1995AJ....110.2665M} {110, 2665}

\bibitem[\protect\citeauthoryear{{Meurer}, {Heckman}  \& {Calzetti}}{{Meurer} et~al.}{1999}]{1999ApJ...521...64M}
{Meurer} G.~R.,  {Heckman} T.~M.,   {Calzetti} D.,  1999, \mn@doi [\apj] {10.1086/307523}, \href {https://ui.adsabs.harvard.edu/abs/1999ApJ...521...64M} {521, 64}

\bibitem[\protect\citeauthoryear{{Miller} et~al.,}{{Miller} et~al.}{2013}]{2013ApJS..205...13M}
{Miller} N.~A.,  et~al., 2013, \mn@doi [\apjs] {10.1088/0067-0049/205/2/13}, \href {https://ui.adsabs.harvard.edu/abs/2013ApJS..205...13M} {205, 13}

\bibitem[\protect\citeauthoryear{{Moutard}, {Sawicki}, {Arnouts}, {Golob}, {Coupon}, {Ilbert}, {Yang}  \& {Gwyn}}{{Moutard} et~al.}{2020}]{2020MNRAS.494.1894M}
{Moutard} T.,  {Sawicki} M.,  {Arnouts} S.,  {Golob} A.,  {Coupon} J.,  {Ilbert} O.,  {Yang} X.,   {Gwyn} S.,  2020, \mn@doi [\mnras] {10.1093/mnras/staa706}, \href {https://ui.adsabs.harvard.edu/abs/2020MNRAS.494.1894M} {494, 1894}

\bibitem[\protect\citeauthoryear{{Murphy} et~al.,}{{Murphy} et~al.}{2011}]{2011ApJ...737...67M}
{Murphy} E.~J.,  et~al., 2011, \mn@doi [\apj] {10.1088/0004-637X/737/2/67}, \href {https://ui.adsabs.harvard.edu/abs/2011ApJ...737...67M} {737, 67}

\bibitem[\protect\citeauthoryear{{Narayanan}, {Dav{\'e}}, {Johnson}, {Thompson}, {Conroy}  \& {Geach}}{{Narayanan} et~al.}{2018}]{2018MNRAS.474.1718N}
{Narayanan} D.,  {Dav{\'e}} R.,  {Johnson} B.~D.,  {Thompson} R.,  {Conroy} C.,   {Geach} J.,  2018, \mn@doi [\mnras] {10.1093/mnras/stx2860}, \href {https://ui.adsabs.harvard.edu/abs/2018MNRAS.474.1718N} {474, 1718}

\bibitem[\protect\citeauthoryear{{Nguyen} et~al.,}{{Nguyen} et~al.}{2010}]{2010A&A...518L...5N}
{Nguyen} H.~T.,  et~al., 2010, \mn@doi [\aap] {10.1051/0004-6361/201014680}, \href {https://ui.adsabs.harvard.edu/abs/2010A&A...518L...5N} {518, L5}

\bibitem[\protect\citeauthoryear{{Nordon} et~al.,}{{Nordon} et~al.}{2013}]{2013ApJ...762..125N}
{Nordon} R.,  et~al., 2013, \mn@doi [\apj] {10.1088/0004-637X/762/2/125}, \href {https://ui.adsabs.harvard.edu/abs/2013ApJ...762..125N} {762, 125}

\bibitem[\protect\citeauthoryear{{Oesch} et~al.,}{{Oesch} et~al.}{2010}]{2010ApJ...725L.150O}
{Oesch} P.~A.,  et~al., 2010, \mn@doi [\apjl] {10.1088/2041-8205/725/2/L150}, \href {https://ui.adsabs.harvard.edu/abs/2010ApJ...725L.150O} {725, L150}

\bibitem[\protect\citeauthoryear{{Oesch} et~al.,}{{Oesch} et~al.}{2018}]{2018ApJS..237...12O}
{Oesch} P.~A.,  et~al., 2018, \mn@doi [\apjs] {10.3847/1538-4365/aacb30}, \href {https://ui.adsabs.harvard.edu/abs/2018ApJS..237...12O} {237, 12}

\bibitem[\protect\citeauthoryear{{Oke} \& {Gunn}}{{Oke} \& {Gunn}}{1983}]{1983ApJ...266..713O}
{Oke} J.~B.,  {Gunn} J.~E.,  1983, \mn@doi [\apj] {10.1086/160817}, \href {https://ui.adsabs.harvard.edu/abs/1983ApJ...266..713O} {266, 713}

\bibitem[\protect\citeauthoryear{{Oliver} et~al.,}{{Oliver} et~al.}{2012}]{2012MNRAS.424.1614O}
{Oliver} S.~J.,  et~al., 2012, \mn@doi [\mnras] {10.1111/j.1365-2966.2012.20912.x}, \href {https://ui.adsabs.harvard.edu/abs/2012MNRAS.424.1614O} {424, 1614}

\bibitem[\protect\citeauthoryear{Overzier et~al.,}{Overzier et~al.}{2010}]{Overzier_2010}
Overzier R.~A.,  et~al., 2010, \mn@doi [The Astrophysical Journal] {10.1088/2041-8205/726/1/l7}, 726, L7

\bibitem[\protect\citeauthoryear{{Page} et~al.,}{{Page} et~al.}{2021}]{2021MNRAS.506..473P}
{Page} M.~J.,  et~al., 2021, \mn@doi [\mnras] {10.1093/mnras/stab1638}, \href {https://ui.adsabs.harvard.edu/abs/2021MNRAS.506..473P} {506, 473}

\bibitem[\protect\citeauthoryear{{Parsa}, {Dunlop}, {McLure}  \& {Mortlock}}{{Parsa} et~al.}{2016}]{2016MNRAS.456.3194P}
{Parsa} S.,  {Dunlop} J.~S.,  {McLure} R.~J.,   {Mortlock} A.,  2016, \mn@doi [\mnras] {10.1093/mnras/stv2857}, \href {https://ui.adsabs.harvard.edu/abs/2016MNRAS.456.3194P} {456, 3194}

\bibitem[\protect\citeauthoryear{{Pilbratt} et~al.,}{{Pilbratt} et~al.}{2010}]{2010A&A...518L...1P}
{Pilbratt} G.~L.,  et~al., 2010, \mn@doi [\aap] {10.1051/0004-6361/201014759}, \href {https://ui.adsabs.harvard.edu/abs/2010A&A...518L...1P} {518, L1}

\bibitem[\protect\citeauthoryear{{Poglitsch} et~al.,}{{Poglitsch} et~al.}{2010}]{2010A&A...518L...2P}
{Poglitsch} A.,  et~al., 2010, \mn@doi [\aap] {10.1051/0004-6361/201014535}, \href {https://ui.adsabs.harvard.edu/abs/2010A&A...518L...2P} {518, L2}

\bibitem[\protect\citeauthoryear{{Reddy}, {Steidel}, {Erb}, {Shapley}  \& {Pettini}}{{Reddy} et~al.}{2006}]{2006ApJ...653.1004R}
{Reddy} N.~A.,  {Steidel} C.~C.,  {Erb} D.~K.,  {Shapley} A.~E.,   {Pettini} M.,  2006, \mn@doi [\apj] {10.1086/508851}, \href {https://ui.adsabs.harvard.edu/abs/2006ApJ...653.1004R} {653, 1004}

\bibitem[\protect\citeauthoryear{{Reddy}, {Erb}, {Pettini}, {Steidel}  \& {Shapley}}{{Reddy} et~al.}{2010}]{2010ApJ...712.1070R}
{Reddy} N.~A.,  {Erb} D.~K.,  {Pettini} M.,  {Steidel} C.~C.,   {Shapley} A.~E.,  2010, \mn@doi [\apj] {10.1088/0004-637X/712/2/1070}, \href {https://ui.adsabs.harvard.edu/abs/2010ApJ...712.1070R} {712, 1070}

\bibitem[\protect\citeauthoryear{{Reddy} et~al.,}{{Reddy} et~al.}{2012}]{2012ApJ...744..154R}
{Reddy} N.,  et~al., 2012, \mn@doi [\apj] {10.1088/0004-637X/744/2/154}, \href {https://ui.adsabs.harvard.edu/abs/2012ApJ...744..154R} {744, 154}

\bibitem[\protect\citeauthoryear{{Reddy} et~al.,}{{Reddy} et~al.}{2018}]{2018ApJ...853...56R}
{Reddy} N.~A.,  et~al., 2018, \mn@doi [\apj] {10.3847/1538-4357/aaa3e7}, \href {https://ui.adsabs.harvard.edu/abs/2018ApJ...853...56R} {853, 56}

\bibitem[\protect\citeauthoryear{{Rosati} et~al.,}{{Rosati} et~al.}{2002}]{2002ApJ...566..667R}
{Rosati} P.,  et~al., 2002, \mn@doi [\apj] {10.1086/338339}, \href {https://ui.adsabs.harvard.edu/abs/2002ApJ...566..667R} {566, 667}

\bibitem[\protect\citeauthoryear{{Roseboom} et~al.,}{{Roseboom} et~al.}{2012}]{2012MNRAS.419.2758R}
{Roseboom} I.~G.,  et~al., 2012, \mn@doi [\mnras] {10.1111/j.1365-2966.2011.19827.x}, \href {https://ui.adsabs.harvard.edu/abs/2012MNRAS.419.2758R} {419, 2758}

\bibitem[\protect\citeauthoryear{{Salpeter}}{{Salpeter}}{1955}]{1955ApJ...121..161S}
{Salpeter} E.~E.,  1955, \mn@doi [\apj] {10.1086/145971}, \href {https://ui.adsabs.harvard.edu/abs/1955ApJ...121..161S} {121, 161}

\bibitem[\protect\citeauthoryear{{Sanders} \& {Mirabel}}{{Sanders} \& {Mirabel}}{1996}]{1996ARA&A..34..749S}
{Sanders} D.~B.,  {Mirabel} I.~F.,  1996, \mn@doi [\araa] {10.1146/annurev.astro.34.1.749}, \href {https://ui.adsabs.harvard.edu/abs/1996ARA&A..34..749S} {34, 749}

\bibitem[\protect\citeauthoryear{{Schiminovich} et~al.,}{{Schiminovich} et~al.}{2005}]{2005ApJ...619L..47S}
{Schiminovich} D.,  et~al., 2005, \mn@doi [\apjl] {10.1086/427077}, \href {https://ui.adsabs.harvard.edu/abs/2005ApJ...619L..47S} {619, L47}

\bibitem[\protect\citeauthoryear{{Schreiber}, {Elbaz}, {Pannella}, {Ciesla}, {Wang}  \& {Franco}}{{Schreiber} et~al.}{2018}]{2018A&A...609A..30S}
{Schreiber} C.,  {Elbaz} D.,  {Pannella} M.,  {Ciesla} L.,  {Wang} T.,   {Franco} M.,  2018, \mn@doi [\aap] {10.1051/0004-6361/20173150610.48550/arXiv.1710.10276}, \href {https://ui.adsabs.harvard.edu/abs/2018A&A...609A..30S} {609, A30}

\bibitem[\protect\citeauthoryear{{Seibert} et~al.,}{{Seibert} et~al.}{2005}]{2005ApJ...619L..55S}
{Seibert} M.,  et~al., 2005, \mn@doi [\apjl] {10.1086/427843}, \href {https://ui.adsabs.harvard.edu/abs/2005ApJ...619L..55S} {619, L55}

\bibitem[\protect\citeauthoryear{{Sharma}, {Page}  \& {Breeveld}}{{Sharma} et~al.}{2022}]{2022MNRAS.511.4882S}
{Sharma} M.,  {Page} M.~J.,   {Breeveld} A.~A.,  2022, \mn@doi [\mnras] {10.1093/mnras/stac356}, \href {https://ui.adsabs.harvard.edu/abs/2022MNRAS.511.4882S} {511, 4882}

\bibitem[\protect\citeauthoryear{Sharma, Page, Ferreras  \& Breeveld}{Sharma et~al.}{2023}]{sharma2023brightend}
Sharma M.,  Page M.~J.,  Ferreras I.,   Breeveld A.~A.,  2023, {} (\mn@eprint {arXiv} {2212.00215})

\bibitem[\protect\citeauthoryear{{Skelton} et~al.,}{{Skelton} et~al.}{2014}]{2014ApJS..214...24S}
{Skelton} R.~E.,  et~al., 2014, \mn@doi [\apjs] {10.1088/0067-0049/214/2/24}, \href {https://ui.adsabs.harvard.edu/abs/2014ApJS..214...24S} {214, 24}

\bibitem[\protect\citeauthoryear{{Soifer}, {Neugebauer}  \& {Houck}}{{Soifer} et~al.}{1987}]{1987ARA&A..25..187S}
{Soifer} B.~T.,  {Neugebauer} G.,   {Houck} J.~R.,  1987, \mn@doi [\araa] {10.1146/annurev.aa.25.090187.001155}, \href {https://ui.adsabs.harvard.edu/abs/1987ARA&A..25..187S} {25, 187}

\bibitem[\protect\citeauthoryear{Steidel, Shapley, Pettini, Adelberger, Erb, Reddy  \& Hunt}{Steidel et~al.}{2004}]{Steidel_2004}
Steidel C.~C.,  Shapley A.~E.,  Pettini M.,  Adelberger K.~L.,  Erb D.~K.,  Reddy N.~A.,   Hunt M.~P.,  2004, \mn@doi [The Astrophysical Journal] {10.1086/381960}, 604, 534

\bibitem[\protect\citeauthoryear{{Sullivan}, {Treyer}, {Ellis}, {Bridges}, {Milliard}  \& {Donas}}{{Sullivan} et~al.}{2000}]{2000MNRAS.312..442S}
{Sullivan} M.,  {Treyer} M.~A.,  {Ellis} R.~S.,  {Bridges} T.~J.,  {Milliard} B.,   {Donas} J.,  2000, \mn@doi [\mnras] {10.1046/j.1365-8711.2000.03140.x}, \href {https://ui.adsabs.harvard.edu/abs/2000MNRAS.312..442S} {312, 442}

\bibitem[\protect\citeauthoryear{{Swinbank} et~al.,}{{Swinbank} et~al.}{2014}]{2014MNRAS.438.1267S}
{Swinbank} A.~M.,  et~al., 2014, \mn@doi [\mnras] {10.1093/mnras/stt2273}, \href {https://ui.adsabs.harvard.edu/abs/2014MNRAS.438.1267S} {438, 1267}

\bibitem[\protect\citeauthoryear{{Symeonidis} \& {Page}}{{Symeonidis} \& {Page}}{2019}]{2019MNRAS.485L..11S}
{Symeonidis} M.,  {Page} M.~J.,  2019, \mn@doi [\mnras] {10.1093/mnrasl/slz022}, \href {https://ui.adsabs.harvard.edu/abs/2019MNRAS.485L..11S} {485, L11}

\bibitem[\protect\citeauthoryear{{Symeonidis} \& {Page}}{{Symeonidis} \& {Page}}{2021}]{2021MNRAS.503.3992S}
{Symeonidis} M.,  {Page} M.~J.,  2021, \mn@doi [\mnras] {10.1093/mnras/stab598}, \href {https://ui.adsabs.harvard.edu/abs/2021MNRAS.503.3992S} {503, 3992}

\bibitem[\protect\citeauthoryear{{Symeonidis}, {Page}, {Seymour}, {Dwelly}, {Coppin}, {McHardy}, {Rieke}  \& {Huynh}}{{Symeonidis} et~al.}{2009}]{2009MNRAS.397.1728S}
{Symeonidis} M.,  {Page} M.~J.,  {Seymour} N.,  {Dwelly} T.,  {Coppin} K.,  {McHardy} I.,  {Rieke} G.~H.,   {Huynh} M.,  2009, \mn@doi [\mnras] {10.1111/j.1365-2966.2009.15040.x}, \href {https://ui.adsabs.harvard.edu/abs/2009MNRAS.397.1728S} {397, 1728}

\bibitem[\protect\citeauthoryear{{Symeonidis} et~al.,}{{Symeonidis} et~al.}{2013}]{2013MNRAS.431.2317S}
{Symeonidis} M.,  et~al., 2013, \mn@doi [\mnras] {10.1093/mnras/stt330}, \href {https://ui.adsabs.harvard.edu/abs/2013MNRAS.431.2317S} {431, 2317}

\bibitem[\protect\citeauthoryear{{Takeuchi}, {Buat}  \& {Burgarella}}{{Takeuchi} et~al.}{2005}]{2005A&A...440L..17T}
{Takeuchi} T.~T.,  {Buat} V.,   {Burgarella} D.,  2005, \mn@doi [\aap] {10.1051/0004-6361:200500158}, \href {https://ui.adsabs.harvard.edu/abs/2005A&A...440L..17T} {440, L17}

\bibitem[\protect\citeauthoryear{{Thomson} et~al.,}{{Thomson} et~al.}{2017}]{2017ApJ...838..119T}
{Thomson} A.~P.,  et~al., 2017, \mn@doi [\apj] {10.3847/1538-4357/aa61a6}, \href {https://ui.adsabs.harvard.edu/abs/2017ApJ...838..119T} {838, 119}

\bibitem[\protect\citeauthoryear{{Tress} et~al.,}{{Tress} et~al.}{2018}]{2018MNRAS.475.2363T}
{Tress} M.,  et~al., 2018, \mn@doi [\mnras] {10.1093/mnras/stx3334}, \href {https://ui.adsabs.harvard.edu/abs/2018MNRAS.475.2363T} {475, 2363}

\bibitem[\protect\citeauthoryear{{Viero} et~al.,}{{Viero} et~al.}{2012}]{2012MNRAS.421.2161V}
{Viero} M.~P.,  et~al., 2012, \mn@doi [\mnras] {10.1111/j.1365-2966.2012.20456.x}, \href {https://ui.adsabs.harvard.edu/abs/2012MNRAS.421.2161V} {421, 2161}

\bibitem[\protect\citeauthoryear{{Viero} et~al.,}{{Viero} et~al.}{2013}]{2013ApJ...779...32V}
{Viero} M.~P.,  et~al., 2013, \mn@doi [\apj] {10.1088/0004-637X/779/1/32}, \href {https://ui.adsabs.harvard.edu/abs/2013ApJ...779...32V} {779, 32}

\bibitem[\protect\citeauthoryear{{Wilkins}, {Bunker}, {Stanway}, {Lorenzoni}  \& {Caruana}}{{Wilkins} et~al.}{2011}]{2011MNRAS.417..717W}
{Wilkins} S.~M.,  {Bunker} A.~J.,  {Stanway} E.,  {Lorenzoni} S.,   {Caruana} J.,  2011, \mn@doi [\mnras] {10.1111/j.1365-2966.2011.19315.x}, \href {https://ui.adsabs.harvard.edu/abs/2011MNRAS.417..717W} {417, 717}

\bibitem[\protect\citeauthoryear{{Wilkins}, {Gonzalez-Perez}, {Lacey}  \& {Baugh}}{{Wilkins} et~al.}{2012a}]{2012MNRAS.424.1522W}
{Wilkins} S.~M.,  {Gonzalez-Perez} V.,  {Lacey} C.~G.,   {Baugh} C.~M.,  2012a, \mn@doi [\mnras] {10.1111/j.1365-2966.2012.21344.x}, \href {https://ui.adsabs.harvard.edu/abs/2012MNRAS.424.1522W} {424, 1522}

\bibitem[\protect\citeauthoryear{{Wilkins}, {Gonzalez-Perez}, {Lacey}  \& {Baugh}}{{Wilkins} et~al.}{2012b}]{2012MNRAS.427.1490W}
{Wilkins} S.~M.,  {Gonzalez-Perez} V.,  {Lacey} C.~G.,   {Baugh} C.~M.,  2012b, \mn@doi [\mnras] {10.1111/j.1365-2966.2012.22092.x}, \href {https://ui.adsabs.harvard.edu/abs/2012MNRAS.427.1490W} {427, 1490}

\bibitem[\protect\citeauthoryear{{Wilkins}, {Bunker}, {Coulton}, {Croft}, {di Matteo}, {Khandai}  \& {Feng}}{{Wilkins} et~al.}{2013}]{2013MNRAS.430.2885W}
{Wilkins} S.~M.,  {Bunker} A.,  {Coulton} W.,  {Croft} R.,  {di Matteo} T.,  {Khandai} N.,   {Feng} Y.,  2013, \mn@doi [\mnras] {10.1093/mnras/stt096}, \href {https://ui.adsabs.harvard.edu/abs/2013MNRAS.430.2885W} {430, 2885}

\bibitem[\protect\citeauthoryear{{Witt} \& {Gordon}}{{Witt} \& {Gordon}}{2000}]{2000ApJ...528..799W}
{Witt} A.~N.,  {Gordon} K.~D.,  2000, \mn@doi [\apj] {10.1086/308197}, \href {https://ui.adsabs.harvard.edu/abs/2000ApJ...528..799W} {528, 799}

\bibitem[\protect\citeauthoryear{{Wyder} et~al.,}{{Wyder} et~al.}{2005}]{2005ApJ...619L..15W}
{Wyder} T.~K.,  et~al., 2005, \mn@doi [\apjl] {10.1086/424735}, \href {https://ui.adsabs.harvard.edu/abs/2005ApJ...619L..15W} {619, L15}

\bibitem[\protect\citeauthoryear{{Xu} \& {Buat}}{{Xu} \& {Buat}}{1995}]{1995A&A...293L..65X}
{Xu} C.,  {Buat} V.,  1995, \aap, \href {https://ui.adsabs.harvard.edu/abs/1995A&A...293L..65X} {293, L65}

\bibitem[\protect\citeauthoryear{{Xu} et~al.,}{{Xu} et~al.}{2007}]{2007ApJS..173..432X}
{Xu} C.~K.,  et~al., 2007, \mn@doi [\apjs] {10.1086/516641}, \href {https://ui.adsabs.harvard.edu/abs/2007ApJS..173..432X} {173, 432}

\bibitem[\protect\citeauthoryear{{Zheng}, {Dole}, {Bell}, {Le Floc'h}, {Rieke}, {Rix}  \& {Schiminovich}}{{Zheng} et~al.}{2007}]{2007ApJ...670..301Z}
{Zheng} X.~Z.,  {Dole} H.,  {Bell} E.~F.,  {Le Floc'h} E.,  {Rieke} G.~H.,  {Rix} H.-W.,   {Schiminovich} D.,  2007, \mn@doi [\apj] {10.1086/520529}, \href {https://ui.adsabs.harvard.edu/abs/2007ApJ...670..301Z} {670, 301}

\makeatother
\end{thebibliography}

\appendix

\section{Deblended maps and stacks}
\label{sec:deb_maps_stacks}
\begin{figure*}
    \centering
    \hspace*{-0.4cm}\includegraphics[width=1.06\textwidth]{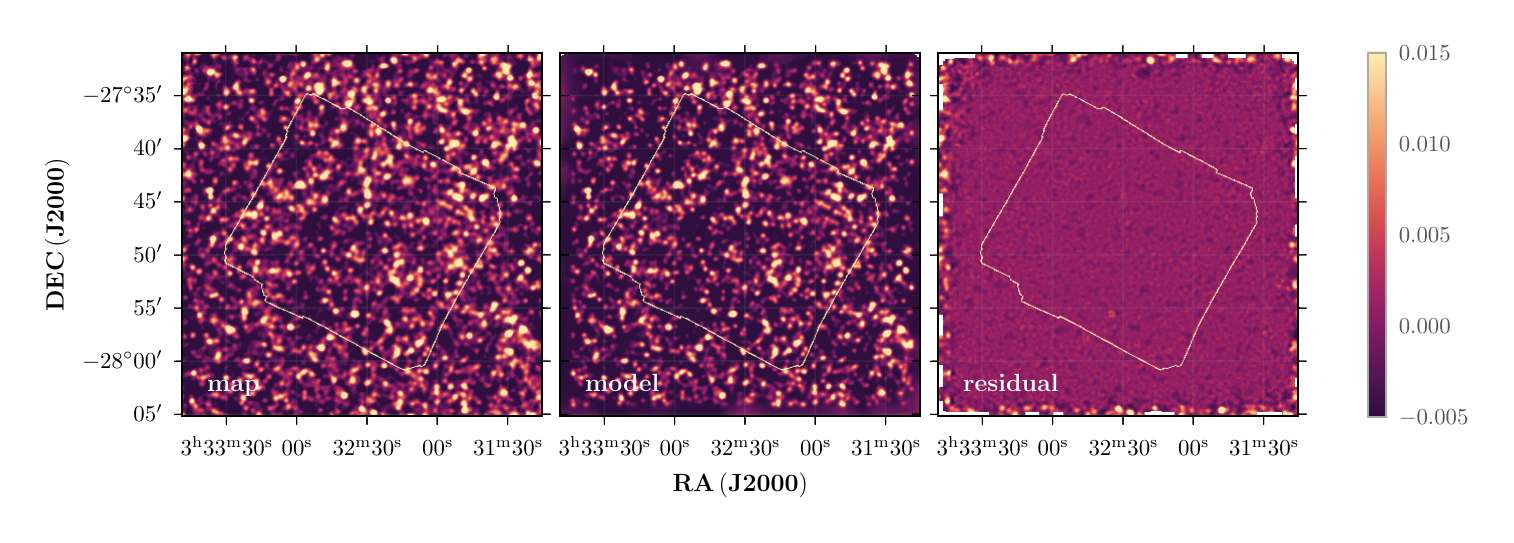}
    \hspace*{-0.4cm}\includegraphics[width=1.06\textwidth]{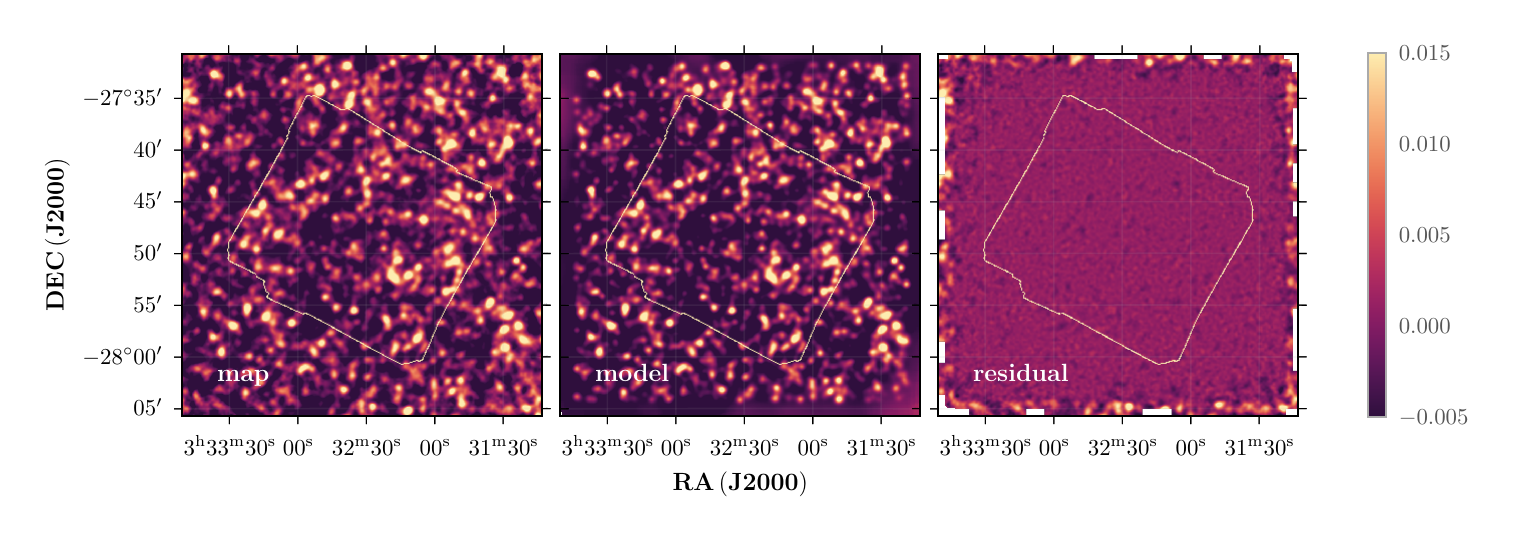}
    \caption{The deblending process as described in Section \ref{sec:3} for the 350 and 500 $\mu \mathrm{m}$ maps from \textit{Herschel} SPIRE. The arrangement of the panels is the same as in Figure \ref{fig:deb250}.}
    \label{fig:ap1}
\end{figure*}

Here we show the deblending process for 350 and 500 $\mu$m maps from \textit{Herschel} SPIRE in
Figure \ref{fig:ap1}.
These maps are produced as explained in Section \ref{sec:3}.

\bsp
\label{lastpage}
\end{document}